\documentclass[11pt,hidelinks]{article}
\usepackage[body={17.5cm, 22cm},right=2cm]{geometry}
\usepackage{color}
\usepackage{xcolor}
\usepackage{amssymb,graphicx}
\usepackage{amsmath}
\usepackage{epsfig}
\usepackage[bookmarks=true,pdfborder={0 0 0}]{hyperref} 
\usepackage{cite}
\pdfoutput=1
\usepackage{subcaption}
\usepackage{xcolor}
\usepackage{pifont}

\usepackage[utf8]{inputenc}
\usepackage{verbatim}   
\usepackage{amsfonts}
\usepackage{amssymb}
\usepackage{graphicx}
\usepackage{slashed}
\usepackage[normalem]{ulem}

 \usepackage[multiple]{footmisc}
\usepackage{enumitem}

\usepackage{orcidlink}
 \usepackage{tikz}
 \usetikzlibrary{arrows}
\usetikzlibrary{decorations.pathmorphing,decorations.markings,trees,shapes}

\newcommand{\yr}{{\, {\rm yr}}}
\newcommand{\Gyr}{{\, {\rm Gyr}}}

\newcommand{\km}{{\, {\rm km}}}

\newcommand{\Mpc}{{\, {\rm Mpc}}}

\newcommand{\AU}{{\, {\rm AU}}}
\newcommand{\eV}{{\, {\rm eV}}}
\newcommand{\keV}{{\, {\rm keV}}}

\newcommand{\GeV}{{\, {\rm GeV}}}

\newcommand{\s}{{\, {\rm s}}}

\def\beq{\begin{equation}}
\def\eeq{\end{equation}}
\def\bea{\begin{eqnarray}}
\def\eea{\end{eqnarray}}
\def\bitem{\begin{itemize}}
	\def\eitem{\end{itemize}}
\newcommand{\bec}{\begin{center}}
	\newcommand{\eec}{\end{center}}
\newcommand{\ba}{\begin{array}}
	\newcommand{\ea}{\end{array}}

\def\bar#1{\overline{#1}}

\def\bra#1{\left\langle #1\right|}
\def\ket#1{\left| #1\right\rangle}

\def\inv{^{\raise.15ex\hbox{${\scriptscriptstyle -}$}\kern-.05em 1}}
\def\lbar{{\lower.35ex\hbox{$\mathchar'26$}\mkern-10mu\lambda}} %

\def\to{\rightarrow}

\let\<=\langle
\let\>=\rangle

\let\+=\uparrow

\def\Mpl{M_{\rm Pl}}

\begin{document}

\begin{titlepage}
\noindent \makebox[15.5cm][l]{\small \hspace*{-.2cm} }{\small 
IPMU23-0022}  \\  [-1mm]

	~\vspace{1cm}
	\begin{center}

{\LARGE \bf A Generic Formation Mechanism of Ultralight
\\ \vspace{0.1mm}
			 {\LARGE   Dark Matter Solar Halos }
	\\
	\vspace{3.1mm}
		}

		\vspace{0.7cm}

		{\large
            Dmitry Budker\,\orcidlink{0000-0002-7356-4814},$^{a,b,c}$\,	      Joshua~Eby\,\orcidlink{0000-0003-0562-9177},$^d$\,		Marco~Gorghetto\,\orcidlink{0000-0002-5479-485X},$^e$\,
	Minyuan~Jiang\,\orcidlink{0000-0002-7564-3145},$^e$\,
			 Gilad~Perez\,\orcidlink{0000-0002-3878-1821}$\,{}^e$}
		
		\vspace{.6cm}

        {\normalsize { \sl $^{a}$}
                Johannes Gutenberg-Universit\"at Mainz, 55128 Mainz, Germany}
                
		\vspace{.2cm}
        {\normalsize { \sl $^{b}$}
                Helmholtz-Institut, GSI Helmholtzzentrum f\"ur Schwerionenforschung, 55128 Mainz, Germany}
    
        \vspace{.2cm}
		{\normalsize { \sl $^{c}$ 
				Department of Physics, University of California at Berkeley, Berkeley, California 94720-7300, USA}}
                
		\vspace{.2cm}
		{\normalsize { \sl $^{d}$ 
				Kavli Institute for the Physics and Mathematics of the Universe (WPI), The University of Tokyo Institutes for Advanced Study, The University of Tokyo, Kashiwa, Chiba 277-8583, Japan}}
		
		\vspace{.2cm}
		{\normalsize { \sl $^{e}$ Department of Particle Physics and Astrophysics, Weizmann Institute of Science,\\
				Herzl St 234, Rehovot 761001, Israel}}

	\end{center}
	\vspace{1cm}

\begin{abstract}

As-yet undiscovered light bosons may constitute all or part of the dark matter (DM) of our Universe, and are expected to have (weak) self-interactions. We show that the quartic self-interactions generically induce the capture of dark matter from the surrounding halo by external gravitational potentials such as those of stars, including the Sun. This leads to the subsequent formation of dark matter bound states supported by such external potentials, resembling gravitational atoms (e.g. a \emph{solar halo} around our own Sun). 
 Their growth is governed by the ratio $\xi_{\rm foc} \equiv \lambda_{\rm dB}/R_\star$ between the de Broglie wavelength of the incoming DM waves, $\lambda_{\rm dB}$, and the radius of the ground state $R_\star$.
For $\xi_{\rm foc}\lesssim 1$, the gravitational atom grows to an (underdense) steady state that balances the capture of particles and the inverse (stripping) process. For  $\xi_{\rm foc}\gtrsim 1$, a significant gravitational-focusing effect leads to exponential accumulation of mass from the galactic DM halo into the gravitational atom. 
 For instance, a dark matter axion with mass of the order of $10^{-14}$\,eV and decay constant between $10^{7}$ and $10^8$\,GeV would  form a dense halo around the Sun on a timescale comparable to the lifetime of the Solar System, leading to a local DM density at the position of the Earth $\mathcal{O}(10^4)$ times larger than that expected in the standard halo model. For attractive self-interactions, after its formation, the gravitational atom is destabilized at a large density, which leads to its collapse; this is likely to be accompanied by emission of relativistic bosons (a `Bosenova').

	\end{abstract}

\end{titlepage}

	\thispagestyle{empty}

{\fontsize{12.5}{11.5}
\tableofcontents
}

\newpage

\section{Introduction}

As-yet undiscovered ultralight spin-0 (scalar or pseudoscalar) fields are well-motivated candidates to explain the dark matter (DM) of our Universe. These fields can address open questions of the Standard Model (SM) such as the strong-CP~\cite{Peccei:1977hh,Weinberg:1977ma,Wilczek:1977pj,Kim:1979if,Shifman:1979if,Dine:1981rt,Zhitnitsky:1980tq} and hierarchy problems \cite{Graham:2015cka,Choi:2016kke,Flacke:2016szy,Banerjee:2018xmn}, and are generic predictions of String Theory %
\cite{Svrcek:2006yi,Arvanitaki:2009fg}. Additionally, %
they are automatically produced as cold dark matter relics in the early Universe via the misalignment mechanism \cite{Preskill:1982cy,Abbott:1982af,Dine:1982ah}.

We denote these fields as ultralight if their mass $m$ is smaller than about $1\eV$. This implies that their occupation number in galaxies such as the Milky Way is macroscopic. As a result, rather than behaving as a collection of individual particles, their evolution is better approximated by their %
classical equations of motion (EoM), the Schr\"odinger--Poisson equations, whose free  solutions are scalar waves. See \cite{Ferreira:2020fam} for a recent review of ultralight DM.

The typical parameters assumed in experimental searches for DM, derived from the large-scale properties of the Milky Way halo, are an energy density $\rho_{\rm dm}\simeq 0.4$\,GeV/cm$^3$ and velocity ${v}_{\rm dm}\simeq 240$\,km/s (see, for example, \cite{Weber_2010,Nesti:2012zp,Bovy:2012tw,Read:2014qva,Evans:2018bqy,Necib:2018igl}). %
For ultralight DM (ULDM), the coherence time and other properties related to the ULDM stochasticity are also important for the theoretical interpretation of these searches~\cite{Foster:2017hbq,Centers:2019dyn,Lisanti:2021vij}.

However, overdensities at scales much smaller than the galaxy can 
modify this picture. 
In the literature, two possible scenarios have been considered thus far. First, a large fraction of the ULDM could reside in overdense objects supported by their own gravitational interactions \cite{Kaup:1968zz,Ruffini:1969qy,Kolb:1993zz,Schive:2014dra,Levkov:2018kau,Eggemeier:2019jsu,Veltmaat:2019hou,Chen:2020cef} (see~\cite{Colpi:1986ye,Chavanis:2011zi,Chavanis:2011zm,Eby:2014fya} for discussion of non-gravitational self-interactions).\footnote{Note also Refs.\,\cite{Brito:2015pxa,Minamitsuji:2018kof,Amin:2019ums,Jain:2021pnk,Gorghetto:2022sue} where the case of vector fields is discussed.} %
However, unless the DM-particle mass is close to the eV scale,
these objects have a
small encounter rate with the Earth over the typical experimental timescale of a year, and/or a negligible overdensity, and thus unlikely to be detected via terrestrial experiments~\cite{Banerjee:2019epw}.

In the second scenario, the ULDM is trapped in the gravitational field of a massive astrophysical object, such as a star or a planet, resulting in a halo surrounding the object. The density of such a `solar halo' or `Earth halo', which resembles a `gravitational atom' with the Sun or Earth as its center,\footnote{The atom analogy is especially good because, in both cases, the underlying potential is proportional to $1/r$.} 
could be many orders of magnitudes larger than what is conventionally inferred from the properties of the galactic halo. This can potentially lead to a dramatic change in the way we search for DM~\cite{Banerjee:2019epw,Banerjee:2019xuy}. 

However, the question of how such a halo could have formed remained unanswered.
The obvious challenge stems from the fact that the typical relative velocity between the DM in our galaxy and the Solar System is of the order of a few hundred km/s which, for instance, is somewhat larger than %
the escape velocity from the Sun at e.g. an AU distance. Therefore, an efficient mechanism of energy loss is required for the DM particles to be captured by the solar gravitational potential.

In this work we show that there is in fact a class of scalar/axion theories where, in the background of the gravitational potential of astrophysical sources, DM capture becomes efficient. 
Our basic observation is that there is a regime of masses and quartic self-interactions where processes mediated by such interactions induce the capture of one of two ULDM waves that scatter with each other. The quantity that determines 
the efficiency of the above capture processes is the ratio between the de Broglie  wavelength of an incoming ULDM wave%
, $\lambda_{\rm dB}$,  and the gravitational Bohr radius, %
 $R_\star$, characterizing the size of the halo:%
\begin{equation}
\xi_{\rm foc}\equiv \frac{\lambda_{\rm dB}}{R_\star}\,. \label{xifoc}
\end{equation}
When $\xi_{\rm foc}$ is larger than unity, the density of the DM waves is increased around and %
behind the astrophysical object -- an effect known as gravitational focusing%
, leading to an enhanced capture rate (see the next Section for more details, and~\cite{Kryemadhi:2022vuk,Hoffmann:2003zz,Alenazi:2006wu,Patla:2013vza,Prezeau:2015lxa,Sofue:2020mda,Lee:2013wza,Kim:2021yyo} for other consequences of gravitational focusing). %
In the case of attractive self-interactions%
, as  for axions (see e.g. \cite{Crewther:1979pi,GrillidiCortona:2015jxo}), we identify three phases in the evolution of such a halo, with typical timescale of the order of the so-called \emph{relaxation time}: \emph{(i)} slow growth from DM capture, \emph{(ii)} exponential capture phase, and, after the density has grown beyond some critical value (defined below), \emph{(iii)} collapse leading to a rapid emission of relativistic 
scalars, known as a \emph{Bosenova}.

The paper is organized as follows. In the next Section we discuss the basic dynamics giving rise to the ULDM halo and summarize the main results of the paper. After defining the bound states in Section~\ref{sec:boson_stars_waves}%
, we discuss their formation via DM capture in Section~\ref{sec:analytic}, and check and refine these results with numerical simulations in Section~\ref{sec:simulation}. Some phenomenological implications of %
a gravitational atom in the Solar System are discussed in Section~\ref{sec:solarsoliton}, and constraints on the relevant values of mass and self-interaction couplings (mostly from structure formation) are discussed in Section~\ref{sec:lss}. We conclude in Section~\ref{sec:conclusion}.

\section{Basic mechanism}
\label{sec:basic}
To understand the formation of the ULDM halo, we consider an astrophysical body of mass $M$, say our Sun, moving in the background of virialized DM. In the rest frame of the body and far from it, the DM has a velocity distribution with mean $\mathbf{v}_{\rm dm}$ and variance of order $\sigma^2\simeq v^2_{\rm dm}/2$. %
The equation of motion (EoM) of the ULDM field is the Schr\"odinger equation with an external gravitational potential $\propto 1/r$. 
As a result, the stationary solutions include bound states corresponding to a gravitational atom%
, with ground state %
radius $R_\star \equiv (m\alpha)^{-1}$, where $\alpha\equiv GMm$ is the gravitational coupling of the DM to the body, and is the analogue of the fine-structure constant. %
Explicitly, the radius is %
\begin{equation}\label{eq:Rstar_sun}
R_\star=1\,{\rm AU}\left[\frac{1.3 \cdot 10^{-14}\eV}{m}\right]^2\left[\frac{M_\odot}{M}\right] \, , 
\end{equation}
where $M_\odot$ is the solar mass. Therefore, $R_\star$ is fully determined by $m$ and $M$.%
\begin{figure}%
\begin{center}
\includegraphics[width=0.57\textwidth]{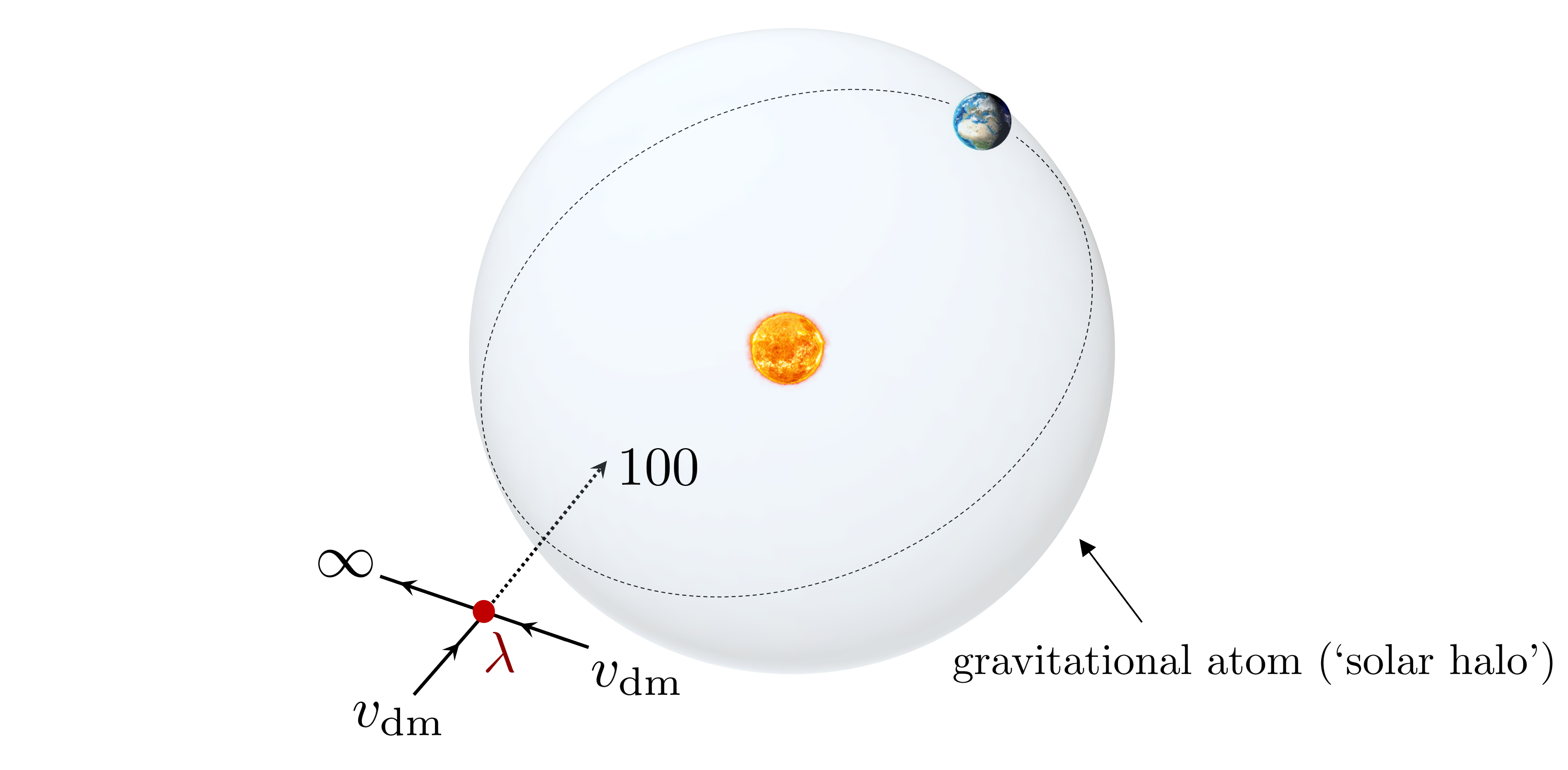} 
	\vspace{2mm}
	\caption{\small{The formation of a gravitational atom bound to the Sun, i.e. a solar halo. The dark matter in our galactic halo, traveling with average velocity $v_{\rm dm}\simeq 240\,\km/\s$  with respect to the Sun (and dispersion $\sigma\simeq v_{\rm dm}/\sqrt{2}$), experiences $2\to2$ scattering processes mediated by the quartic self-coupling $\lambda$ that decrease (or increase) the velocity of the particles. When this occurs in the gravitational potential of the Sun, the dark matter can become trapped into a bound configuration, which resembles the ground state ($nlm=100$) of the hydrogen atom. Excited states ($n>1$) are also populated but are likely to decay into the ground state, see Appendix\,\ref{app:twolevel}. }
	}
	\label{fig:soliton-sun}
\end{center}
	\vspace{-3mm}
\end{figure} 

These initial conditions can be thought of as all of the ULDM occupying the continuum (unbound states of the atom) at early times.  Despite being initially empty, bound states do become populated from the DM in the continuum via processes mediated by the quartic self-interactions, which we parameterize as
\begin{equation}\label{eq:selfint}
V\supset \frac{\lambda}{4!}\phi^4 \, ,
\end{equation}
where $\phi$ is the ULDM field and $\lambda$ the quartic coupling. Indeed, by treating Eq.\,\eqref{eq:selfint} as a perturbation in the EoM, in Section~\ref{sec:analytic} we show that the total mass $M_\star$ of ULDM bound to the Sun changes in time as 
\begin{equation}\label{eq:Mstarintro}
    \dot{M}_{\star}=C+ (\Gamma_1-\Gamma_2) {M}_{\star}\, ,
\end{equation}
where $C$, $\Gamma_1$, and $\Gamma_2$ are positive constants (that can be computed semi-analytically) resulting from the 
self-interactions, and at leading order proportional to $\lambda^2$%
. The first term in right hand side of Eq.\,\eqref{eq:Mstarintro}, here referred to as `capture', contributes positively to the bound mass and can be interpreted as arising from the $2\to2$ process in Figure~\ref{fig:soliton-sun}, where two unbound particles scatter into a bound particle and a (more energetic) unbound one, which escapes to infinity.
The second, `stimulated capture', is a consequence of the Bose enhancement of the indistinguishable bosons and represents the same process,
but is proportional to $M_\star$ itself and so is effective only when the bound state is already populated.
The last term is negative and simply represents the depletion of the bound states via the inverse process, `stripping'. Consequently, the change in $M_\star$ can be understood as the net effect of capture and stripping.%

\begin{figure}%
\begin{center}
    \includegraphics[width=0.3\textwidth]{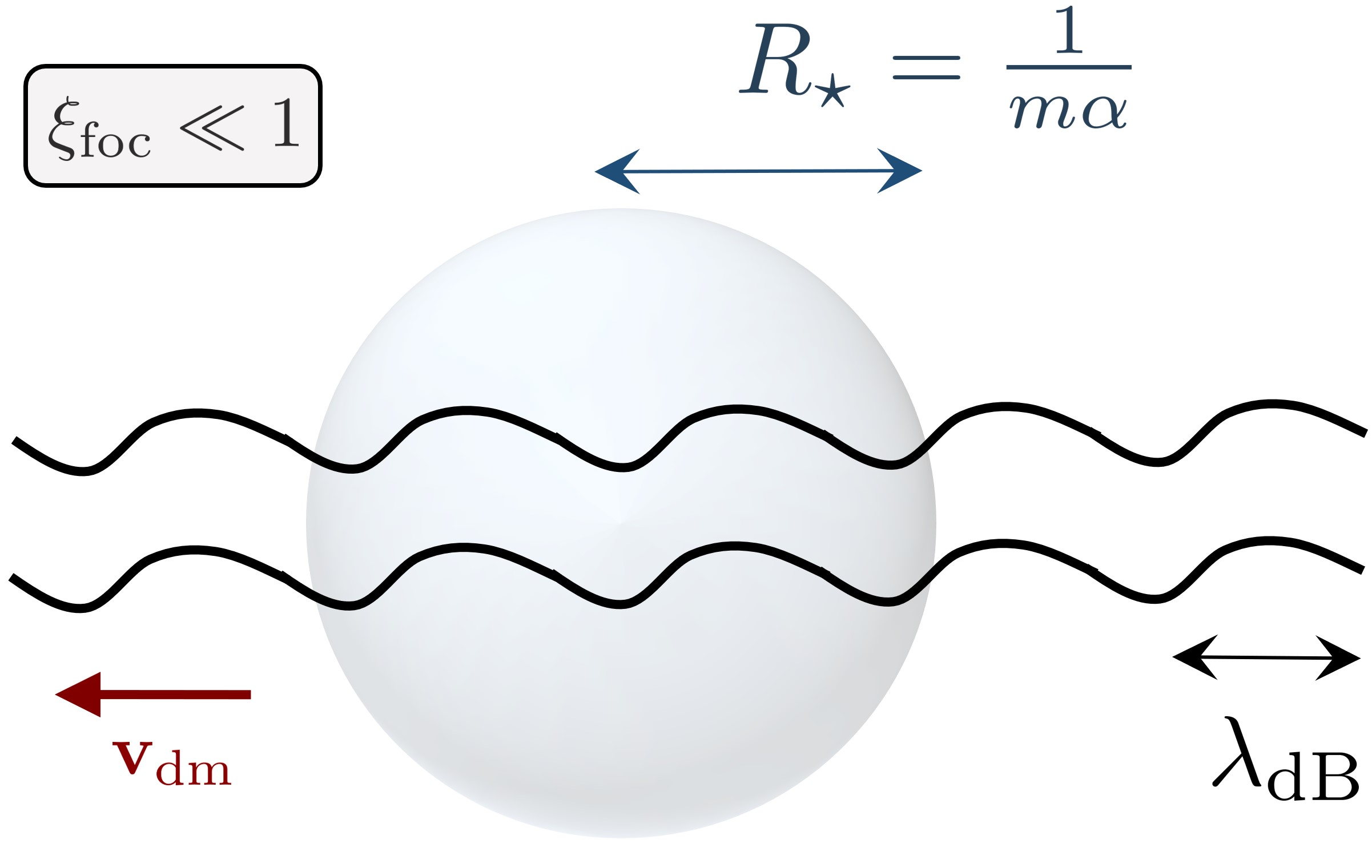} \qquad\qquad\qquad\quad  \includegraphics[width=0.425\textwidth]{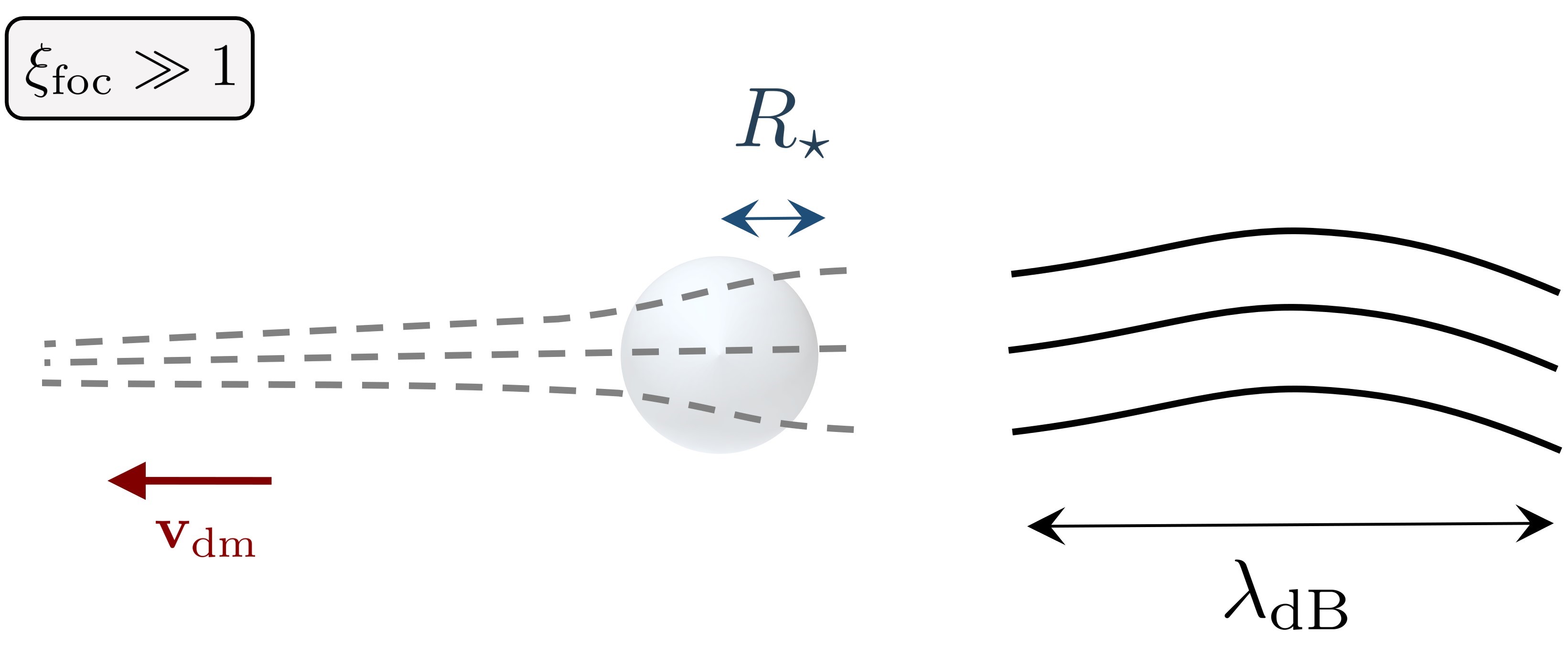}
	\caption{\small{ 
 A schematic representation of the ULDM waves (black lines) with de Broglie wavelength $\lambda_{\rm dB}$, and of the ground state of the atom (spheres), whose radius is $R_\star\equiv (m\alpha)^{-1}$, where $\alpha \equiv G M m$. We show the two limits $\xi_{\rm foc}\equiv \lambda_{\rm dB}/R_\star\ll1$ (left) and $\xi_{\rm foc}\gg1$ (right), which correspond to large/small DM velocity respectively. In the latter, the (otherwise plane) waves are distorted and amplified close to and behind the astrophysical body, an effect known as gravitational focusing. Correspondingly, the dark matter density (dashed lines) is enhanced. %
 In the regime $\xi_{\rm foc}\gg1$, dark matter capture overcomes stripping.
 }}\label{fig:small_large_sol}
	\end{center}
	\vspace{-4mm}
\end{figure}

Substantial DM capture occurs if the rate of stimulated capture exceeds that of stripping, i.e. if $\Gamma\equiv\Gamma_1-\Gamma_2>0$.  Our crucial observation is that this occurs when the dark matter field is significantly gravitationally focused by the external body,
namely if%
\begin{equation}\label{eq:xifoc}
\xi_{\rm foc} \equiv\frac{\lambda_{\rm dB}}{R_\star}=\frac{2\pi\alpha}{v_{\rm dm}}
\simeq\left[\frac{m}{1.7\times10^{-14}\eV}\right]\left[\frac{M}{M_\odot}\right]
                        \left[\frac{240\,{\rm km/s}}{v_{\rm dm}}\right]\,
\end{equation}
is larger than 1 (here we expressed the typical de Broglie wavelength $\lambda_{\rm dB}=2\pi/mv_{\rm dm}$ in terms of the mean DM velocity $v_{\rm dm}$%
).  %
Indeed, if %
$\lambda_{\rm dB}$ is smaller than $R_\star$ ($\xi_{\rm foc}\ll1$, see Figure~\ref{fig:small_large_sol}, left), corresponding to $v_{\rm dm}>2\pi\alpha$, the incoming waves are so fast that %
 the effect of the gravitational potential on them is negligible, and they behave as plane waves everywhere. The kinetic energy $mv_{\rm dm}^2/2$ of the incoming particles is much larger than the binding energy of the ground state, $-m\alpha^2/2$, and capture is inefficient.

Instead, if $\lambda_{\rm dB}$ is larger than $R_\star$ ($\xi_{\rm foc}\gg1$, see Figure~\ref{fig:small_large_sol}, right), the dynamics of the waves close to the Sun is dominated by the Sun itself: Their magnitude increases in the region close and behind it, i.e. they are gravitationally `focused'%
. This effect is  analogous to the Sommerfeld enhancement, which is the increase of the scattering cross section (of a factor $S$) of two quantum mechanical particles in the presence of a central potential~\cite{https://doi.org/10.1002/andp.19314030302}. %
Indeed, the DM density at the center increases as $\xi_{\rm foc}/(1-e^{-\xi_{\rm foc}})$, see also Figure~\ref{fig:psi0}, which is the same as the Sommerfeld enhancement factor $S$.  %

The enhancement is ultimately related to the fact that the (initially slow) particles increase their velocity as a result of the gravitational potential, gaining a %
kinetic energy of the order of their binding energy. %
In this regime, their kinetic energy is small enough that further (order-one) energy changes, resulting from particles scattering through e.g. their self-interactions, have a chance of trapping them in the gravitational potential well. Additionally, incoming particles are not energetic enough (on average) to ionize the ground state, without getting themselves captured. Thus, we expect that stripping is suppressed compared to stimulated capture in this case.  In Section~\ref{ss:light_vs_heavy} we prove that these intuitions are correct by showing that $\Gamma>0$ if $\xi_{\rm foc} \gg 1$ and $\Gamma<0$ if $\xi_{\rm foc} \ll 1$. By simulating the system numerically, in Section~\ref{sec:simulation} we provide evidence that the transition is likely to happen  around $\xi_{\rm foc}\simeq 1$. %

\begin{figure}%
\begin{center}
 	\includegraphics[width=0.5\textwidth]{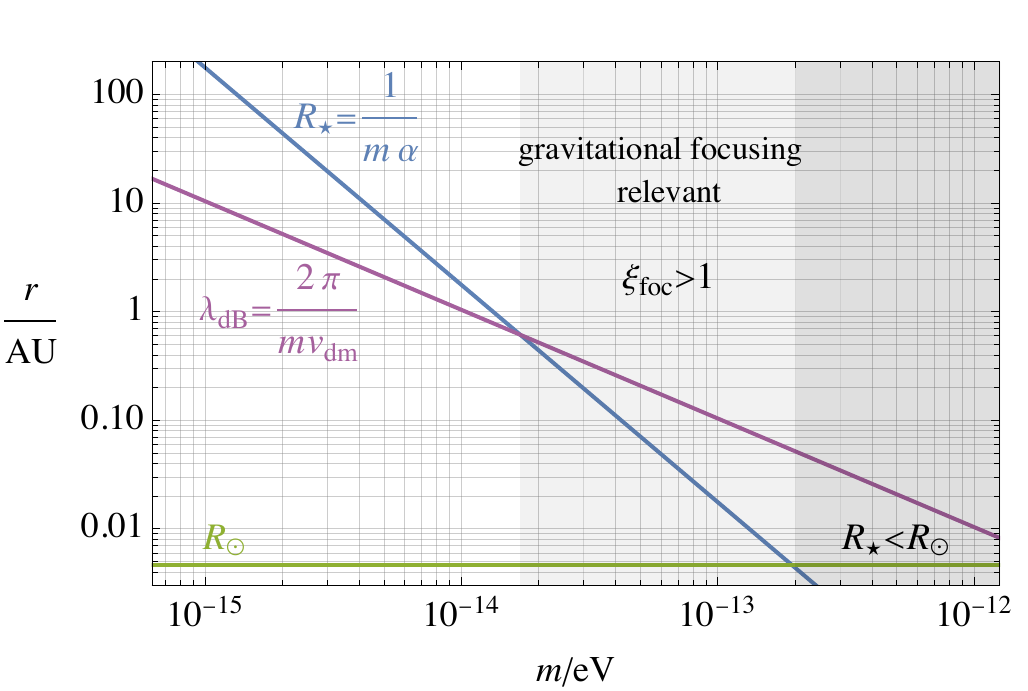}
	\caption{\small{
	A comparison between the typical de Broglie wavelength of the DM waves, $\lambda_{\rm dB}=2\pi/mv_{\rm dm}$ assuming $v_{\rm dm}\simeq 240\km/\s$, and the radius $R_\star=1/m\alpha\propto 1/m^2$, %
 in our Solar System as a function of the boson mass $m$. When $\lambda_{\rm dB}>R_\star$, i.e. $\xi_{\rm foc}>1$, which happens for $m\gtrsim 1.7\cdot 10^{-14}\eV$, gravitational focusing is relevant (light gray region); the corresponding value of the radius $R_\star$ when the two are equal is close to an $\AU$. At large masses, $m\gtrsim2\cdot 10^{-13}\eV$ (darker region), the radius of the Sun $R_\odot$ (green line) is larger than the would-be ground state radius $R_\star$: our treatment of the bound states does not apply, and the effect of the gravitational focusing changes.%
 }
	}\label{fig:focussing_SS}
	\end{center}
	\vspace{-5mm}
\end{figure}

In Figure~\ref{fig:focussing_SS} we compare $\lambda_{\rm dB}$ and $R_\star $ in the Solar System as a function of the mass $m$. From Eq.\,\eqref{eq:xifoc} and Figure~\ref{fig:focussing_SS}, the condition $\xi_{\rm foc}\gtrsim1$ is verified for $\alpha\gtrsim 10^{-4}$ or, equivalently, $m$ larger than approximately $10^{-14}\,\eV$, for which the corresponding ground state turns out to have a radius of order AU or smaller; see Eq.\,\eqref{eq:Rstar_sun}. %
This coincidence can be easily understood by noticing that $\xi_{\rm foc} = 1$ corresponds to a gravitational coupling of size $\alpha= v_{\rm dm}/2\pi$; see Eq.\,\eqref{eq:xifoc}. On the other hand, $\alpha$ can be also interpreted as the effective circular velocity in the bound state at a distance $R_\star$, 
$v_b\equiv\sqrt{GM/R_\star}=%
\alpha$. For $R_\star\simeq \AU$, this is $v_b\simeq 30\km/\s$ and is indeed about of factor of $2\pi$ smaller than $v_{\rm dm}\simeq240\km/\s$. {Note that our approximation of the $1/r$ potential and  treatment of the bound states break down when $R_\star$ is smaller than the solar radius $R_\odot$ (green line), which occurs for $m\gtrsim 2\cdot10^{-13}\,\eV$ (darker region in Figure~\ref{fig:focussing_SS}).%
}%

\vspace{-1mm}
\subsubsection*{Phases of formation}
\vspace{-2mm}
A consequence of Eq.\,(\ref{eq:Mstarintro}) is that the evolution of the system is characterized by two or three phases, depending on whether $\Gamma$ is positive or negative. %
The relevant timescale tracks the \emph{relaxation time} via self-interactions%
\begin{equation}\label{eq:tau_rel0}
\tau_{\rm rel}\equiv \frac{64m^7v_{\rm dm}^2}{\lambda^2\rho_{\rm dm}^2}
            \simeq 9\Gyr\left[\frac{f_a}{10^8 \GeV}\right]^4\left[\frac{m}{10^{-14}\eV}\right]^3
                \left[\frac{0.4 \GeV/{\rm cm}^3}{\rho_{\rm dm}}\right]^{2}\left[\frac{v_{\rm dm}}{240\,{\rm km/s}}\right]^2 \, ,
\end{equation}
where in the last equality we defined $\lambda\equiv -m^2/f_a^2$, valid if the ULDM is an axion with decay constant $f_a$. This represents the typical time a particle in a gas with density $\rho_{\rm dm}$ and average square velocity $v^2_{\rm dm}$ takes to change its velocity by order one via the self-interactions mediated by $\lambda$, in the absence of external gravitational potentials~\cite{Chen:2021oot,Kirkpatrick:2021wwz}. This timescale $\tau_{\rm rel}$ is therefore the analogue of the gravitational relaxation time~\cite{Levkov:2018kau}, with gravitational interactions replaced by the self-interactions. Even extremely weak self-interactions ($\lambda\ll1$), for which the probability of any single $2\to2$ scattering is small, lead a relatively short %
timescale, ultimately because the constants $C,\Gamma_1,\Gamma_2$ 
in Eq.\,\eqref{eq:Mstarintro} are proportional to factors of the occupation number of the bosons in the galactic halo (from Bose enhancement). We summarize the evolution below.

\vspace{-1mm}
\begin{itemize}[leftmargin=0.2in] \setlength\itemsep{0.15em}
\item[\emph{(i)}] \emph{Linear growth} 

\vspace{-1mm}
Since no bound state is initially present ($M_\star=0$ at $t=0$), the stimulated capture and stripping terms in Eq.\,(\ref{eq:Mstarintro}) are negligible. As a result of direct capture only, the bound mass $M_\star$ starts increasing linearly with time, $M_{\star}(t)=Ct$. This linear growth lasts for a time $1/|\Gamma|$, after which %
the $\Gamma_1$ and $\Gamma_2$ terms become important. In Section~\ref{ss:light_vs_heavy} we show that for $\xi_{\rm foc}\gtrsim 1$ this  critical timescale is in fact similar to the relaxation time,  
\begin{equation}
1/\Gamma\simeq 0.3\,\tau_{\rm rel}\,.\label{rel1}\end{equation} 
On the other hand, for $\xi_{\rm foc}\lesssim 1$, both processes of capture and stripping are suppressed and the timescale is longer than $\tau_{\rm rel}$ and reads
\begin{equation}
1/|\Gamma|\simeq  10\,\tau_{\rm rel}/\xi_{\rm foc}^4 \, .\label{rel2}
\end{equation}

\item[\emph{(ii)}] \emph{Exponential growth/saturation} 

\vspace{-1mm}
After the time  $1/|\Gamma|$, enough bound mass $M_\star$ has been accumulated %
that the stimulated capture/stripping terms become relevant.%
\vspace{-1.5mm}
\begin{itemize}[leftmargin=0.2in] \setlength\itemsep{0.15em}
    \item[$\bullet$] For $\xi_{\rm foc}\gtrsim 1$, as mentioned, stimulated capture dominates over stripping and $\Gamma>0$. As a result of stimulated capture only, the bound mass increases exponentially, i.e. $M_\star\propto e^{\Gamma t}$, with an exponential timescale $1/\Gamma\simeq0.3\tau_{\rm rel}$ similar to the relaxation time. This leads to the formation of a `dense' gravitational atom. In Section~\ref{ss:light_vs_heavy} and Appendix~\ref{app:twolevel} we show that also the first few excited states are populated by DM capture at an exponential rate, but then quickly decay (on a timescale fixed approximately by $\tau_{\rm rel}$) into the ground state. It is thus a good approximation to consider only the ground state.

    \item[$\bullet$] On the other hand, for $\xi_{\rm foc}\lesssim 1$, we find that $\Gamma<0$ and the capture and stripping processes %
    rapidly reach equilibrium after $t\simeq1/|\Gamma|$, resulting in a constant bound mass. In this regime, the overdensity
    at the center of the atom (relative to the DM background) is 
    \begin{equation}\label{eq:Mrhoeq}
    \frac{\rho(r=0)%
    }{\rho_{\rm dm}} \simeq \frac{0.4\xi_{\rm foc}^3}{\pi^{5/2}} \simeq 
    4\cdot 10^{-5}\left[\frac{M}{M_\odot}\frac{m}{2\cdot10^{-15}\eV}\frac{240\km/\s}{v_{\rm dm}}\right]^3 \, , %
\end{equation}
which is at most a few percent of the local DM density (when $\xi_{\rm foc}\simeq 1$). The gravitational atom is therefore `dilute'.

\end{itemize}

\item[\emph{(iii)}] \emph{Collapse and Bosenova} 

\vspace{-1mm}
The exponential growth of the dense gravitational atoms taking place for $\xi_{\rm foc}\gtrsim1$ stops (after a few e-foldings) when the DM density in the 
bound state
is so large that the energy in the quartic self-interactions is of the order of the gravitational potential energy that keeps the particles bound. This occurs  at the critical density
\begin{equation}\label{eq:rhocrit0}
\rho_{\rm crit}\simeq 16\frac{\alpha^2m^4}{|\lambda|}\simeq 7\cdot 10^3\rho_{\rm dm}          \left[\frac{f_a}{5\cdot 10^7\,{\rm GeV}}\right]^{2}
                \left[\frac{m}{10^{-14}\,{\rm eV}}\right]^4
                \left[\frac{M}{M_\odot}\right]^2
                \left[\frac{0.4\,{\rm GeV/cm}^3}{\rho_{\rm dm}}\right] \, ,
\end{equation}
which can be orders of magnitude larger than the background density $\rho_{\rm dm}$. For attractive self-interactions, $\lambda<0$, when the density reaches $\rho_{\rm crit}$ the gravitational atom becomes unstable and collapses. A short time after collapse begins, higher-order self-interaction terms become relevant and, for an axion-like potential, the bound state is expected to undergo a \emph{Bosenova} explosion, emitting an order-one fraction of its mass into relativistic ULDM particles (as does its self-gravitating counterpart, a boson star \cite{Eby:2016cnq,Levkov:2016rkk}). The phases \emph{(i)}, \emph{(ii)}, and \emph{(iii)} then repeat again%
. On the other hand, for repulsive self-interactions, $\lambda>0$, the density saturates to (at least) $\rho_{\rm crit}$ and there is no collapse or Bosenova. 
\end{itemize}

\noindent
Summarizing, the role of the two ULDM parameters $m$ and $\lambda$ %
is the following:
\vspace{-1mm}
\begin{itemize}[leftmargin=0.2in] \setlength\itemsep{0.15em}
\item The mass $m$ sets the radius $R_\star$ as in Eq.\,\eqref{eq:Rstar_sun}  and the relevance of gravitational focusing through $\xi_{\rm foc}$, as in Eq.\,\eqref{eq:xifoc}, and thus determines the possibility of exponential increase of the bound states, occurring when $\xi_{\rm foc}\gtrsim1$. Larger values of $m$ lead to larger $\xi_{\rm foc}$, but at the same time to a smaller radius; 
they also lead to a longer capture timescale (for $\lambda$ fixed) as in Eq.\,\eqref{eq:tau_rel0}.%

In the Solar System, for $v_{\rm dm}\simeq240\km/\s$ the exponential increase occurs for scalar masses in the range $(1\div 2)\cdot 10^{-14}\eV\lesssim m\lesssim2\cdot  10^{-13}\eV$. The upper limit correspond to the smallest possible atom (with radius equal to the Sun's radius), for which $\xi_{\rm foc}\simeq 10$%
, and the lower limit to $\xi_{\rm foc}\simeq1$, for which as mentioned $R_\star$ is of order AU. 
However, the lower limit for $m$ decreases for smaller $v_{\rm dm}$, as we discuss further below.

\item On the other hand, $\lambda$ only sets the capture timescale via the relaxation time in Eq.\,\eqref{eq:tau_rel0}. Intuitively, the self-interactions determine 
how quickly the particles can change their energy by order one, but only when $\xi_{\rm foc}\gtrsim1$ this energy change leads to their efficient capture. %

Finally, $\lambda$ also determines the critical density $\rho_{\rm crit}$ in Eq.\,\eqref{eq:rhocrit0}%
: For weaker self-interactions, a larger density is needed for the self-interaction energy to equal the gravitational potential energy.

\end{itemize}

\begin{figure}%
\begin{center}
 \includegraphics[width=0.65\textwidth]{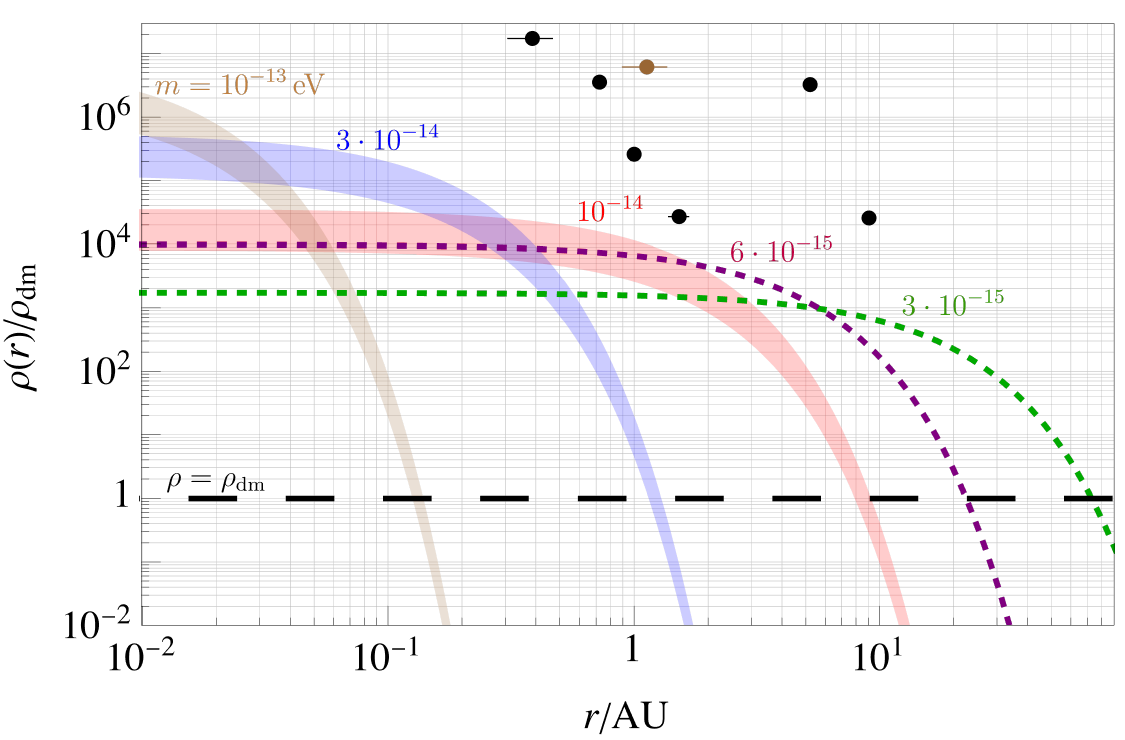} 
	\caption{\small{The density profile of the solar halo $\rho(r)$, normalized to the background DM density $\rho_{\rm dm}$, for different values of $m$. In the shaded regions the DM velocity parameters are varied from $v_{\rm dm}=\sqrt{2}\sigma=240$\,km/s to $50$\,km/s, with the lower edge corresponding to the largest velocity in this range (see text for details; the value $50\km/\s$  might be reasonable if the DM resides in a dark disk).  Dashed lines correspond to the profile for $m < 10^{-14}$\,eV, for which exponential growth of the bound state occurs only when $v_{\rm dm} \ll 240$\,km/s (and results are shown for $v_{\rm dm}=50\km/\s$).   %
 The profile is shown in the best-case scenario, choosing the values of $\lambda$ such that the critical density in Eq.\,\eqref{eq:rhocrit0} is reached at 5 Gyr (roughly the lifetime of the Sun). %
 The black and brown dots are the constraints on the maximum DM mass that can be bound to the Sun, arising from the analysis of Solar System ephemerides using planets~\cite{Pitjev:2013sfa} and asteroids~\cite{Tsai:2022jnv} (respectively) as test masses. For each dot, the horizontal axis indicates the radius within which such DM would be confined, and the vertical axis its maximum possible average density.
}  
	}
	
	\label{fig:rhovsr}
\end{center}
	\vspace{-3mm}
\end{figure}

\vspace{-1mm}
\subsubsection*{Dark matter overdensity}
\vspace{-2mm}
To demonstrate the impact of the dense solar halo ($\xi_{\rm foc}\gtrsim1$) on the local DM density in the Solar System, in Figure~\ref{fig:rhovsr} we show its density profile, which follows $\rho(r)\propto%
e^{-2r/R_\star}$, for different values of the boson mass between $10^{-13}\eV$ and $10^{-15}\eV$. The density is normalized to the background density $\rho_{\rm dm}$.  %
 For each mass we show the profile in the best-case scenario, i.e. at the final stages of the exponential increase, just before the critical density is reached, choosing for each line the value $\lambda$ such that this is reached at 5~Gyr (roughly the lifetime of the Sun). For smaller $|\lambda|$, the maximum density would be reached after a longer time, according to Eq.\,\eqref{eq:tau_rel0}; for larger $|\lambda|$, within 5 Gyr multiple Bosenova cycles would have occurred, for each of which the density is smaller than in Figure~\ref{fig:rhovsr} proportionally to $1/|\lambda|$. The values of $\lambda$ used, of order $10^{-57}\div 10^{-61}$, for an axion correspond to $f_a$ within the range $10^7\GeV$ to $10^8\GeV$ for the highest and lowest masses, respectively. %

Additionally, to show the effect of a lower velocity, over the shaded bands in Figure~\ref{fig:rhovsr} the DM velocity parameters are varied from $v_{\rm dm}=\sqrt{2}\sigma=240$\,km/s to $50$\,km/s. Although speculative, this lower value might be plausible if the DM resides in a dark disk, where it could have a smaller average velocity with respect to the Sun, and a smaller dispersion~\cite{Read:2008fh,2009ApJ...703.2275P,Kim:2021yyo}. The lower/upper edges of the bands correspond respectively to the larger/smaller velocity.\footnote{At smaller $v_{\rm dm}$, $\tau_{\rm rel}\simeq5\Gyr$ is reached for a smaller $\lambda$, so that $\rho_{\rm crit}$ is larger.} For $m \ll 10^{-14}$\,eV, the exponential growth occurs only when $v_{\rm dm} \ll 240$\,km/s; dashed lines show the corresponding profile when the critical density is reached for $v_{\rm dm}=50$\,km/s.

For comparison, in Figure \ref{fig:rhovsr} we show with black dots the constraints on the maximum mass that can be bound to the Sun, derived from Solar System ephemerides~\cite{Pitjev:2013sfa,Banerjee:2019epw}. 
Dark matter in the orbit of a planet would contribute a force $\propto r$ and give rise to anomalous perihelion precession; %
this leads to a point-like constraint on the DM density in the orbits of each planet (see \cite{Tsai:2022jnv}%
).
Although the density in the halo can be orders of magnitude larger than $\rho_{\rm dm}$, it does not violate these direct constraints. Note that the total bound mass is a small fraction of the solar mass for all the plotted lines.

The dilute gravitational atoms ($\xi_{\rm foc}\lesssim 1$) are also novel targets for experimental searches. Although they only constitute a subcomponent of the DM, this %
could have a `coherence' time $\tau_\star$ larger than that of the DM in the galactic halo, which is $\tau_{\rm dm}\equiv 2\pi/(m v_{\rm dm}^2)$. 
Unfortunately, our analytic treatment in Section\,\ref{sec:analytic} only determines the change in the captured mass over times longer than $\tau_{\rm dm}$, so it does not allow to compute $\tau_{\star}$ from first principles. 
If the bound particles correspond to an exact energy eigenstate of the system, then $\tau_\star$ is in principle infinite. If this is not the case, then one could estimate a lower bound on $\tau_\star$
by interpreting the virial velocity $v_b=\alpha$ as a velocity dispersion in the field, i.e. 
\begin{equation} \label{eq:taustar1}
    \tau_\star \gtrsim \frac{2\pi}{m\alpha^2} = \left[\frac{2\pi}{\xi_{\rm foc}}\right]^2\tau_{\rm dm}%
    \simeq 1 \, {\rm year}\left[\frac{1.3\cdot10^{-14}\eV}{m}\right]^3\,.
\end{equation}
This lower bound on $\tau_{\star}$ is larger than $\tau_{\rm dm}$ for all $\xi_{\rm foc}<2\pi$. In the Solar System for $v_{\rm dm}=240$ km/s, this holds for $m \lesssim 10^{-13}\eV$, i.e. for essentially all the halos that can form around the Sun.

\begin{figure}%
\begin{center}
 \includegraphics[width=0.56\textwidth]{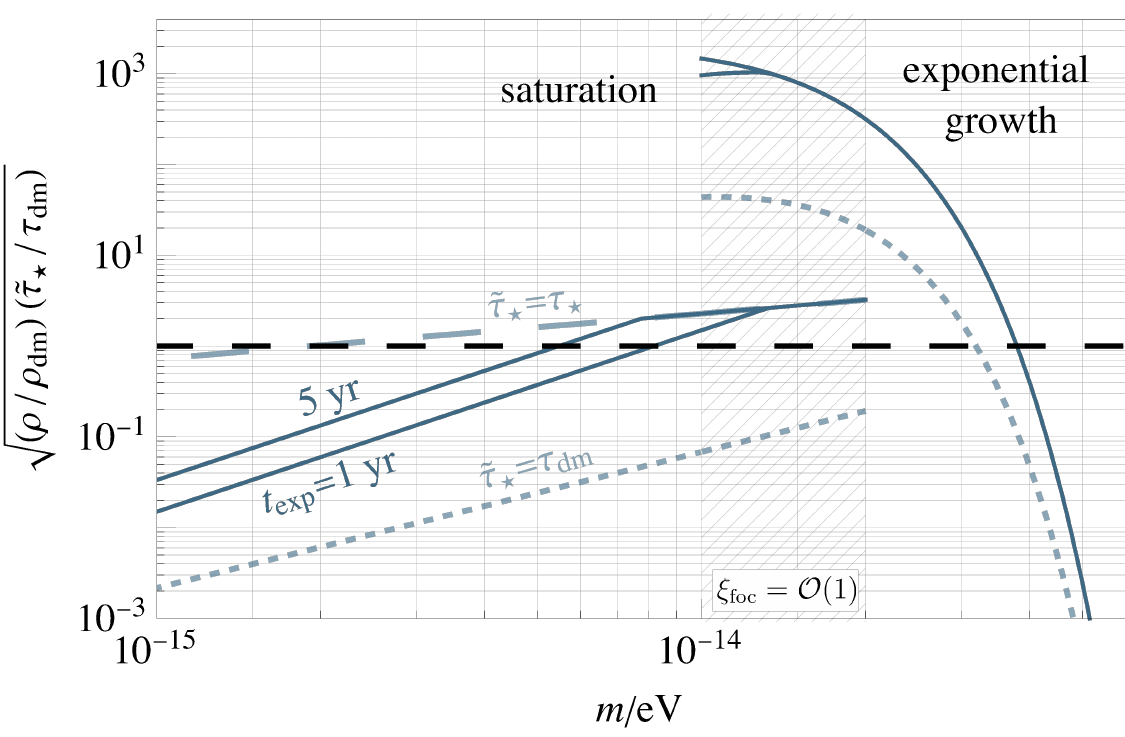} 
	\caption{\small{The ratio $\sqrt{(\rho/\rho_{\rm dm})(\tilde{\tau}_\star/\tau_{\rm dm})}$, which describes the detection reach of ULDM (linearly coupled to the SM) bound into a gravitational atom relative to that in the galactic halo, as a function of $m$, assuming $v_{\rm dm} = 240$~km/sec. The effective coherence time $\tilde{\tau}_\star=\min(\tau_\star,t_{\rm exp})$ is relevant to an experiment lasting $t_{\rm exp}$; $\tau_\star$ in Eq.\,\eqref{eq:taustar1} is a conservative lower bound on the coherence time of the atom. We show $t_{\rm exp}=1\yr,\, 5\yr$ (solid lines). For comparison, we also show the same ratio if the coherence time coincided with that of the DM galactic halo, $\tilde{\tau}_\star = \tau_{\rm dm}$ (dashed line) and also for $\tilde{\tau}_\star=\tau_\star$ (long dashed). Thanks to the larger coherence time, even for $m\lesssim 10^{-14}\eV$, for which the gravitational atom is a subcomponent of the DM, the experimental reach could be better than that of the galactic halo DM.%
}  
}
	\label{fig:signalgain}
\end{center}
	\vspace{-3mm}
\end{figure}

Eq.\,\eqref{eq:taustar1} could be important for experimental searches based on resonance: The sensitivity of these searches to ULDM e.g. linearly coupled to the SM grows with the local DM density $\rho$ and the measurement time $t$ as $\phi%
\sqrt{t}\propto \sqrt{\rho\,t}$, until $t$ approaches the coherence time (see~\cite{Budker:2013hfa,Dror:2022xpi}). In Figure~\ref{fig:signalgain} we illustrate the ratio $\sqrt{(\rho/\rho_{\rm dm})(\tilde{\tau}_\star/\tau_{\rm dm})}$ between the sensitivity reach for a gravitational atom and the DM in the galactic halo; here, $\tilde{\tau}_\star \equiv \min(\tau_{\star},t_{\rm exp})$ is an effective coherence time relevant to a search lasting $t_{\rm exp}$.  For $m\gtrsim (1\div2)\cdot10^{-14}$, where exponential growth occurs, there is a large enhancement of the sensitivity, from both a larger density and a larger coherence time. In addition, even for $m\lesssim 10^{-14}\eV$, where the growth saturates to $\rho<\rho_{\rm dm}$ as in Eq.\,\eqref{eq:Mrhoeq}, the detection reach could be better than that of the virialized halo, owing to the larger coherence time. %

\vspace{-1mm}
\subsubsection*{Implications}
\vspace{-2mm}

The results above imply that, %
after $\tau_{\rm rel}$, dense gravitational atoms are expected to form 
around all the massive objects for which the condition $\xi_{\rm foc}\gtrsim1$ is satisfied (including, for instance, solar-mass stars and black holes), which act as `seeds' for DM capture. %
For astrophysical systems with $M>M_{\rm \odot}$ or smaller DM velocity, the exponential increase occurs also for $m$ smaller than $10^{-14}\eV$, for which $\tau_{\rm rel}$ is shorter.

Over a time of the order of $\tau_{\rm rel}$, dynamical relaxation of the DM in the galactic halo via self-interactions is likely to lead to additional effects. Similarly to what happens to fuzzy dark matter (with $m\simeq 10^{-21}\div10^{-22}\eV$) via the gravitational interactions alone~\cite{Levkov:2016rkk,Chen:2020cef}, after $\tau_{\rm rel}$ a solitonic core is expected to form 
in the galactic center. %
(This would occur even faster than the formation of the solar halo, as the DM abundance close to the galactic center is expected to be larger, corresponding to a smaller $\tau_{\rm rel}$.) For the masses $m$ of interest, such an object may be absorbed by the supermassive black hole~\cite{Bar:2019pnz,Davies:2019wgi}. %
Additionally, self-gravitating boson stars may randomly be produced anywhere in space by %
relaxation of the scalar field to its ground state via self-interactions
\cite{Chen:2020cef,Chen:2021oot}. In contrast to the gravitational atoms around external gravitational potentials considered here, their accretion rate does not follow Eq.\,\eqref{eq:Mstarintro} and their density is not expected to increase exponentially (see also~\cite{Chan:2022bkz}). Thus, these bound objects should accrete more slowly. %

Sufficiently strong ULDM self-interactions alter the cosmological evolution of DM perturbations. As we discuss in Section \ref{sec:lss}, attractive/repulsive self-interactions tend to enhance/suppress the perturbation growth and can therefore be constrained from measurements of the linear matter power spectrum~\cite{Arvanitaki:2014faa,Fan:2016rda,Cembranos:2018ulm}. In any case, as shown in Figure~\ref{fig:structure_form}, the values of $\lambda$ required for the formation of the atom to take place within 5 Gyr are consistent with these constraints over a few orders of magnitude. 
We will also show that %
 these constraints may become considerably stronger if one considers the evolution of the field before matter-radiation equality, an effect not pointed out in previous studies.

Let us now briefly comment about the relic abundance and axions. 
The simplest version of the 
misalignment mechanism underproduces the DM for the parameters $m\simeq 10^{-13}\div10^{-14}\eV$ and $f_a\lesssim10^{8}\GeV$ that lead to the formation of the halo around the Sun within the age of the Solar System (equivalently, the expected size of the self-interactions is too small to have $\tau_{\rm rel}\simeq\Gyr$). This is because, for a conventional axion potential, $f_a$ both sets $\lambda$ %
and the axion field range, of order $f_a$. However this is not necessarily the case for scalars with more generic potentials, where in principle the field range is not related to the self-couplings and can extend up to the Planck scale, or in the presence of monodromies~\cite{Jaeckel:2016qjp}. {For instance, a scalar field with potential containing the mass term and the self-interaction term in Eq.\,\eqref{eq:selfint} could lead to the observed relic abundance via misalignment, provided that the initial field value is chosen appropriately.%
}
In any case, there is a variety of other production mechanisms leading to a cosmologically nontrivial axion relic density (see for instance~\cite{Co:2017mop,Co:2018mho,Cyncynates:2021xzw}), some of which could lead to the correct DM abundance %
for the formation to take place over Gyr timescales.  

\begin{figure}%
\begin{center}
 \includegraphics[width=0.56\textwidth]{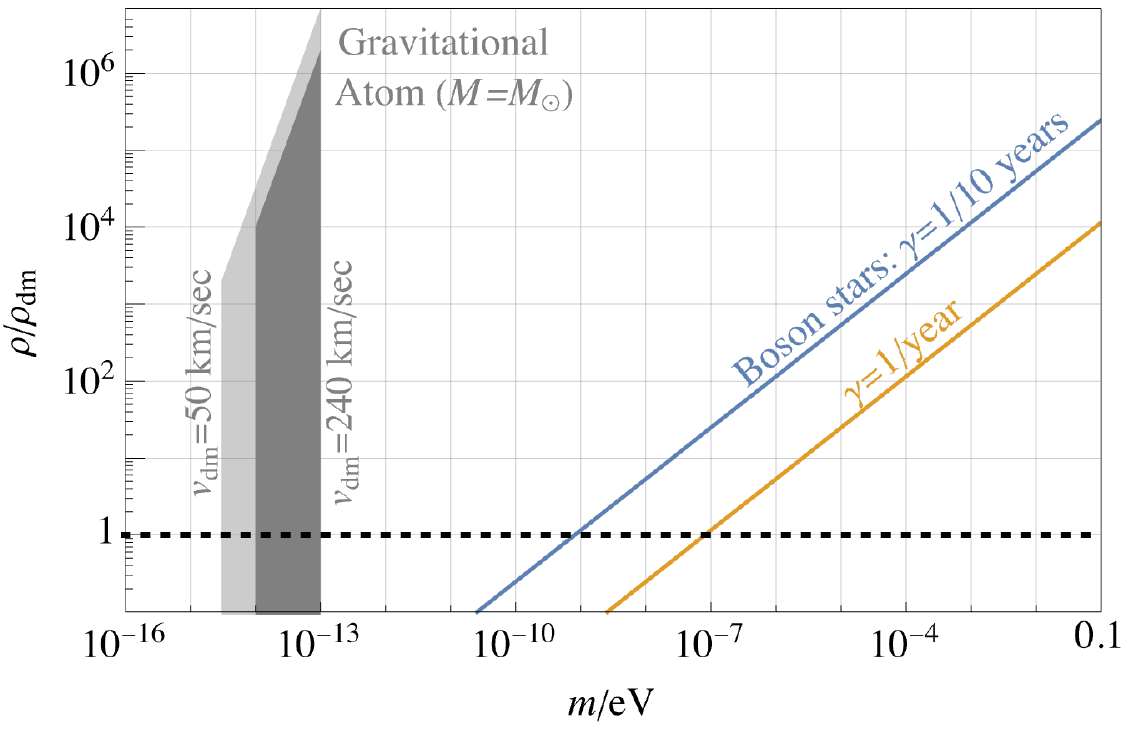} 
	\caption{\small{ A comparison between the overdensity near the center of 
    a gravitational atom, and that of a boson star in the scenario where an order-one fraction of DM resides in such bound objects. For the gravitational atom, we show the density captured within 5\,Gyr around a central mass $M=M_\odot$ for large enough $\lambda$ (see text for details); large overdensities occur for $\xi_{\rm foc}\gtrsim 1$, i.e. for {(a)} $m\simeq 10^{-14}\div10^{-13}\eV$ for mean velocity $v_{\rm dm}=240\km/\s$ (dark gray) and {(b)} down to $m=3\cdot 10^{-15}\eV$ for $v_{\rm dm}=50$\,km/s. For the boson stars, we show their central density assuming they constitute $100\%$ of the DM in the galaxy, with the encounter rate to be fixed to $\gamma=1/$year (yellow) and $1/10$ year (blue). %
	}	}
	\label{fig:boson_stars}
\end{center}
	\vspace{-4mm}
\end{figure} 

The axion coupling to photons is usually parameterized by the Lagrangian $\mathcal{L}\supset \frac14g_{\phi\gamma\gamma}\phi F_{\mu\nu}\tilde{F}^{\mu\nu}$, with $g_{\phi\gamma\gamma}\equiv c_{\phi\gamma\gamma}\alpha_{em}/(2\pi f_a)$, 
where $F_{\mu\nu}$ is the photon field strength, $\alpha_{em}$ is the electromagnetic fine-structure constant, and $c_{\phi\gamma\gamma}$ is a model-dependent constant. For masses $m\lesssim 10^{-13}\eV$, the strongest bounds currently come from helioscope 
measurements~\cite{CAST:2017uph} and several astrophysical observations (see e.g.~\cite{FermiLAT:2016nkz,Reynolds:2019uqt,Dolan:2022kul}) as $g_{\phi\gamma\gamma}\lesssim10^{-10}\div 10^{-12}\GeV^{-1}$.\footnote{Across a broader range of $m$, note also many other ongoing searches for axion fields, including: microwave-cavity-based haloscopes~\cite{Du:2018uak,Asztalos:2003px,Asztalos:2009yp}, experiments exploiting magnetic resonance~\cite{Budker:2013hfa,Graham:2013gfa,Ruoso:2015ytk}, interferometry~\cite{Liu:2018icu,Nagano:2019rbw,Obata:2018vvr,DeRocco:2018jwe}, precision magnetometers and LC circuits~\cite{Sikivie:2013laa,Chaudhuri:2014dla,Kahn:2016aff,Silva-Feaver:2016qhh,Ouellet:2018beu,DMRadio:2023igr}, atomic transitions~\cite{Sikivie:2014lha,Kim:2022ype}, searches for an oscillating neutron electric dipole moment~\cite{Abel:2017rtm}, and many others (a more complete list can be found in Refs.~\cite{Kim:2008hd,Graham:2015ouw} and references therein).} %
Fast enough capture requires $f_a\lesssim10^8\GeV$, so that the unknown constant $c_{\phi\gamma\gamma}$ needs to be somewhat smaller than $\mathcal{O}(1)$.  
Additionally, 
as discussed further in Section~\ref{sec:SMcouplings}, SM couplings can in principle disrupt the formation of the bound state. However, their effect is not amplified by Bose enhancement, and so weak and not expected to affect the formation. %

Finally, in Figure~\ref{fig:boson_stars} we compare, as a function of $m$, the DM overdensity of the solar halo with that occurring if an order-one fraction of the DM is in boson stars (as a boson star passes through the Solar System). As illustrated, the encounter rate  $\gamma$ of boson stars with the Earth is of experimental relevance (at least one encounter per $\sim\ $few years, with overdensity of at least unity) only for relatively large particle masses $m\gtrsim 10^{-9}$\,eV (see for instance \cite{Banerjee:2019xuy} for details). On the other hand, a gravitational atom formed around the Sun gives rise to a persistent overdensity, making it a prime target for experimental searches. %

The results in this paper open up several signatures and opportunities for  discovery of (or constraints on) ultralight dark matter, which we briefly discuss in Section~\ref{sec:conclusion} and will explore in future work.%

\section{Bound states and waves} 
\label{sec:boson_stars_waves}

Consider a scalar field $\phi$ with potential $V(\phi)$ (we will use the terms `scalar' and `axion' interchangeably, because in what follows nothing will depend on the parity of the field). We are interested in cases where the amplitude of $\phi$ is small compared to the typical field range, as expected in our virialized DM halo. Thus, it is appropriate to expand $V$ around its minimum as
\begin{equation}\label{eq:pot}
V(\phi)=\frac{1}{2}m^2\phi^2+\frac{\lambda}{4!}\phi^4+\dots \, ,
\end{equation}
where the dots include higher orders in the expansion.\footnote{In principle, a cubic coupling $\propto \phi^3$ is also possible in CP-violating theories. However, to leading order it only mediates $2\to1$ or $1\to 2$ processes (for which energy changes are of order $m$) which are suppressed in the the nonrelativistic limit, so we do not include this term below.}
The sign of $\lambda$ (negative or positive) determines whether the quartic self-interactions are attractive or repulsive. In the following it will be useful to express $\lambda$ in terms of a dimensionful coupling $g$ (to be used interchangeably with $\lambda$) as $$\lambda\equiv 8gm^2.$$
If $\phi$ is an axion with mass arising from weakly-coupled instantons~\cite{Gross:1980br,Svrcek:2006yi}%
, its potential is a periodic function dominated by a single harmonic, $V(\phi)=-m^2f_a^2\cos(\phi/f_a)\,.$ In this case, the attractive self-interactions have $g=-1/(8f_a^2)$ and  %
$-\lambda \simeq 
10^{-61}(m/10^{-14}{\eV})^2(10^8{\GeV}/f_a)^2$.
However, 
the only requirement on $V(\phi)$ is that it can be approximated as in Eq.\,\eqref{eq:pot}. 

Given its large occupation number, the DM follows the classical equations of motion (EoM) corresponding to the potential of Eq.\,\eqref{eq:pot}, i.e. 
\begin{equation}
(g^{\mu\nu}D_\mu\partial_\nu+m^2)\phi=-(1/6)\lambda\phi^3\,,
\end{equation}
where the metric is $g_{00}=1+2\Phi$, $g_{ij}=-(1-2\Phi)\delta_{ij}$, $D_\mu$ is a covariant derivative, and $\Phi$ is the gravitational potential. The (general-)relativistic corrections will be small in the following. Thus,  the EoM can be rewritten in a nonrelativistic form, i.e. working in terms of a nonrelativistic field $\psi$ defined by
\begin{equation}
\phi\equiv\frac{1}{\sqrt{2m}}(\psi e^{-imt}+{\rm c.c.}) \, ,
\end{equation}
in the limit where $\psi$ is slowly varying, $\dot{\psi}\ll m\psi$%
. In terms of $\psi$, the EoM reduces to the Gross--Pitaevskii (GP) equation
$(i\partial_t+\nabla^2/2m-m\Phi)\psi=g|\psi|^2\psi$. In the vicinity of a massive object of mass $M$ the gravitational potential is well approximated by that generated by the object itself, i.e. $\Phi=\Phi_{\rm ex}\equiv-GM/r$, with $r\equiv|\mathbf{x}|$.\footnote{\label{footnotePoisson} In reality, $\Phi=\Phi_{\rm se}+\Phi_{\rm ex
}$, where $\Phi_{\rm se}$ is the self-potential, which satisfies the Poisson equation $\nabla^2\Phi_{\rm se}=4\pi G m(|\psi|^2-\langle|\psi|^2\rangle)$. However, we are interested in the case where the mass of the external body is much larger than that of the enclosed DM, thus $\Phi_{\rm se}$ can be safely neglected.} As a result, the GP equation simplifies to
\begin{equation}\label{eq:EoM}
\left(i\partial_t+\frac{\nabla^2}{2m}+\frac{\alpha}{r}\right)\psi=g|\psi|^2\psi \, .
\end{equation}
The gravitational coupling $\alpha\equiv GMm$ measures the strength with which the particles are pulled towards the potential well. Eq.\,\eqref{eq:EoM} is nonlinear due to the self-interaction term.%

In the nonrelativistic limit, the mass density of the field is $\rho=m|\psi|^2$, and the number density of particles is $n\equiv \rho/m=|\psi|^2$. The (nonrelativistic) energy density, defined as $\epsilon\equiv T^{00}-\rho$, with $T^{\mu\nu}$ being the stress-energy tensor of $\phi$,
is given by
\begin{equation}\label{eq:energy_density}
\epsilon=\frac{|\nabla\psi|^2}{2m}+%
m\Phi_{\rm ex}|\psi|^2+\frac12g|\psi|^4\, ,
\end{equation}
see Appendix\,\ref{app:bound_states} for the derivation. The terms in the decomposition of Eq.\,\eqref{eq:energy_density} correspond, respectively, to the kinetic, gravitational potential, and self-interaction energies. The latter is negative for attractive self-interactions, as is the gravitational potential energy.

If the density $\rho$ is (locally) smaller than a critical density $\rho_{\rm crit}\equiv 2|\Phi_{\rm ex}| m^2/|g|$, %
the self-interaction energy is negligible compared to that of gravity (at $r\simeq R_\star$, this reproduces Eq.\,\eqref{eq:rhocrit0}). In this limit the interaction term in the right-hand side of Eq.\,\eqref{eq:EoM} can be treated as a perturbation. At the zeroth order, the EoM are the same as those of the hydrogen atom, with $\alpha$ and the number density $|\psi|^2$ playing the roles of the fine-structure constant and the quantum-mechanical probability density, respectively. 

In analogy with the hydrogen atom, Eq.\,\eqref{eq:EoM} admits quasi-stationary time-periodic solutions of the form $\psi= e^{-i\omega t}\Psi(\mathbf{x})$, where $\Psi$ solves the eigenvalue equation $(-\nabla^2/2m+m\Phi_{\rm ex})\Psi=\omega\Psi$. %
Their energy is $E\equiv\int d^3x\epsilon=N\omega$, where $N\equiv\int d^3x|\psi|^2$ is the total number of particles, so that $\omega$ has the interpretation of energy per particle. Such solutions can be divided into two classes: bound and unbound states of the external potential, depending on whether $\omega$ (or $E$) is negative or positive.

\subsection*{Bound solutions}
Bound-state solutions of Eq.\,\eqref{eq:EoM} are the equivalent of hydrogen atom orbitals. 
They are discrete states, $\Psi\propto\psi_{nlm}$, labelled by the integers $n,l,m$ and have negative (binding) energy $\omega_n\equiv-m\alpha^2/(2n^2)$.\footnote{{We use the symbol $m$ both for the particle mass and the quantum number of the bound state, but it will be always obvious which of the two meaning applies.}}
The normalized lowest-energy solution (i.e. the ground state) is
\begin{equation}\label{eq:psi100}
\psi_{100}=\frac{1}{\sqrt{\pi R_\star^3}} e^{-r/R_\star} \, , \qquad R_\star\equiv\frac{1}{m\alpha} =\frac{1}{Gm^2M}\ .%
\end{equation}
This corresponds to a spherically-symmetric density field, which is maximized at the center $r=0$. The parameter $R_\star$, given in  Eq.\,\eqref{eq:Rstar_sun}, is the typical radius. We refer the reader to Appendix \ref{app:bound_states} for the form of the excited states%
, where for example $e^{-r/R_\star} \to e^{-r/(n R_\star)}$.

The bound mass is $M_\star\equiv\int d^3x\rho = m \int d^3x|\Psi|^2$, so that $\Psi=\sqrt{M_\star/m}\,\psi_{100}$ if only the ground state is populated. In contrast to the hydrogen atom, the occupation number and therefore $M_\star$ can be macroscopic because of the bosonic nature of the particles. In particular, in the 
limit of $g\to 0$ (and of negligible self-gravity), the mass can grow arbitrarily (until relativistic corrections become relevant)%
. Note that $R_\star$ only depends on $m$ and $M$, and not on $M_\star$, which enters only in the normalization of $\Psi$.\footnote{We stress that this is true only in the limit $M\gg M_\star$. The situation is different in the case of a self-gravitating boson star (or `soliton'), where instead the star radius depends on the star mass itself \cite{Kaup:1968zz,Ruffini:1969qy,BREIT1984329,Colpi:1986ye}. Here, since the self-gravity is negligible, it is $M$ rather than $M_\star$ that determines the radius.} The gravitational atom can be interpreted as a collection of particles bound to the external body, with typical `virial' velocity $v_b=\sqrt{G M/(n^2R_\star)}= \alpha/n$, set by $\alpha$ (this is consistent  with the semi-classical virial relation $|\omega_1|\approx mv_b^2/2$).

\subsection*{Unbound solutions}

The unbound solutions are \emph{traveling waves}, which correspond to the Coulomb scattering states. They form a continuum of states, $\Psi\propto\psi_\mathbf{k}$, parametrized by a three-vector $\mathbf{k}$ that represents their (asymptotic) momentum, and have positive energy $\omega_k\equiv k^2/2m$. 
The normalized solution reads \cite{Hui:2016ltb,Kim:2021yyo}
\begin{equation}\label{eq:psik}
    \psi_{\mathbf{k}}=e^{i\mathbf{k}\cdot\mathbf{x}}\,\Gamma\left[1-\frac{i}{k R_\star}\right]
                e^{\frac{\pi }{2 k R_\star}} \, _1F_1\left[\frac{i}{k R_\star},1,i (k r-\mathbf{k}\cdot\mathbf{x})\right] \, ,
\end{equation}
where $\Gamma[a]$ is the Gamma function and  $_1F_1[a,b,c]$ is the confluent hypergeometric function.\footnote{The bound-state wavefunctions are normalized such that $\int d^3x \psi^*_{n'l'm'}\psi^{}_{nlm}=\delta_{nn'}\delta_{ll'}\delta_{mm'}$, whereas the unbound ones satisfy $\int d^3x \psi_{{\bf k}}^*\psi^{}_{\bf k'} = (2\pi)^3\delta^3({\bf k}-{\bf k}')$; see also Eq.\,\eqref{eq:orthonormality}.} For later convenience, in Eq.\,\eqref{eq:psik} we expressed $m$ and $\alpha$ in terms of $R_\star=(m\alpha)^{-1}$, although these solutions %
do not depend on bound-state properties per se.

The momentum $\mathbf{k}$ of the waves can be also written in terms of their velocity $\mathbf{v}$ (with respect to the body) as $\mathbf{k}=m\mathbf{v}$. Importantly, $m$ and $\alpha$ appear in Eq.\,\eqref{eq:psik} only through
\begin{equation} \label{eq:2pikRstar}
 \xi_{\rm foc}(k)\,\equiv \frac{2\pi}{kR_\star}= \frac{2\pi\alpha}{v}=\frac{2\pi v_b}{v}\, .
\end{equation}
As mentioned in the Introduction, this measures the ratio between the de Broglie wavelength $
2\pi/k$ and $R_\star$.
Given than $\alpha=v_b$ for the ground state, this is also $\propto v_b/v$, and therefore controls the relative importance of the virial velocity 
and that of the wave, modulo a $2\pi$ factor. 
The limits $\xi_{\rm foc}(k) \ll 1$ and $\xi_{\rm foc}(k)\gg 1$, shown in Figure~\ref{fig:small_large_sol}, correspond respectively to large/small $R_\star$ compared to $2\pi/k%
$, or waves that are fast/slow with respect to the virial velocity ($v\ll 2\pi\alpha$ and $v\gg 2\pi\alpha$). %
In these two regimes the scattering states in Eq.\,\eqref{eq:psik} have simpler expressions. 
\vspace{-1mm}
\begin{itemize}[leftmargin=0.2in] \setlength\itemsep{0.15em}
	\item If %
        $\xi_{\rm foc}(k) \ll 1$, %
        $\psi_\mathbf{k}$ in Eq.\,\eqref{eq:psik} 
	reduces to a plane wave everywhere, as if no potential were present:
	\begin{equation}\label{eq:psik_plane}
	\psi_{\mathbf{k}}\to e^{i\mathbf{k}\cdot\mathbf{x}} \, , \qquad\qquad 
            \xi_{\rm foc}(k) \ll 1 \, .
	\end{equation}
	In this limit, the density field $\propto|\psi|^2$ is homogeneous, with (maximum) value independent of %
    $k$ %
    (see blue line in Figure~\ref{fig:psi0}, left).

	\item If %
        $\xi_{\rm foc}(k)\gg 1$, %
        the plane wave is distorted around the body at distances smaller than the wavelength, $r\lesssim 2\pi/k$%
        . Within this region, an excellent approximation of Eq.\,\eqref{eq:psik} is
	\begin{equation}\label{eq:psik_bessel}
	\psi_{\mathbf{k}}\to e^{i\varphi(k)}\sqrt{\frac{2\pi}{kR_\star}}J_0\left[2\sqrt{\frac{r}{R_\star}(1-\hat{k}\cdot\hat{x})}~\right]\, , \qquad\qquad 
                \xi_{\rm foc}(k) \gg 1 \, ,
	\end{equation}
	where $J_0$ is the spherical Bessel function.\footnote{The phase $\varphi(k)\equiv-\pi/4+(1+\log(kR_\star))/kR_\star$ will be irrelevant in our discussions.} Instead, %
	for $r\gg 1/k$, $\psi_\mathbf{k}\to e^{i\mathbf{k}\cdot\mathbf{x}}$ and recovers its plane wave form, at least in the region where $\hat{k}\cdot\hat{x}<1$.\footnote{Indeed, on the line $\hat{k}\cdot\hat{x}=1$, even very far away from the body, $r\gtrsim 1/k$, a `wake' of overdensities persists; see below.}%

\begin{figure}%
\begin{center}
	\includegraphics[width=0.5\textwidth]{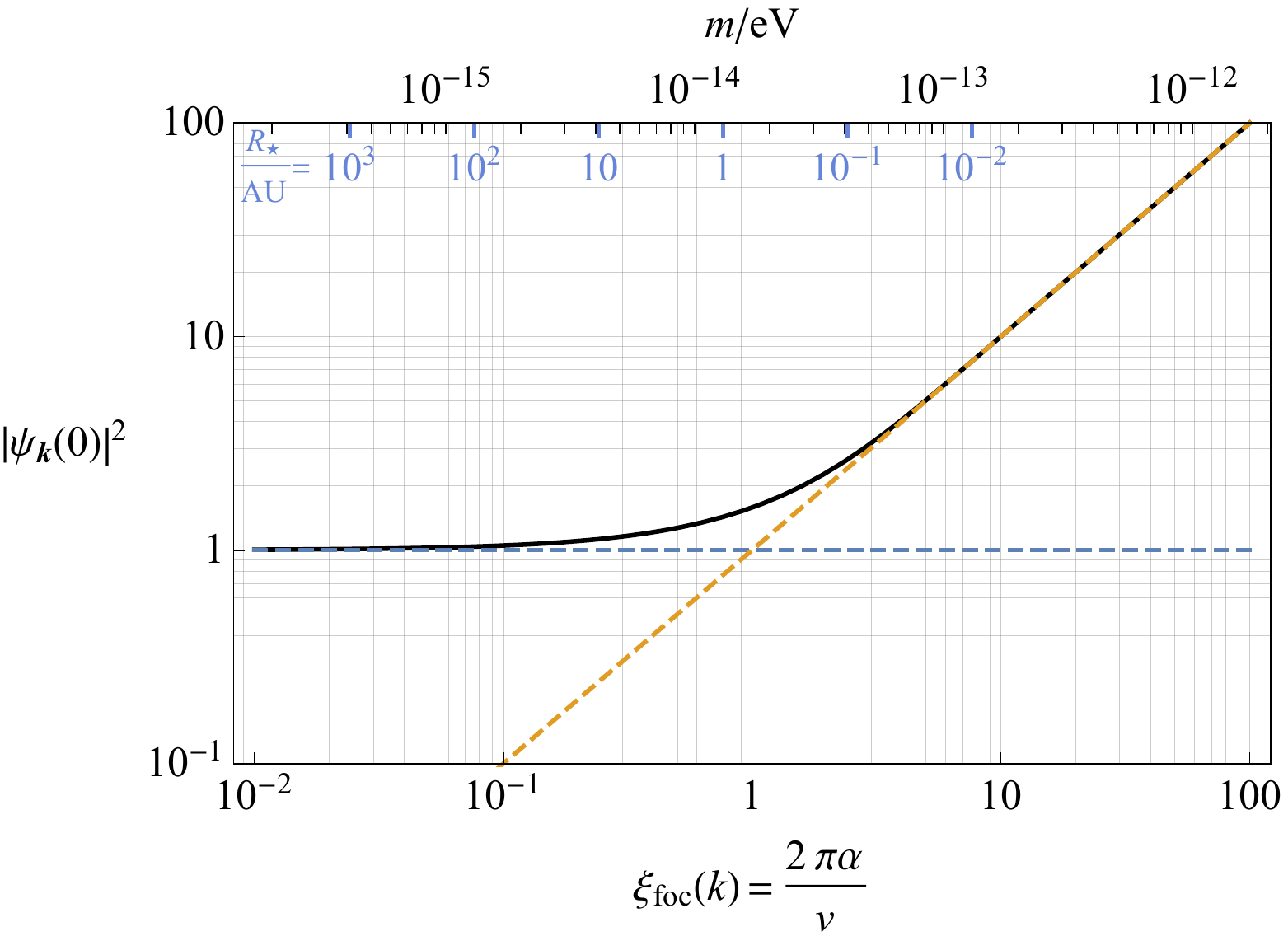}  
	\ \ \ \ \ \ \ 	\includegraphics[width=0.40\textwidth]{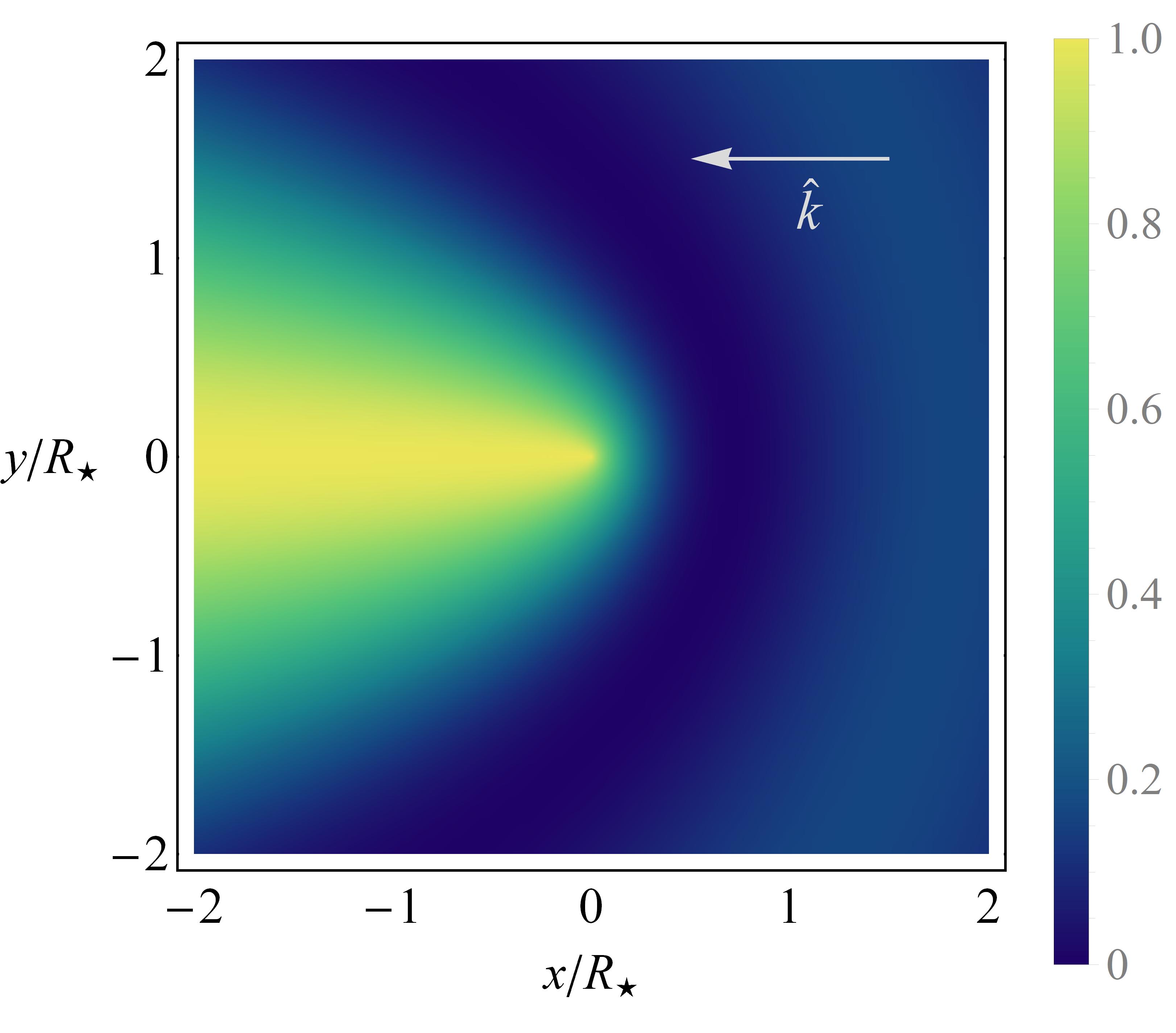}
	\caption{\small{
	\textbf{\emph{Left:}}  The maximum value of the density field $|\psi_\mathbf{k}|^2$, %
 which occurs at $r=0$, for the scattering states $\psi_\mathbf{k}$, as a function of their asymptotic momentum $k$ expressed in terms of $\xi_{\rm foc}(k)\equiv 2\pi/kR_\star=2\pi\alpha/v$. 
	The black line represents the exact result in Eq.\,\eqref{eq:psik0exact}, while the orange and blue lines are the limits $|\psi_\mathbf{k}(0)|^2\to1$ and $|\psi_\mathbf{k}(0)|^2\to\xi_{\rm foc}(k)$ for $\xi_{\rm foc}(k)\ll1$ and $\xi_{\rm foc}(k)\gg1$ (i.e. large/small wave momenta) respectively, where the field is well approximated by a plane wave and a Bessel function, Eqs.~\eqref{eq:psik_plane} and\,\eqref{eq:psik_bessel}, respectively. The turning point where gravitational focusing becomes relevant is at $\xi_{\rm foc}(k)\simeq 1$.
    In the upper axis we also show the corresponding value of $m$ and $R_\star$ if $v=240\km/\s$ and $M=M_\odot$. \textbf{\emph{Right:}} The density field $|\psi_\mathbf{k}|^2$  in the limit $\xi_{\rm foc}(k)\gg1$, normalized to the density at the center $|\psi_\mathbf{k}(0)|^2$ shown in the left panel. Gravitational focusing manifests itself as an overdensity around the object in a region of size $R_\star$, together with a line of overdensities along the direction of $\hat{k}$.
	}
	}\label{fig:psi0}
	\end{center}
	\vspace{-3mm}
\end{figure} 

To better understand the spatial distribution of Eq.\,\eqref{eq:psik_bessel}, in Figure~\ref{fig:psi0} (right) we plot the density normalized to that at the center, $|\psi_\mathbf{k}|^2/|\psi_\mathbf{k}(0)|^2$, for $\hat{k}=-\hat{x}$. The function $J_0$ satisfies $J_0[0]=1$, where it is maximized, and decays as $J_0[a]\to\sin{a}/a$ for $a\to\infty$. As a result, the maximum density occurs in the line $r(1-\hat{k}\cdot\hat{x})=0$. This corresponds to a `tail' of overdensities behind the body along the direction of $\hat{k}$. 

The absolute value $k$ of the momentum appears only as an overall multiplier in Eq.\,\eqref{eq:psik_bessel}. Thus, in this limit $k$ only controls the maximum density (as opposed to its full spatial distribution), which increases as $|\psi_\mathbf{k}(0)|^2=2\pi/kR_\star=\xi_{\rm foc}(k)$ at small momenta, as shown by the orange line of Figure~\ref{fig:psi0} (left).  Additionally, $r$ only enters as the ratio $r/R_\star$. Given that $J_0[a]$ changes by order one between $a=0$ and $a\simeq1$, the amplification of the amplitude occurs around the origin over a region \emph{always} of size $R_\star$, irrespective of $k$; see Figure~\ref{fig:psi0} (right). This amplification is the gravitational focusing discussed in Section~\ref{sec:basic}. Importantly for the growth of the bound state in Section \ref{sec:analytic}, the amplified region tracks the size of the gravitational atom for all $k$ in this limit.
\end{itemize}

The transition between the two regimes is encoded into the hypergeometric function of Eq.\,\eqref{eq:psik}. A good measure to distinguish between them %
is the maximum value of $|\psi_\mathbf{k}|^2$, which always occurs at $r=0$ also for the full scattering states in Eq.\,\eqref{eq:psik} and reads 
\begin{equation}\label{eq:psik0exact}
|\psi_\mathbf{k}(0)|^2=\frac{\xi_{\rm foc}(k)}{1-e^{-{\xi_{\rm foc}(k)}}} \, .
\end{equation}
This quantity is shown in Figure~\ref{fig:psi0} (left) together with its limits $|\psi_\mathbf{k}|^2\to1$ and $|\psi_\mathbf{k}(0)|^2\to\xi_{\rm foc}(k)$ in the two regimes. As mentioned, the turning point between these occurs at around 
$\xi_{\rm foc}(k)=1$%
.  We stress that this corresponds to the wavelength $2\pi/k$ being equal to $R_\star$, rather than the velocity $v$ equal to $\alpha$. In fact, waves with velocity $v= 2\pi \alpha$ still get distorted at the order-one level, even though one could naively guess that they are not affected because their speed exceeds the escape velocity at $r=R_\star$ (which is $\sqrt{2}\alpha$). %

\subsection{DM in the galactic halo}\label{ss:ourDM}

In the vicinity of the Sun, the DM in our galactic halo is a superposition of scattering states. In the reference frame of the Sun, the field takes the form %
\begin{equation}\label{eq:psi_w}
\psi_w(t,\mathbf{x})\equiv\int \frac{d^3k}{(2\pi)^3}a(\mathbf{k})e^{-i\omega_kt}\psi_\mathbf{k}(\mathbf{x}) \, ,
\end{equation}
where $\omega_k=k^2/2m$, $\psi_\mathbf{k}$ is as in Eq.\,\eqref{eq:psik} and $a(\mathbf{k})$ %
encodes the statistical properties of the DM distribution. Over timescales longer than the `coherence' time $\tau_{\rm dm}\equiv 2\pi/mv_{\rm dm}^2$, $a(\mathbf{k})$ can be effectively treated as a random variable~\cite{Foster:2017hbq}. Its norm describes the DM number density present at each momentum $\mathbf{k}$ and, at a given $\mathbf{k}$, it follows a Rayleigh distribution. Its phase should be uniformly distributed in $[0,2\pi)$. In particular, if we indicate by $\langle\cdot\rangle$ the statistical average, which corresponds to averaging over times much longer than $\tau_{\rm dm}$, we expect $\langle a(\mathbf{k})\rangle=0$ and 
\begin{equation}\label{eq:ff}
\langle a^*(\mathbf{k})a(\mathbf{k}')\rangle=(2\pi)^3f(\mathbf{k})\delta(\mathbf{k}-\mathbf{k}') \ .
\end{equation}
More generally, all such averages vanish unless the same number of $a$ and $a^*$ appear (so that the uniformly distributed phases cancel), and in this case they can be reconstructed from Eq.\,\eqref{eq:ff} via Wick's theorem~\cite{PhysRev.80.268}. The function $f(\mathbf{k})$ in Eq.\,\eqref{eq:ff} is the DM occupation number at momentum $\mathbf{k}$ and can be connected to local background DM density $\rho_{\rm dm}$ by requiring that $\rho_{\rm dm}=\langle\rho_w\rangle$ for $|\mathbf{x}|\gg2\pi/k$ (i.e. far from the Sun), which leads to
\begin{equation}
\rho_{\rm dm}=\langle\rho_w\rangle=m\langle |\psi_w|^2\rangle=m\int \frac{d^3k}{(2\pi)^3} f(\mathbf{k})
\,,
\end{equation}
where in the last equality we used the fact that $|\psi_\mathbf{k}(\mathbf{x})|^2\to 1$ in the limit $|\mathbf{x}|\gg2\pi/k$. Thus $f(\mathbf{k})$ can be also seen as the average DM number density per phase space volume, with corresponding number of particles $d N= f(\mathbf{k})d^3kd^3x$.

An approximate form of $f(\mathbf{k})$ is a Gaussian with average momentum $m\mathbf{v}_{\rm dm}\equiv-m\mathbf{v}_\odot$ 
and variance $(m\sigma)^2$, 
where $\mathbf{v}_\odot$
is the Sun's velocity with respect to the galactic rest frame~\cite{Schoenrich:2009bx,Eilers_2019} and $\sigma$ is the DM velocity dispersion in the same frame. In other words, 
\begin{equation}\label{eq:rhok}
f(\mathbf{k})=(2\pi)^{3/2}\frac{\rho_{\rm dm}}{\sigma^3m^4}e^{-(\mathbf{k}-m\mathbf{v}_{\rm dm})^2/(2m^2\sigma^2)} \, .
\end{equation}
We will assume ${v}_{\odot} =v_{\rm dm}= 240$ km/s~(see e.g. \cite{Schoenrich:2009bx,Eilers_2019}). 
In the standard virialized halo model, the density near the Sun is $\rho_{\rm dm} %
\simeq 0.4$ GeV/cm$^3$ \cite{Weber_2010,Nesti:2012zp,Bovy:2012tw,Read:2014qva}, and the DM velocity dispersion is $\sigma=v_c/\sqrt{2}$ where $v_c\simeq230\km/\s$
is the circular velocity of the Milky Way at the position of the Sun~\cite{Eilers_2019}. 
Note that although $v_\odot$ slightly differs from 
$v_c$ 
(because the Sun has a small relative velocity with respect to the average Milky Way motion), we will assume that $v_\odot\simeq v_c$, so that $\sigma\simeq v_{\odot}/\sqrt{2}=v_{\rm dm}/\sqrt{2}$ (in reality, $\sigma$ is slightly smaller than this). In any case, the DM distribution is by no means precisely known, and the above should therefore be treated as a benchmark only; see e.g. \cite{Evans:2018bqy,Necib:2018igl} for further discussion. 

In the following, as in Eq.\,\eqref{eq:xifoc}, we define the typical focusing parameter in the Solar System as 
\begin{equation}
\xi_{\rm foc}\equiv \xi_{\rm foc}(mv_{\rm dm})=\frac{2\pi/(m v_{\rm dm}) }{ R_\star}
=\frac{2\pi \alpha}{v_{\rm dm}}\, ,
\end{equation}
i.e. when $\xi_{\rm foc}$ does not appear with an argument, it is evaluated at $k=mv_{\rm dm}$.

\section{Bound state formation: analytic approach}
\label{sec:analytic}

In the absence of self-interactions, the dark matter of our galactic halo is described the superposition $\psi_w$ of unbound solutions in Eq.\,\eqref{eq:psi_w}. If $g\neq0$, $\psi_w$ is however only an approximate solution of the EoM, Eq.\,\eqref{eq:EoM}. 
We now show that, starting from initial conditions provided by $\psi_w$, the self-interactions lead to capture processes resulting in the growth of the bound-state population: %
The final result describing this evolution is given in Eq.\,\eqref{eq:Ndotavg}. This is derived in two ways: In Section~\ref{ss:classical_p} by solving Eq.\,\eqref{eq:EoM} perturbatively in the self-interaction term, and in Section~\ref{ss:quantum_scattering} using a much simpler quantum-mechanical scattering calculation.

\subsection{Perturbation theory}\label{ss:classical_p}

As long as $g|\psi|^2<m\Phi_{\rm ex}$ (i.e. $\rho<\rho_{\rm crit}$%
), the spectrum of bound states is well-approximated by %
$\psi_{nlm}$. In this regime it makes sense to calculate the rate of change $\dot{N}_{nlm}(t)$ in the number of particles bound to the $nlm$ level, by solving Eq.\,\eqref{eq:EoM} perturbatively.  At the end of this subsection we discuss the limitations of this calculation.
\vspace{-3mm}
\subsubsection*{Direct capture}
\vspace{-2mm}

At $t=0$, we assume that the bound states are not populated and the initial condition is $\psi=\psi^{(0)}\equiv\psi_w$%
. We write the full solution of Eq.\,\eqref{eq:EoM} as $\psi=\psi^{(0)}+\psi^{(1)}+\psi^{(2)}+\dots$, where $\psi^{(i>0)}$ indicates the perturbations of $\psi^{(0)}$ due to the self-interactions. By comparing the left- and right-hand sides of Eq.\,\eqref{eq:EoM}, the effective perturbative parameter of this expansion is $g_{\rm eff}=g\rho/(m\omega)$ where $\rho\simeq \rho_{\rm dm}$ is the local density and $\omega\simeq mv_{\rm dm}^2$ is the typical frequency of the field, and $\psi^{(i)}$ are defined so that $\psi^{(i)}\propto g_{\rm eff}^i$.\footnote{The condition $g_{\rm eff}\ll1$ will be verified for the parameters of interest%
.} %
This expansion is valid over times $0<t<\Delta t$ short enough that the perturbation remains smaller in magnitude than the initial condition, i.e. $|\psi^{(0)}|\gg|\psi^{(1)}|\gg|\psi^{(2)}|\gg\cdots$, 
which is sufficient for the purpose of determining $\dot{N}_{nlm}(t=0)$. 

Since the bound and scattering states $\psi_{nlm}$ and $\psi_{\mathbf{k}}$ 
constitute a complete basis%
, it is convenient to expand the field as 
\begin{equation}\label{eq:deltapsii}
\psi^{(i)}=\sum_{nlm}c^{(i)}_{nlm}(t)e^{-i\omega_nt}\psi_{nlm} +\int \frac{d^3k}{(2\pi)^3} c^{(i)}_{\mathbf{k}}(t)\,e^{-i\omega_{{k}}t}\,\psi_{\mathbf{k}}\,.
\end{equation}
In this way, the total field can be interpreted as a superposition of bound and wave components: The number of particles bound to the $nlm$ level is the bound part of $\int d^3x |\psi|^2$ and at the leading order in the perturbative expansion is simply $N_{nlm}(t)=|c^{(1)}_{nlm}(t)|^2$.%

The coefficient $c^{(1)}_{nlm}(t)$ is obtained by solving Eq.\,\eqref{eq:EoM} at first order in $g_{\rm eff}$, which reads
\begin{equation}\label{eq:EoM1o}
\left(i\partial_t+\frac{\nabla^2}{2m}+\frac{\alpha}{r}\right)\psi^{(1)}=g|\psi^{(0)}|^2\psi^{(0)} \, .
\end{equation}
This is a source equation for $\psi^{(1)}$: Via their self-interactions, the waves induce a smaller field with both bound and wave components. Diagrammatically, this field can be interpreted as arising from $2\to2$ scattering processes mediated by the quartic self-interactions. Indeed, %
a generic `source' term on the right-hand side of Eq.\,\eqref{eq:EoM1o} has the form $g\psi_i\psi_j\psi_k^*e^{-i(\omega_i+\omega_j-\omega_k)t}$ and `induces' a component of $\psi^{(1)}$ with frequency $\omega_{\rm ind}=\omega_i+\omega_j-\omega_k$. Such a term can be associated to the process $i+j\to k+{\rm ind}$.%

Specifically, substituting $\psi^{(0)}=\psi_w$ into Eq.\,\eqref{eq:EoM1o}, multiplying %
by $\psi_{nlm}^* e^{i \omega_n t}$ and using the othonormality of $\psi_{nlm}$, we obtain that the bound-state components are generated with coefficients
\begin{equation}\label{eq:cnlm1}
c^{(1)}_{nlm}(t)=-ig\int[dk_1][dk_2][dk_3]a^*(\mathbf{k}_3)a(\mathbf{k}_1)a(\mathbf{k}_2)\mathcal{M}_{k_1+k_2\to k_3+nlm} \int_{0}^{t} dt'e^{-i(\omega_{k_1}+\omega_{k_2}-\omega_{k_3}-\omega_{n})t'}\, ,
\end{equation}
where we used the short-hand notation $[dk]\equiv d^3k/(2\pi)^3$, and defined the `matrix element' for the $2\to2$ %
process $k_1+k_2\to k_3+nlm$ as 
\begin{equation}\label{eq:Melement}
\mathcal{M}\equiv\mathcal{M}_{k_1+k_2\to k_3+nlm}\equiv\int d^3x\psi_{nlm}^*\psi_{\mathbf{k}_3}^*\psi_{\mathbf{k}_1}\psi_{\mathbf{k}_2} \,.
\end{equation}
Note that $c_{nlm}$ grows, and the corresponding state gets populated, only for momenta for which the energy difference
\begin{equation}\label{eq:Deltaomega}
\Delta\omega\equiv\omega_{k_1}+\omega_{k_2}-\omega_{k_3}-\omega_{n}
\end{equation}
vanishes, i.e. if energy is conserved in this scattering%
. The relation $\langle a^*(\mathbf{k})a(\mathbf{k}')\rangle=(2\pi)^3f(\mathbf{k})\delta(\mathbf{k}-\mathbf{k}')$ in Eq.\,\eqref{eq:ff} allows us to easily compute the average $\langle \dot{N}_{nlm}\rangle$ over times longer than the DM coherence time $\tau_{\rm dm}$%
. Wick's theorem leads to
\begin{equation}\label{eq:dN1}
\langle\dot{N}_{nlm}\rangle=\frac{d}{dt}\langle |c_{nlm}^{(1)}(t)|^2 \rangle =2 g^2  \int [d k_1] [d k_2][d k_3] f(\mathbf{k}_1)f(\mathbf{k}_2)f(\mathbf{k}_3) (2\pi) \delta(\Delta \omega) |\mathcal{M}%
|^2 \, ,
\end{equation}
where we used the identity $\frac{d}{dt}|\int_{0}^tdt'e^{i\Delta\omega t'}|^2=2\pi\delta(\Delta\omega)$ valid in the limit $t\gg 1/\Delta\omega$. %

It is useful to represent the results in Eqs.\,\eqref{eq:cnlm1} and\,\eqref{eq:dN1} via the following diagrams: %
\begin{equation}\label{eq:diagrams1}
\begin{aligned}
\hspace*{-10mm}
\begin{tikzpicture}
[decoration={markings, 
    mark= at position 0.5 with {\arrow{stealth}}}
]

  \node at (-1.3,1.) {$\mathbf{k}_1$};
  \node at (-1.3,-1.) {$\mathbf{k}_2$};
  \node at (1.3,1.) {$\mathbf{k}_3$};
  \node at (1.4,-1.1) {$nlm$};
\node at (-2.,0) {$c^{(1)}_{nlm}=-i$};
\node at (1.6,0) {$,$}; 
\draw[thick,postaction={decorate}] (0,0) to (1,1);
\draw[thick,dotted,postaction={decorate}] (0,0) to (1,-1);
\draw[thick,postaction={decorate}] (-1,-1) to (0,0) ;
\draw[thick,postaction={decorate}] (-1,1) to (0,0);
  
  \begin{scope}
    [xshift=5cm]
\node at (-1.3,1.) {$\mathbf{k}_3$};
  \node at (-1.4,-1.1) {$nlm$};
  \node at (1.3,1.) {$\mathbf{k}_1$};
  \node at (1.3,-1.) {$\mathbf{k}_2$};
\node at (-2.,0) {$c^{(1)*}_{nlm}=i$};
\draw[thick,postaction={decorate}] (0,0) to (1,1);
\draw[thick,postaction={decorate}] (0,0) to (1,-1);
\draw[thick,dotted,postaction={decorate}] (-1,-1) to (0,0) ;
\draw[thick,postaction={decorate}] (-1,1) to (0,0);
\node at (1.6,0) {$,$};  
   \end{scope}
  
   \begin{scope}
  [xshift=11.4cm]
  \node[draw,circle,fill=black,inner sep=2pt] at (0,-0.05) {};
  \node at (-1.3,1.) {$\mathbf{k}_1$};
  \node at (-1.3,-1.) {$\mathbf{k}_2$};
  \node at (1.3,1.) {$\mathbf{k}_3$};
  \node at (1.4,-1.1) {$nlm$};
\node at (-2.5,0) {$\frac{d}{dt}\langle |c_{nlm}^{(1)}|^2\rangle=2\ \times\ $};

\node at (1.8,0) {.};
\draw[thick] (0,0) to (1,1);
\draw[thick] (1.05,0.95) to (0.05,-0.05) ;
\draw[thick,dotted] (0,0) to  (1,-1) ;
\draw[thick,dotted] (0.95,-1.05) to (-0.05,-0.05) ;
\draw[thick] (0,0) to (-1,-1) ;
\draw[thick] (-0.95,-1.05) to (0.05,-0.05) ;
\draw[thick] (0,0) to  (-1,1) ;
\draw[thick] (-1.05,0.95) to (-0.05,-0.05) ;
\end{scope}
\end{tikzpicture}
\end{aligned}
\end{equation}
In the first two diagrams, incoming (outgoing) lines represent incoming (outgoing) states in the process $k_1+k_2\to k_3+nlm$ and are related to the factors $e^{-i \omega_i t}$ ($e^{i \omega_i t}$) in Eq.\,\eqref{eq:cnlm1}. %
Solid lines indicate the (three) factors of the coefficients $a(\mathbf{k})$ (and their conjugate for outgoing states), which come from the zeroth-order field %
from the right-hand side of Eq.\,\eqref{eq:EoM1o}. %
 On the other hand, dotted lines represent the remaining `sourced' state. The last diagram can be thought as the `product' of the first two after the statistical average of Eq.\,\eqref{eq:ff}. In this average, an incoming solid line in the first diagram has been `contracted' with an outgoing solid line in the second diagram: This provides a double line, associated with the occupation number factor $f(\mathbf{k})$ in Eq.\,\eqref{eq:dN1}. The two ways of performing such a contraction give a factor of 2 for the last diagram, as in Eq.\,\eqref{eq:dN1}.

\vspace{-3mm}
\subsubsection*{Stimulated capture and stripping}
\vspace{-2mm}
Eq.\,\eqref{eq:dN1} shows that the number of bound particles grows (linearly) in time via the the capture one of two scattered DM particles. When enough particles have been accumulated in the bound states, %
they could stimulate further capture through Bose enhancement, and  be simulteneously depleted via scattering with the background DM waves. 
To understand these effects, we consider a generic later time $t=t_0>0$, when the bound states are populated with occupation number $N^{(0)}_{nlm}$. At this time the unperturbed solution takes the form $\psi^{(0)}\equiv \psi_w+\psi_b$, with 
\begin{equation}\label{eq:psi_s_ini}
\psi_b\equiv\sum_{nlm}\sqrt{N^{(0)}_{nlm}}e^{-i\omega_{n}t}\psi_{nlm}\,
\end{equation}
representing the bound component. %
As before, we write $\psi=\psi^{(0)}+\psi^{(1)}+\psi^{(2)}+\dots$ and determine the rate $\dot{N}_{nlm}(t_0)$ by solving Eq.\,\eqref{eq:EoM} perturbatively in a time interval $t_0<t<t_0+\Delta t$ short enough that the 
perturbation of the field is smaller than $\psi^{(0)}$. Note that in regions where the bound-state density dominates, the perturbative parameter is $g_{\rm eff}=g\rho/(m\omega)\simeq \rho/\rho_{\rm crit}$ %
(and is smaller than unity by assumption), since $\omega\simeq m\alpha^2$ is the typical field's frequency in the region.

Using Eq.\,\eqref{eq:deltapsii}, $N_{nlm}(t)$ reads %
\begin{equation}\label{eq:Nnlm}
     \Big|\sqrt{N^{(0)}_{nlm}}+c^{(1)}_{nlm}(t)+c^{(2)}_{nlm}(t)+\dots\Big|^2=N_{nlm}^{(0)}+2\sqrt{N^{(0)}_{nlm}}\, {\rm Re} \left[c_{nlm}^{(1)}(t)+c_{nlm}^{(2)}(t)\right] +|c_{nlm}^{(1)}(t)|^2%
     +\dots \, ,
\end{equation}
where in the last expression the dots stand for terms of order $g_{\rm eff}^3$. The change in  $N_{nlm}$ has, as before, a positive contribution from $|c_{nlm}^{(1)}|^2$, of order $g_{\rm eff}^2$, which can be read from Eq.\,\eqref{eq:dN1}. However, an additional time-dependent term -- the second in the right-hand side of Eq.\,\eqref{eq:Nnlm} -- %
appeared from the `interference' of the induced field with the initial bound component. This term is of order $g_{\rm eff}$. In any case, $c_{nlm}^{(1)}\propto aaa^*$ from Eq.\,\eqref{eq:cnlm1}, and if we are interested only in the variation of $N_{nlm}$ over times longer than the DM coherence time $\tau_{\rm dm}$, the contribution proportional to $c_{nlm}^{(1)}$ %
vanishes given the relation $\langle a^*(\mathbf{k}_3)a(\mathbf{k}_1)a(\mathbf{k}_2)\rangle=0$, see Section~\ref{ss:ourDM}. As a result, the full leading contribution to $\langle\dot{N}_{nlm}\rangle$ arises at second order and reads%
\begin{equation}\label{eq:Nnlm_avg}
\langle\dot{N}_{nlm}\rangle=\frac{d}{dt}\langle |c_{nlm}^{(1)}(t)|^2 \rangle+2\sqrt{N^{(0)}_{nlm}}\, {\rm Re} \langle \dot{c}_{nlm}^{(2)}(t)\rangle
\end{equation}
The second term of Eq.\,\eqref{eq:Nnlm_avg} requires calculating the field's perturbation at order $g_{\rm eff}^2$, %
which follows%
\begin{equation} 
\left(i\partial_t+\frac{\nabla^2}{2m}+\frac{\alpha}{r}\right) \psi^{(2)}= g\left(\psi^{(0)2} \psi^{(1)*}+2|\psi^{(0)}|^2 \psi^{(1)}\right).
\label{eq:2ordermain}
\end{equation}
This can be solved similarly to the first-order equation, as explained in %
 Appendix~\ref{app:onelevel}. We give here only a diagrammatic representation of the result. Eq.\,\eqref{eq:2ordermain} is a source equation for $\psi^{(2)}$, which is generated by two powers of $\psi^{(0)}$ and one power of $\psi^{(1)}$. Thus, $c^{(2)}_{nlm}$ is represented by a $2\to2$ diagram where one of legs contains the first-order field $\psi^{(1)}$, and, similarly to Eq.\,\eqref{eq:diagrams1}, this is incoming/outgoing depending on whether $\psi^{(1)}$ appears without/with the complex conjugate in the right hand side of Eq.\,\eqref{eq:2ordermain}. The non-vanishing terms are:

\begin{equation}\label{eq:diagrams2}
\begin{aligned}
\hspace*{-10mm}
\begin{tikzpicture}
[decoration={markings, 
    mark= at position 0.5 with {\arrow{stealth}}}
]

\node at (-4.,0) {$c^{(2)}_{nlm}(t)=$};
  \node at (-2.2,0) {$\mathbf{k}_1$};
   \node at (-2.2,-2) {$\mathbf{k}_2$};
 \node at (-1.4,1.) {$nlm$};
  \node at (-0.9,-0.3) {$\mathbf{k}_3$};
  \node at (1.3,1.) {$\mathbf{p}_1$};
  \node at (1.4,-1) {$\mathbf{p}_2$};
   \node at (0.4,-2) {$nlm$};
\node at (-1.6,-1.) {$-i$};
\node at (-.4,0.) {$i$};

\draw[thick,postaction={decorate}] (0,0) to (1,1);
\draw[thick,postaction={decorate}] (0,0) to (1,-1);
\draw[thick,dotted,postaction={decorate}] (-1,-1) to (0,0) ;
\draw[thick,postaction={decorate}] (-1,1) to (0,0);
\draw[thick,dotted,postaction={decorate}] (-1,-1) to [out=-45,in=135] (0,-2) ;
\draw[thick,postaction={decorate}] (-2,-2) to (-1,-1);
\draw[thick,postaction={decorate}] (-2,0) to (-1,-1);

  \begin{scope}[xshift=5.cm]

\node at (-2.5,0) {$+\,\,\,\, 4\times$};
  \node at (-1.3,1.) {$\mathbf{k}_3$};
  \node at (-1.4,-1.) {$nlm$};
  \node at (1.3,1.) {$\mathbf{k}_2$};
  \node at (0.8,-0.2) {$\mathbf{k}_1$};
    \node at (2.3,0) {$\mathbf{p}_3$};
    \node at (2.3,-2.) {$nlm$};
   \node at (-0.3,-2.) {$\mathbf{p}_2$};
   \node at (3.,0)  {.};
     \node at (-.6,0)  {$-i$};
    \node at (0.4,-1.)  {$-i$};
\draw[thick,postaction={decorate}] (0,0) to  (1,1);
\draw[thick,dotted,postaction={decorate}] (0,0) to  (1,-1);
\draw[thick,postaction={decorate}] (-1,-1) to (0,0) (-1,-1);
\draw[thick,postaction={decorate}] (-1,1) to (0,0);
\draw[thick,postaction={decorate}] (1,-1) to (2,0) ;
\draw[thick,dotted,postaction={decorate}] (1,-1) to (2,-2) ;
\draw[thick,postaction={decorate}] (0,-2) to (1,-1);

\end{scope}
\end{tikzpicture}
\end{aligned}
\end{equation}
These are generated from the first and second terms in Eq.\,\eqref{eq:2ordermain} respectively. In the first (second) diagram, an outgoing (incoming) wave marked with $\mathbf{k}_3$ ($\mathbf{k}_1$) is replaced with the (unbound component of) first-order field, %
$c_{\mathbf{k}_3}^{(1)*}(t)$ ($%
c_{\mathbf{k}_1}^{(1)}(t)$), which is produced by a $2\to2$ scattering processes involving the bound state $nlm$, discussed in Appendix~\ref{app:onelevel}. %

As before, solid lines represent factors of $a(\mathbf{k})$ or $\sqrt{N_{nlm}^{(0)}}$, so $c^{(2)}_{nlm}$ is proportional to $\sqrt{N_{nlm}^{(0)}}$.\footnote{For the second diagram there is an extra combinatoric factor of 4: a factor of 2 from picking one incoming leg in the lower sub-diagram to be dotted, and another factor of 2 from picking one incoming leg in the upper diagram to be the $nlm$ state.} 
One can also view these diagrams as a combination of two first-order sub-diagrams connected by a `propagator'.  
When calculating the average $\langle c_{nlm}^{(2)}\rangle$, 
an outgoing wave is contracted with an incoming one and in the first diagram there are two choices for this. %
After the contraction, the propagator is `cut' and provides the square of the amplitude $\mathcal{M}$ in Eq.\,\eqref{eq:Melement}, in similar fashion to the optical theorem. A straightforward derivation, see Appendix~\ref{app:onelevel}, gives %
\begin{equation}\label{eq:Ndot2contr}
\begin{aligned}
\hspace*{-10mm}
\begin{tikzpicture}

 \node at (-1.3,1.) {$\mathbf{k}_1$};
  \node at (-1.3,-1.) {$\mathbf{k}_2$};
  \node at (1.4,1.) {$nlm$};
  \node at (1.3,-1.1) {$\mathbf{k}_3$};
\node at (-4.,0) {$2\sqrt{N_{nlm}^{(0)}}{\rm Re}\langle \dot{c}_{nlm}^{(2)}\rangle=2\times$};
  \node[draw,circle,fill=black,inner sep=2pt] at (0,-0.05) {};

\draw[thick] (0,0) to (1,1);
\draw[thick] (1.05,0.95) to (0.05,-0.05) ;
\draw[thick,dotted] (0,0) to  (1,-1) ;
\draw[thick,dotted] (0.95,-1.05) to (-0.05,-0.05) ;
\draw[thick] (0,0) to (-1,-1) ;
\draw[thick] (-0.95,-1.05) to (0.05,-0.05) ;
\draw[thick] (0,0) to  (-1,1) ;
\draw[thick] (-1.05,0.95) to (-0.05,-0.05) ;
  \begin{scope}[xshift=5cm]
    \node[draw,circle,fill=black,inner sep=2pt] at (0,-0.05) {};
  \node at (-1.3,1.) {$\mathbf{k}_3$};
  \node at (-1.4,-1.) {$nlm$};
  \node at (1.3,1.) {$\mathbf{k}_2$};
  \node at (1.4,-1.1) {$\mathbf{k}_1$};
\node at (-2.5,0) {$-\,\,\,4\times$};

\node at (2.2,0) {,};
\draw[thick] (0,0) to (1,1);
\draw[thick] (1.05,0.95) to (0.05,-0.05) ;
\draw[thick,dotted] (0,0) to  (1,-1) ;
\draw[thick,dotted] (0.95,-1.05) to (-0.05,-0.05) ;
\draw[thick] (0,0) to (-1,-1) ;
\draw[thick] (-0.95,-1.05) to (0.05,-0.05) ;
\draw[thick] (0,0) to  (-1,1) ;
\draw[thick] (-1.05,0.95) to (-0.05,-0.05) ;
\end{scope}
\end{tikzpicture}
\end{aligned}
\end{equation}
where, as before, double lines are associated to $f(\mathbf{k})$ or $N_{nlm}^{(0)}$. %

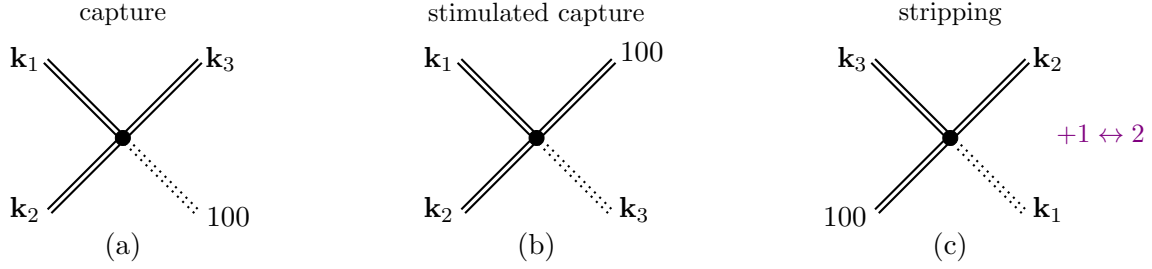
\begin{figure}%
\begin{center}

\begin{tikzpicture}
  \node[draw,circle,fill=black,inner sep=2pt] at (0,-0.05) {};
   \node at (0,1.6) {\small{capture}};
   \node at (0,-1.5) {(a)};
  \node at (-1.3,1.) {$\mathbf{k}_1$};
  \node at (-1.3,-1.) {$\mathbf{k}_2$};
  \node at (1.3,1.) {$\mathbf{k}_3$};
  \node at (1.4,-1.1) {$100$};
\draw[thick] (0,0) to (1,1);
\draw[thick] (1.05,0.95) to (0.05,-0.05) ;
\draw[thick,dotted] (0,0) to  (1,-1) ;
\draw[thick,dotted] (0.95,-1.05) to (-0.05,-0.05) ;
\draw[thick] (0,0) to (-1,-1) ;
\draw[thick] (-0.95,-1.05) to (0.05,-0.05) ;
\draw[thick] (0,0) to  (-1,1) ;
\draw[thick] (-1.05,0.95) to (-0.05,-0.05) ;
  \begin{scope}
    [xshift=5.5cm]
     \node at (0,1.6) {\small{stimulated capture}};
   \node at (0,-1.5) {(b)};
\node[draw,circle,fill=black,inner sep=2pt] at (0,-0.05) {};
  \node at (-1.3,1.) {$\mathbf{k}_1$};
  \node at (-1.3,-1.) {$\mathbf{k}_2$};
  \node at (1.3,-1.) {$\mathbf{k}_3$};
  \node at (1.4,1.1) {$100$};
\draw[thick] (0,0) to (1,1);
\draw[thick] (1.05,0.95) to (0.05,-0.05) ;
\draw[thick,dotted] (0,0) to  (1,-1) ;
\draw[thick,dotted] (0.95,-1.05) to (-0.05,-0.05) ;
\draw[thick] (0,0) to (-1,-1) ;
\draw[thick] (-0.95,-1.05) to (0.05,-0.05) ;
\draw[thick] (0,0) to  (-1,1) ;
\draw[thick] (-1.05,0.95) to (-0.05,-0.05) ;
   \end{scope}
   
   \begin{scope}
  [xshift=11.cm]
  \node[draw,circle,fill=black,inner sep=2pt] at (0,-0.05) {};
   \node at (0,1.6) {\small{stripping}};
   \node at (0,-1.5) {(c)};
   \node at (2.,0) {\small{\color{violet}{$+1\leftrightarrow 2$}}};
  \node at (-1.3,1.) {$\mathbf{k}_3$};
  \node at (1.3,-1.) {$\mathbf{k}_1$};
  \node at (1.3,1.) {$\mathbf{k}_2$};
  \node at (-1.4,-1.1) {$100$};
\draw[thick] (0,0) to (1,1);
\draw[thick] (1.05,0.95) to (0.05,-0.05) ;
\draw[thick,dotted] (0,0) to  (1,-1) ;
\draw[thick,dotted] (0.95,-1.05) to (-0.05,-0.05) ;
\draw[thick] (0,0) to (-1,-1) ;
\draw[thick] (-0.95,-1.05) to (0.05,-0.05) ;
\draw[thick] (0,0) to  (-1,1) ;
\draw[thick] (-1.05,0.95) to (-0.05,-0.05) ;
\end{scope}
\end{tikzpicture}
	\caption{\small{The self-interaction-mediated processes that contribute to the change $\langle\dot{N}_{100}\rangle$ in the number of particles in the ground state of Eq.\,\eqref{eq:Ndotavg}, where $\langle\cdot\rangle$ is the statistical average over times longer than the DM coherence time $\tau_{\rm dm}$. Diagram~(a) corresponds to the capture of the DM particles. It is governed by a 2$\,\to\,$2 process, where two incoming unbound waves scatter into the ground state ($nlm=100$) of the gravitational atom, and an outgoing `energetic' unbound state. Stimulated capture, (b), corresponds to a similar process, but is active when the 100 (or other) states are already populated, and leads to a rate proportional to the occupation number in that state as a result of Bose enhancement. Stripping, (c), depletes the number of particles in the bound state%
	. Double solid lines correspond to the occupation number factors that appear in the rate of Eq.\,\eqref{eq:Ndotavg}. }\label{fig:processes}}
	\end{center}
	\vspace{-3mm}
\end{figure}

By collecting the contributions in Eqs.\,\eqref{eq:dN1} and \eqref{eq:Ndot2contr} and noting that $N_{nlm}^{(0)}=\langle N_{nlm}(t_0)\rangle$, we obtain
\begin{equation}\label{eq:Ndotavg}
\begin{aligned}
\langle\dot{N}_{nlm}\rangle=2g^2\int[dk_1][dk_2][dk_3]\big\{%
f(\mathbf{k}_1)f(\mathbf{k}_2)f(\mathbf{k}_3)%
+\langle N_{nlm}\rangle\left[f(\mathbf{k}_1)f(\mathbf{k}_2)-2f(\mathbf{k}_2)f(\mathbf{k}_3)\right]\big\}(2\pi)\delta(\Delta\omega)|\mathcal{M}|^2 \, ,
\end{aligned}
\end{equation}
where $N_{nlm}$ is evaluated at a generic $t=t_0$ and the second and third terms come from Eq.\,\eqref{eq:Ndot2contr}, where indeed each diagram is proportional to two powers of $f$ and one of $N_{nlm}^{(0)}$. When $N^{(0)}_{nlm}=0$, the first term of Eq.\,\eqref{eq:Ndotavg} dominates and this equation becomes Eq.\,\eqref{eq:dN1}. Thus, at any time until when the critical density is reached, %
the evolution %
follows Eq.\,\eqref{eq:Ndotavg}. We dub the three terms in this equation `capture', `stimulated capture', and `stripping', respectively, and summarize the corresponding physical processes %
in the diagrams in Figure~\ref{fig:processes}.

\vspace{-2mm}
\begin{itemize}[leftmargin=0.2in] \setlength\itemsep{0.15em}
\item
In the capture process, diagram~(a) in Figure \ref{fig:processes}, %
one of the incoming particles transfers enough energy to the other that its velocity decreases and it gets trapped into the gravitational potential, therefore becoming a bound state. The capture term is independent of particles already present in the bound states and thus constitutes an unavoidable positive contribution to the bound state occupation (this corresponds to the first term in Eq.\,\eqref{eq:Mstarintro}). Additionally, it is proportional to the occupation numbers in the initial state, $f(\mathbf{k}_1)$ and $f(\mathbf{k}_2)$, but also to that in the final state $f(\mathbf{k}_3)$, given the three powers of $a$ in Eq.\,\eqref{eq:cnlm1}. %
 
\item Conversely, the stimulated capture and stripping term, diagrams (b) and (c) respectively,  arise from the interference term in Eq.\,\eqref{eq:Nnlm} and vanish if no bound state is initially present. Thus, they are contributions `stimulated' by the (same) bound state being already populated, and are indeed proportional to $N_{nlm}$. Stimulated capture contributes positively to the bound-state occupation, whereas %
stripping contributes negatively and corresponds to the depletion of a bound state by the scattering of a wave, producing another wave with smaller energy. This term has a multiplier of 2 relative to the others because diagram (c) provides a double contribution, by exchanging $\mathbf{k}_1\leftrightarrow \mathbf{k}_2$, as in 
 Eq.\,\eqref{eq:Ndot2contr}. 
Observe that, if only the ground state is populated, i.e. $N_{nlm}^{(0)}=0$ for $nlm\neq100$, stimulated capture and stripping would be irrelevant for all levels that are not $100$.

\end{itemize}

Since $\langle\dot{N}_{nlm} \rangle\propto g^2$, the efficiency of these processes, and the resulting changes to the number of bound particles, do not depend on whether self-interactions are attractive or repulsive. 

\vspace{-1mm}
\subsubsection*{Expected validity}
\vspace{-2mm}

Eq.\,\eqref{eq:Ndotavg} describes the instantaneous variation in the number of bound particles, averaged over times longer than the DM coherence time $\tau_{\rm dm}$. For simplicity of presentation, in Eq.\,\eqref{eq:Ndotavg} we only wrote the contributions to $\dot{N}_{nlm}$ that are linear with $N_{nlm}$. In reality, the changes in occupation number of different $nlm$ levels are not independent from each other, because of additional (more complicated) terms that appear in the right-hand side of Eq.\,\eqref{eq:Ndotavg}. These represent the exchange of particles from one bound state to another, which can be positive or negative, and are nonlinear with $N_{nlm}$. They 
 become more important 
as the number of bound particles
grows,
leading in principle to an infinite set of coupled equations for all the levels. For instance, these include excitation and relaxation terms, 
where bound particles transition to higher/lower $nlm$ states, respectively.
In Appendix~\ref{app:twolevel} we discuss the effect of these terms by studying a simplified two-level system consisting of two bound states, $nlm=100$ and $200$, and show that, if populated, the higher level %
does not deplete the 100 state and can in fact enhance it through de-excitation, e.g. $k_1 + 200\to k_2 + 100$. 
Therefore the captured density described in what follows may be a lower bound to the true result.%

Note that the perturbative approach requires the corrections to the energy levels, typically of order $g\rho/m$, to be much smaller than the unperturbed ones, $\sim m\alpha^2$ (in Appendix~\ref{app:onelevel} we give a precise derivation of such corrections). As anticipated, this holds when $\rho\ll m^2\alpha^2/g \simeq \rho_{\rm crit}$, and as a result this requirement is equivalent to $g_{\rm eff}\lesssim1$.

\subsection{Quantum scattering and Bose enhancement}\label{ss:quantum_scattering}

It is interesting to note that the change in the number of bound particles in Eq.\,\eqref{eq:Ndotavg} can be  understood as arising from $2\to2$ scattering processes at the quantum level, upon adding the expected Bose enhancement factors. First, observe that the action of the scalar field in the nonrelativistic limit reduces to%
\begin{equation}\label{eq:S_nonrel}
S=-\int dt\, d^3x\left[\frac{i}{2}(\dot{\psi}^*\psi-\psi^*\dot{\psi})+\frac{1}{2m}|\nabla\psi|^2+m\Phi_{\rm ex}|\psi|^2+\frac12g|\psi|^4\right]\,,
\end{equation}
where $\psi$ is the propagating degree of freedom, and $\Phi_{\rm ex}=-GM/r$ is the external background field. %
In particular, the EoM in Eq.\,\eqref{eq:EoM} immediately follows by extremizing the action of Eq.\,\eqref{eq:S_nonrel}.

The action $S$ can be quantized by following standard canonical quantization for a non-relativisitic field theory. 
The main difference is that the quadratic part of $S$ implies that the single-particle states are bound states, with wave functions $\psi_{nlm}$ described by Eq.\,\eqref{eq:psi_nlm} of Appendix \ref{app:bound_states}, or waves $\psi_{\mathbf{k}}$ described by Eq.\,\eqref{eq:psik}, which at large distance behave as plane waves, $e^{i\mathbf{k}\cdot\mathbf{x}}$\,. 
These single-particle states are obtained by acting with the corresponding creation operators on the vacuum $\ket{0}$ as $\ket{nlm}=a^\dagger_{nlm}\ket{0}$ and $\ket{k}=a^\dagger_k\ket{0}$. (The free-theory field operator $\hat{\psi}$ is 
constructed from such creation and annihilation operators, with coefficients given by $\psi_{nlm}e^{-i\omega_nt }$ etc.)

The process $k_1+k_2\to k_3+nlm$ is mediated by the (perturbative) interaction Hamiltonian $H_{\rm int}\equiv\int dtd^3x\frac12g|\psi|^4$. Its amplitude is $\mathcal{A}=\bra{k_3\,nlm}T[\hat{H}_{\rm int}]\ket{k_1k_2}=2 g(2\pi)\delta(\Delta\omega)\mathcal{M}$ where $T$ is the Dyson time-ordering operator, $\mathcal{M}$ is as in Eq.\,\eqref{eq:Melement}, and we used 
\begin{equation}
\hat{\psi}(x)\hat{\psi}(x)\ket{k_1k_2}=2\psi_{\mathbf{k}_1}(\mathbf{x})\psi_{\mathbf{k}_2}(\mathbf{x})e^{-i(\omega_{k_1}+\omega_{k_2})t}\ket{0} \, ,
\end{equation}
and similarly for $\bra{k_3\,nlm}\hat{\psi}^\dagger\hat{\psi}^\dagger$.  The probability per unit time of the process $k_1+k_2\to k_3+nlm$ is obtained by squaring $\mathcal{A}$ (stripping off $2\pi\delta(\Delta\omega)$), and reads $(2\pi)\delta(\Delta\omega)4g^2|\mathcal{M}|^2$.

To calculate the total rate of change in the particle number in the $nlm$ state, one needs to subtract the inverse process, which has (in modulus) the same matrix element, multiply by the occupation number of the particles in the initial state, %
and integrate over all possible values of the momenta. This leads to conclude that $\dot{N}_{nlm}$ solves a `Boltzmann' equation, which 
effectively describes the rate of change of particle-number per phase-space volume, and equals
\begin{equation}\label{eq:Ndot_quantum}
\frac{4g^2}{2}\int[dk_1][dk_2][dk_3]\left[(f(\mathbf{k}_3)+1)(N_{nlm}+1)f(\mathbf{k}_1)f(\mathbf{k}_2)-(f(\mathbf{k}_1)+1)(f(\mathbf{k}_2)+1)f(\mathbf{k}_3)N_{nlm}\right](2\pi)\delta(\Delta\omega)|\mathcal{M}|^2\, ,
\end{equation}
where a factor of $1/2$ has been added in order to avoid double-counting from the exchange of the identical particles 1 and 2. The factors $f+1$ and $N_{nlm}+1$ have been inserted in the final states in Eq.\,\eqref{eq:Ndot_quantum} because of  Bose enhancement: The transition rate of indistinguishable particles $P_{a\to b}$ from the state $a$ to $b$ is given by $(N_b+1)P_{a\to b}$, where $P_{a\to b}$ is the rate for distinguishable particles.  %

Owing to the Bose-enhancement factors, at the leading order in $f$ and $N_{nlm}$  Eq.\,\eqref{eq:Ndot_quantum} coincides with Eq.\,\eqref{eq:Ndotavg}, and reproduces the capture, stimulated capture, and stripping terms derived from the classical EoM. In particular, the presence of a macroscopic $\dot{N}_{nlm}$ can be understood as resulting from the Bose enhancement of identical quantum-mechanical particles.%

\subsection{Dilute \emph{vs} dense gravitational atoms}\label{ss:light_vs_heavy}
We can multiply Eq.\,\eqref{eq:Ndotavg} by the boson mass %
to obtain the instantaneous rate of change in the mass $M_{nlm}$ bound to the $nlm$ level, averaged over times longer than the coherence time. This reads
\begin{equation}\label{eq:Mdot}
    \dot{M}_{nlm}=C+ \Gamma {M}_{nlm} +\dots \, ,
\end{equation}
where the dots represent additional terms (not reported in Eq.\,\eqref{eq:Ndotavg}, and not crucial for the main point of this discussion for the 100 level) that are nonlinear with $M_{nlm}$ and describe the exchange of particles between different levels; see Appendix~\ref{app:twolevel} for further details. %
The total bound mass is $M_\star\equiv\int d^3x\rho=\sum_{nlm} M_{nlm}$. Eq.\,\eqref{eq:Mdot} reproduces and generalizes Eq.\,\eqref{eq:Mstarintro} of Section~\ref{sec:basic}, with 
the $nlm$-dependent coefficients $C>0$ and 
 $\Gamma\equiv\Gamma_1-\Gamma_2$ given by %
\begin{equation}\label{eq:CGamma}
    \{\,C,\,\Gamma\,\}=2g^2\int[dk_1][dk_2][dk_3](2\pi)\delta(\Delta\omega)|\mathcal{M}|^2 \big\{\ \ m\,f(\mathbf{k}_1)f(\mathbf{k}_2)f(\mathbf{k}_3) \,,\ \ f(\mathbf{k}_1)f(\mathbf{k}_2)-2f(\mathbf{k}_2)f(\mathbf{k}_3)\ \ \big\} \ ,
\end{equation}
with $\mathcal{M}$ and $\Delta\omega=\omega_{k_1}+\omega_{k_2}-\omega_{k_3}-\omega_{n}$ as in Eqs.~\eqref{eq:Melement} and~\eqref{eq:Deltaomega}. %

\begin{figure}%
\begin{center}
	\includegraphics[width=0.45\textwidth]{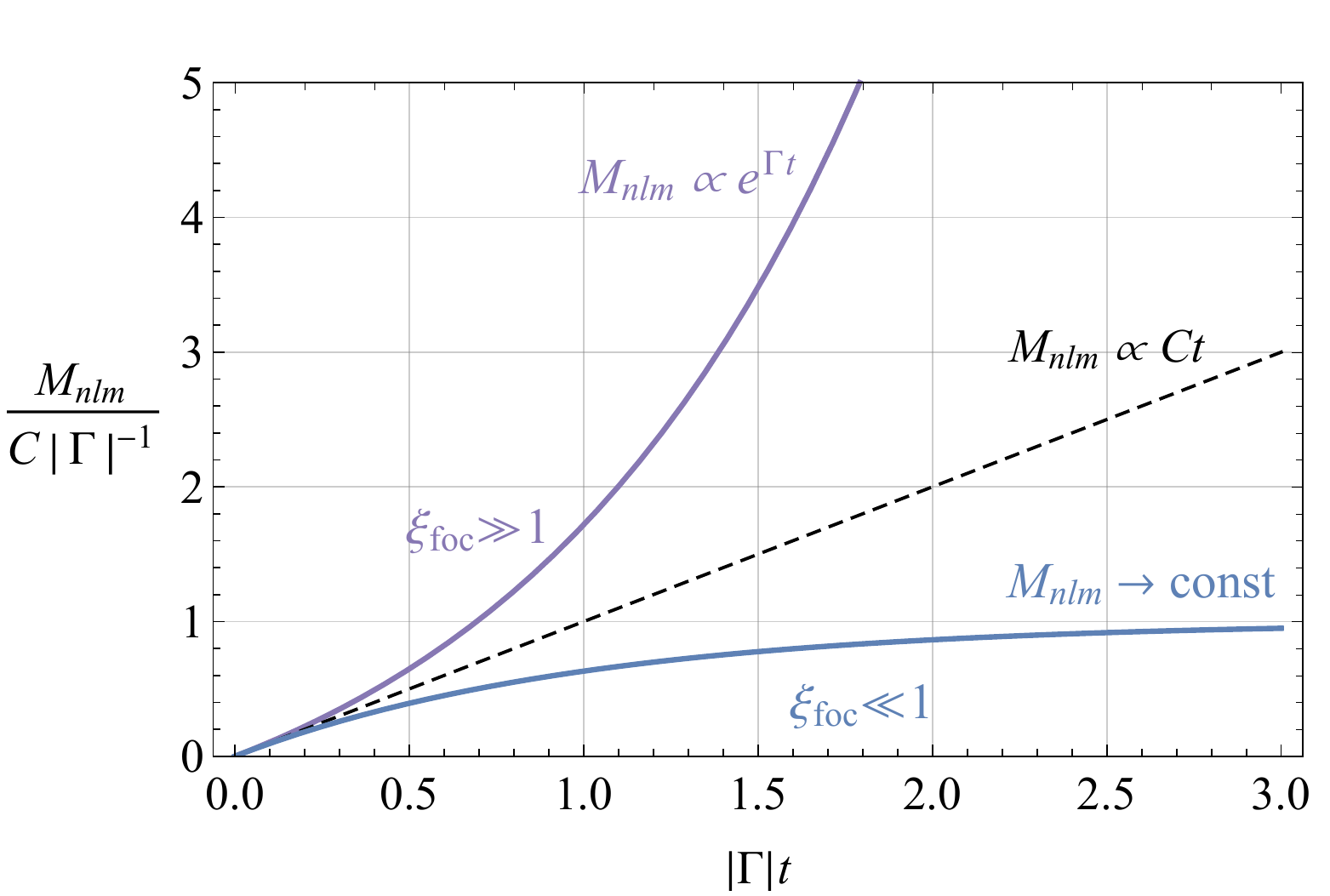}
	\caption{\small{ The time evolution of the mass $M_{nlm}$ bound to the $nlm$ level (neglecting exchange between energy levels), which follows Eq.\,\eqref{eq:Mdot}. %
 The bound mass initially grows as $M_{nlm}=Ct$ (dashed line), as a result of the capture process; this behaviour is valid at $t\ll1/|\Gamma|$. If $\Gamma<0$, which occurs in the limit $\xi_{\rm foc}\ll1$, the mass saturates at the time $t\simeq 1/|\Gamma|$ to the asymptotic value $C/|\Gamma|$ (blue line).
	 If $\Gamma>0$, which occurs in the opposite limit $\xi_{\rm foc}\gg1$ due to to gravitational focusing and the dominance of the stimulated-capture term, the linear growth is instead superseded by the exponential growth $M_{nlm}\propto e^{\Gamma t}$ (purple line). %
  When $M_{nlm}$ is large enough, excited states are likely to decay into the 100 level via terms in the dots of Eq.\,\eqref{eq:Mdot}, see Appendix~\ref{app:twolevel}. 
	}
	}\label{fig:Mt_sat_exp}
	\end{center}
	\vspace{-3mm}
\end{figure}

The matrix element $\mathcal{M}$ is, and therefore $C$ and $\Gamma$ are, independent of the bound mass $M_\star$. Thus, the solution of Eq.\,\eqref{eq:Mdot} %
is straightforward: As mentioned in Section~\ref{sec:basic}, at early times the bound mass increases linearly, $M_{nlm}(t)=Ct$ (dashed line in Figure~\ref{fig:Mt_sat_exp}), with $C$ representing the mass captured per unit time. At $t\simeq1/|\Gamma|$ the bound-state-stimulated processes become relevant. If $\Gamma<0$ the bound mass saturates to the constant value $M_{\star}^{\rm eq}\equiv C/|\Gamma|$ (blue line in Figure~\ref{fig:Mt_sat_exp}), while if $\Gamma>0$ it increases exponentially as $M_{nlm}\propto e^{\Gamma t}$ (purple line%
), starting from the value $C/\Gamma$.\footnote{This behavior would be different for a self-gravitating soliton, studied e.g. in ref.\,\cite{Chan:2022bkz}, for which $C$ and $\Gamma$ do depend on $M_\star$, which provides the (self-)gravitational potential that supports the bound state. In particular, the equation that describes $M_\star(t)$ is nonlinear, resulting in a different time-increase of $M_\star$.%
}

$C$ and $\Gamma$ depend on the gravitational coupling $\alpha$ (e.g. through $\psi_\mathbf{k}$ and $\psi_{nlm}$) and on the DM velocity and dispersion, $v_{\rm dm}$ and $\sigma$ (through $f(\mathbf{k})$). We now focus on the ground state and calculate these two quantities in the limits $\xi_{\rm foc} \equiv 2\pi\alpha/v_{\rm dm}\ll1$ and $\xi_{\rm foc}\gg1$, where the integrals in Eq.\,\eqref{eq:CGamma} are tractable. In the latter limit $1/\Gamma$ is positive and  will track the relaxation time introduced in Section~\ref{sec:basic} (reproduced here for simplicity),
\begin{equation}\label{eq:tau_rel}
\tau_{\rm rel}\equiv \frac{m^3v_{\rm dm}^2}{g^2\rho_{\rm dm}^2}
            \simeq 9\Gyr\left[\frac{f_a}{10^8 \GeV}\right]^4\left[\frac{m}{10^{-14}\eV}\right]^3
               \left[\frac{0.4 \GeV/{\rm cm}^3}{\rho_{\rm dm}}\right]^{2}\left[\frac{v_{\rm dm}}{240\,{\rm km/s}}\right]^2 \, .
\end{equation}
 Eq.\,\eqref{eq:tau_rel} can be easily derived by calculating the inverse of the rate of the process $k_1+k_2\to k_3+k_4$, similarly to Section~\ref{ss:quantum_scattering}. The result is that $1/\tau_{\rm rel}\simeq n_g\Sigma v_{\rm dm}\times f_b$, where $n_g=\rho_{\rm dm}/m$ is the number density of the gas, $\Sigma\simeq m^2g^2$ is the cross section of the $2\to2$ scattering mediated by the self-interactions, and $f_b\simeq n_g/(mv_{\rm dm})^3$ is the occupation number enhancement factor.
 
 Some of the $nlm\neq 100$ states are also populated, but, as discussed below and in Appendix~\ref{app:twolevel}, they are either subdominant or likely to decay to the ground state (when $M_{nlm}$ is large enough) over a time comparable to $\tau_{\rm rel}$,  via the terms in the dots in Eq.\,\eqref{eq:Mdot}, and %
 lead to a faster increase of the 100 level.

%
%
%

\subsubsection*{Dilute gravitational atoms: $\xi_{\rm foc}%
\ll 1$}%

We consider the standard DM halo with $v_{\rm dm}=\sqrt{2}\sigma$ (see Section~\ref{ss:ourDM}). %
Although the integrals in Eq.\,\eqref{eq:CGamma} extend over all values of $\mathbf{k}_1,\mathbf{k}_2$ and $\mathbf{k}_3$, the occupation number  $f(\mathbf{k})$ in Eq.\,\eqref{eq:rhok} is a peaked at $%
m\mathbf{v}_{\rm dm}$ with dispersion $m\sigma$. 
Thus, the dominant contribution to $C$ and $\Gamma$ comes from momenta of order $m v_{\rm dm}$. %
For $C$, this is ensured by the factors of $f$ alone%
 , while for $\Gamma$ this is also a consequence of the $\delta$ function, which sets all the $k_i$ to be of the order $mv_{\rm dm}$ in the limit  $2\pi\alpha/v_{\rm dm}\ll1$%
 .\footnote{More explicitly, energy conservation requires $k_3=\sqrt{k_1^2+k_2^2+m^2\alpha^2/n^2}\simeq\sqrt{k_1^2+k_2^2}%
 $. } 
In other words, in this limit the processes in Figure~\ref{fig:processes} primarily involve waves with momenta of order $m v_{\rm dm}$, because the binding energy of the bound states, $-m\alpha^2/2n^2$, is negligible with respect to the kinetic energy $mv_{\rm dm}^2/2$ of the DM waves.

Consequently, as discussed in Section~\ref{sec:boson_stars_waves}, it is a good approximation to assume that \emph{all} scattering states $\psi_\mathbf{k}$ -- that enter in Eq.\,\eqref{eq:CGamma} through $\mathcal{M}$ -- are plane waves $e^{i\mathbf{k}\cdot\mathbf{x}}$, and the matrix element $\mathcal{M}$ in Eq.\,\eqref{eq:Melement} can be computed analytically.  This matrix element represents the spatial superposition of three plane waves with a bound state, and for the $100$ state it reads
\begin{equation}\label{eq:M_plane_wave}
\mathcal{M}_{k_1+k_2\to k_3+100}\ =\ \int d^3x \,e^{i(\mathbf{k}_1+\mathbf{k}_2-\mathbf{k}_3)\cdot\mathbf{x}}\,\frac{e^{-r/R_\star}}{\sqrt{\pi R^3_\star}}\ =\ \frac{8\sqrt{\pi R^3_\star}}{\left[1+(\Delta k \,R_\star)^2\right]^2} \, .
\end{equation}
This %
depends on the momenta only through the combination $\Delta k R_\star=\Delta v/\alpha$, where $\Delta k \equiv|\mathbf{k}_1+\mathbf{k}_2-\mathbf{k}_3|=m\Delta v$ is the modulus of the momentum `transferred' from the waves to the 
bound state. %

Over most of the integration region in Eq.\,\eqref{eq:CGamma}, $\Delta v$ is of order $v_{\rm dm}$. Thus, $\Delta k R_\star\simeq v_{\rm dm}/\alpha\gg1$ 
(since $\xi_{\rm foc} = 2\pi\alpha/v_{\rm dm}\ll1$)  
and the matrix element in Eq.\,\eqref{eq:M_plane_wave} is suppressed in this limit. This suppression can be understood by the rapidly oscillating phase $\Delta k\gg R_\star^{-1}$ of the plane waves, which averages out in Eq.\,\eqref{eq:M_plane_wave} when integrated over the effective domain fixed by the 
ground-state radius $R_\star$. The matrix elements of higher-$n$ states are even more suppressed%
, with asymptotic behaviour at $\Delta k R_\star\to\infty$ given by $\mathcal{M}_{k_1+k_2\to k_3+n00}\to 8\sqrt{\pi R^3_\star/n^3}/(\Delta k R_\star)^4$. (For these states the cancellation from the oscillating phase is stronger because of the larger region of support for the spatial integration%
.) %

As a result of this suppression, in the limit $2\pi\alpha/v_{\rm dm}\to0$ all the processes of capture, stimulated capture and stripping are inefficient and both the $C$ and $\Gamma$ are suppressed. Intuitively, the particles in the DM waves %
are so fast (relative to those in the 
bound state) that it is improbable that they lose enough energy via $2\to2$ scatterings to acquire a velocity of order of the 
ground-state virial velocity $v_b=\alpha$, and thereby become trapped.
This is 
even more convincing for higher-$n$ states, 
whose
virial velocity $v_b = \alpha/n$ is smaller. Therefore, in the following we will consider only the 100 level; see Appendix \ref{app:twolevel} for the effect of the excited states.

The coefficients $C$ and $\Gamma$ can be calculated by rewriting Eq.\,\eqref{eq:CGamma} in terms of the momenta %
normalized to the inverse %
ground-state radius, $\mathbf{p}_i\equiv\mathbf{k}_i/(m\alpha)$, 
and switching to spherical coordinates $\mathbf{p}_i=p_i \hat{r}+\theta_i \hat{\theta}+\phi_i \hat{\phi}$. Choosing $\mathbf{v}_{\rm dm}$ to point e.g. towards $\hat{z}$, $\Gamma$ simplifies to
\begin{equation}\label{eq:Gamma_partial}
\frac{64 g^2 \rho_{\rm dm}^2\alpha^4}{(\pi mv^2_{\rm dm})^3}\int dp_1dp_2\, d\cos\theta_1\,d\cos\theta_2\,d\cos\theta_3\,d\phi_1d\phi_2\frac{p_1^2p_2^2p_3}{{\left(1+\Delta p^2\right)^4}}\left[e^{-\left(\frac{\alpha}{v_{\rm dm}}\mathbf{p}_1-\hat{z}\right)^2}-2e^{-\left(\frac{\alpha}{v_{\rm dm}}\mathbf{p}_3-\hat{z}\right)^2}\right]e^{-\left(\frac{\alpha}{v_{\rm dm}}\mathbf{p}_2-\hat{z}\right)^2}\, .  
\end{equation}
In the previous equation $\Delta p\equiv |\mathbf{p}_1+\mathbf{p}_2-\mathbf{p}_3|$ and it is understood that %
$p_3=\sqrt{p_1^2+p_2^2+1}$, which arises from the integration over $p_3$ due to the $\delta(\Delta\omega)$ factor.\footnote{We used the identity 
$\delta(p_1^2+p_2^2-p_3^2+1)=(2p_3)^{-1}\sum_{s=\pm1}\delta\left(p_3+s\sqrt{p_1^2+p_2^2+1}\right)$
.} Additionally, %
 the integral over $\phi_3$ has been performed explicitly,\footnote{We took advantage of the freedom in choosing e.g. the $x$-$z$ plane to coincide with the $p_3$-$z$ plane (the $y$ axis is fixed by this choice, and all remaining integrations are therefore nontrivial).} so it is understood that $\phi_3=0$ in the integrand of Eq.\,\eqref{eq:Gamma_partial}. 

\begin{figure}%
\begin{center}
	\includegraphics[width=0.48\textwidth]{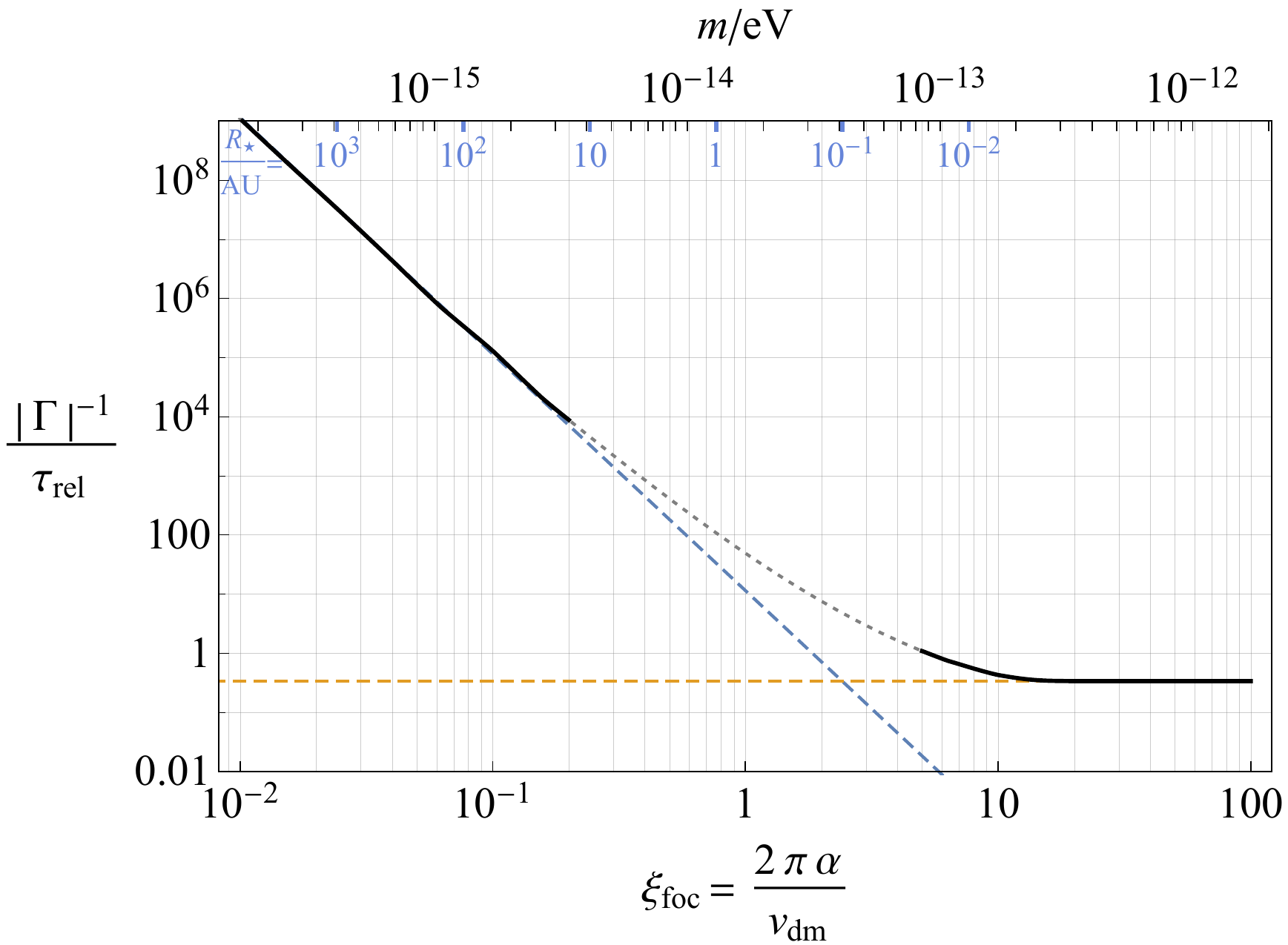}  
	\ \ \ \ 	\includegraphics[width=0.48\textwidth]{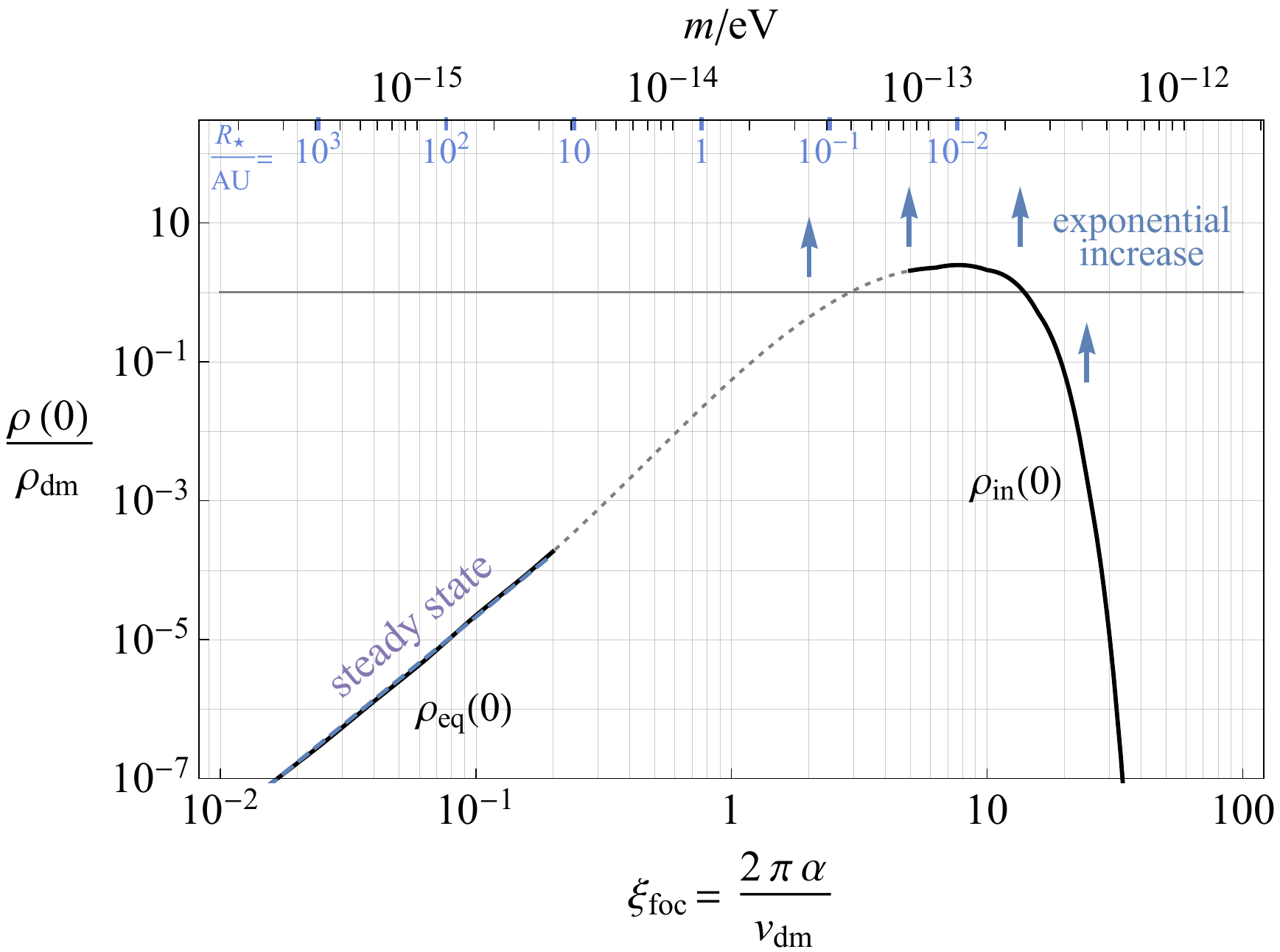}
	\caption{\small{ \textbf{\emph{Left:}} The timescale $1/|\Gamma|$ for the $nlm=100$ level in the limits $\xi_{\rm foc}\ll1$ and $\xi_{\rm foc}\gg1$ (solid lines), in units of the relaxation time via self-interactions $\tau_{\rm rel}=m^3v_{\rm dm}^2/g^2\rho^2_{\rm dm}$, in Eq.\,\eqref{eq:tau_rel}. If $\xi_{\rm foc}\ll1$, $\Gamma$ is negative and suppressed, so the bound mass saturates over a time $1/|\Gamma|\propto (\alpha/v_{\rm dm})^{-4}$ that is parametrically longer than $\tau_{\rm rel}$; see Eq.\,\eqref{eq:gammaInv_1}. If $\xi_{\rm foc}\gg1$, $1/\Gamma$ is positive and coincides with $\tau_{\rm rel}$: the bound mass increases exponentially with timescale $\simeq \tau_{\rm rel}$. 
 \textbf{\emph{Right:}} The central density of the atom in units of the background density at the time $1/|\Gamma|$, when it reaches its equilibrium value for $\xi_{\rm foc}\ll1$ or starts increasing exponentially for $\xi_{\rm foc}\gg1$. In the first case, the equilibrium density is $\rho_{\rm eq}(0)/\rho_{\rm dm}\propto (\alpha/v_{\rm dm})^3$ and is a subpercentage of the DM, see Eq.\,\eqref{eq:Mrhoeq}. In the second case, it is of order one, except for very large $\alpha/v_{\rm dm}$, where it is suppressed because the direct capture process is kinematically disallowed. These small values will in any case exponentially increase in time with rate $\Gamma$. The behavior of the system in transition region $\xi_{\rm foc}\simeq 1$ (dotted lines) is uncertain, and studied with numerical simulations in Section~\ref{sec:simulation}. We show with dotted lines naive interpolations of the two regimes, which likely overestimate $1/|\Gamma|$. %
	}
	}\label{fig:Gamma_rho0}
	\end{center}
	\vspace{-3mm}
\end{figure}

In the limit $2\pi\alpha/v_{\rm dm}\ll1$, the terms dependent on $p_i$ in the exponential factors of Eq.\,\eqref{eq:Gamma_partial} are negligible, because the exponentials are small unless $p_i\simeq v_{\rm dm}/\alpha\gg1$, but for such large values of $p_i$ the matrix element suppresses the integral %
(as mentioned before, for typical momenta in the DM distribution the matrix element is small). Thus, the two addends in Eq.\,\eqref{eq:Gamma_partial} %
coincide except for a factor of $-2$, ultimately due to the multiplicity of the diagram in Figure~\ref{fig:processes}(c). As a result, in this limit $\Gamma_2\simeq 2\Gamma_1$,
and stripping dominates over stimulated capture. %
At the leading order in $\alpha/v_{\rm dm}$, the integral in Eq.\,\eqref{eq:Gamma_partial} is a constant %
and can be estimated as $-A(2\pi)^2 2^3$; %
a numerical evaluation gives $A\simeq0.21$. The corresponding saturation time is %
\begin{align}\label{eq:gammaInv_1}
-1/\Gamma&%
\simeq\frac{2.4}{A}\frac{\tau_{\rm rel}}{\xi_{\rm foc}^4} \nonumber \\
        &\simeq 7\Gyr\left[\frac{f_a}{2\cdot 10^7 \GeV}\right]^4\left[\frac{2\cdot 10^{-15}\eV}{m}\right]
            \left[\frac{M_\odot}{M}\right]^4\left[\frac{0.4 \GeV/{\rm cm}^3}{\rho_{\rm dm}}\right]^{2}\left[\frac{v_{\rm dm}}{240\,{\rm km/s}}\right]^6 \ .
\end{align}
 (The benchmark value of $m$ is smaller than $10^{-14}\eV$, which is more appropriate in this regime.) Due to the $\xi_{\rm foc}^4\ll 1$ factor at the denominator, it takes a time parametrically longer than $\tau_{\rm rel}$ to reach the steady-state regime (roughly a factor of $10^4$ longer, at the benchmark above). In Figure~\ref{fig:Gamma_rho0} (left) we show the full calculation of $\Gamma$ in Eq.\,\eqref{eq:Gamma_partial}, where we can see the %
(small) corrections to Eq.\,\eqref{eq:gammaInv_1}.

Similarly, $C$ reads %
\begin{equation}\label{eq:C_partial}
\frac{512 g^2 \rho_{\rm dm}^3\alpha^4}{\pi^{3/2} m^6  v_{\rm dm}^9}\int dp_1dp_2\, d\cos\theta_1\,d\cos\theta_2\,d\cos\theta_3\,d\phi_1d\phi_2\frac{p_1^2p_2^2p_3}{{\left(1+\Delta p^2\right)^4}}e^{-\left(\frac{\alpha}{v_{\rm dm}}\mathbf{p}_1-\hat{z}\right)^2-\left(\frac{\alpha}{v_{\rm dm}}\mathbf{p}_2-\hat{z}\right)^2-\left(\frac{\alpha}{v_{\rm dm}}\mathbf{p}_3-\hat{z}\right)^2}\, ,
\end{equation}
where the integral can be estimated as 
$B(2\pi)^2 2^3$  with $B\simeq0.08$. %
The mass at equilibrium is $M_{\star}^{\rm eq}=-C/\Gamma\simeq 0.4\cdot 8 \pi ^{3/2} \rho_{\rm dm}/(m v_{\rm dm})^3$ with central density  $\rho_{\rm eq}(0)=M_\star^{\rm eq}/\pi R_\star^3\simeq 0.4\rho_{\rm dm}\xi_{\rm foc}^3/\pi^{5/2}$, as in Eq.\,\eqref{eq:Mrhoeq}.  
The bound state only constitutes a small fraction of the DM density, which can be also understood from the fact that the bound mass $M_\star^{\rm eq}$ is similar that of the background DM contained in a de Broglie wavelength-sized volume $\lambda_{\rm dB}^3$, but $R_\star\gg\lambda_{\rm dB}$ in this limit. 
 In Figure~\ref{fig:Gamma_rho0} (right) we show the overdensity with the full calculation of the integrals, which matches well with the estimate in Eq.\,\eqref{eq:Mrhoeq}.

Since the ratio between the capture and stripping terms is independent of $g$, the overdensity at equilibrium is only a function of $\alpha/v_{\rm dm}$%
. On the other hand, as noticed in Section~\ref{sec:basic}, the timescale to reach equilibrium $|\Gamma|^{-1}$ is strongly dependent on $g$; see Eq.\,\eqref{eq:gammaInv_1}. %
As we show in Appendix~\ref{app:twolevel}, higher levels are also populated at much later times.  The mass bound to the higher levels is the same (when they saturate), which means that the density at the center and at $r\simeq R_\star$ will be dominated by the 100 level (because the excited states have radius larger than $R_\star$)
so that the estimate in Eq.\,\eqref{eq:Mrhoeq} is a good approximation of the full density.

\subsubsection*{Dense gravitational atoms: $\xi_{\rm foc}\gg 1$}%

In the limit $\xi_{\rm foc} =2\pi\alpha/v_{\rm dm}\gg 1$,
the processes of capture and stripping of Figure~\ref{fig:processes}, i.e.\,\,$C$ and $\Gamma_2$ in Eq.\,\eqref{eq:CGamma}, are kinematically suppressed, %
but stimulated-capture ($\Gamma_1$) is not. Indeed, given the density factors in Eq.\,\eqref{eq:CGamma}, $C$ obtains 
its dominant contribution from momenta of order $k_1\simeq k_2\simeq k_3\simeq mv_{\rm dm}\ll 2\pi m \alpha$, and larger momenta are exponentially suppressed. However $\delta(\Delta\omega)$ sets $k_1^2+k_2^2=k_3^2-\alpha^2 m^2/n^2$, forcing $k_3$ to be of order $m\alpha/n$, therefore it is in the UV tail of $f(\mathbf{k}_3)$ (except for  very large $n$, in which case, as described later, the matrix element is suppressed). %
In other words, as evident from Figure~\ref{fig:processes}(a), the particle `3' must be emitted in the final state with (large) energy $k_3^2/2m\simeq m\alpha^2/2n^2$, to compensate for the energy of order $-m\alpha^2/2n^2$ that has been `lost' to the bound state. As a result, the factor $f(\mathbf{k}_3)$ provides a suppression in $C$: direct capture is not enhanced by the Bose enhancement.

 Similarly, in the process of stripping, energy conservation %
 forces the incoming particle, `3' in Figure~\ref{fig:processes}(c), %
 to have (large) momentum of order $m\alpha/n$, in order to strip a bound state while still providing an outgoing state with positive energy. As before, the factor $f(\mathbf{k}_3)$ suppresses $\Gamma_2$ in Eq.\,\eqref{eq:CGamma}, i.e. particles in the DM halo are typically not energetic enough to strip out a bound state. %
 On the other hand, for stimulated capture %
energy conservation requires the \emph{outgoing} momentum, $k_3$ %
in Figure~\ref{fig:processes}(b), to be of order $m\alpha/n$. Since no density factor enters for $k_3$, this term is not suppressed. This leads us to conclude that $\Gamma>0$. %

From the discussion above, the relevant momenta for all the integrals are either $m \alpha/n$ or $m v_{\rm dm}\ll m \alpha$. As a result, the velocities $v$ at play are $\alpha/n$ or $v_{\rm dm}$ and all satisfy $2\pi\alpha/v\gg 1$. Thus, as explained in Section~\ref{sec:boson_stars_waves} and Figure~\ref{fig:psi0} (left), %
 we can approximate \emph{all} scattering states in Eq.\,\eqref{eq:CGamma} with their Bessel limit of Eq.\,\eqref{eq:psik_bessel} {(including those with `large' velocity $v\simeq\alpha$)}. Note that as $2\pi\alpha/v_{\rm dm}$ approaches $1$, for instance $2\pi\alpha/v_{\rm dm}= 3\div4$, $C$ and $\Gamma_2$ are not %
 exponentially suppressed anymore; however, $\delta(\Delta\omega)$ still implies that $k_1,k_2,k_3$ satisfy $2\pi/v_i\gtrsim3\div4$, and the Bessel approximation still holds for the integrands. %

The Bessel form in Eq.\,\eqref{eq:psik_bessel} dramatically simplifies the matrix element in Eq.\,\eqref{eq:Melement}. For instance, for the 100 level this becomes
\begin{equation}\label{eq:M_bessel}
\mathcal{M}_{k_1+k_2\to k_3+100}\ = \frac{2\sqrt{2}\pi}{\sqrt{k_1k_2k_3}}\int d^3\zeta \,e^{-\zeta}\, 
J_0\left[2\zeta(1-\hat{k}_1\cdot\hat{\zeta})\right]
J_0\left[2\zeta(1-\hat{k}_2\cdot\hat{\zeta})\right]
J_0\left[2\zeta(1-\hat{k}_3\cdot\hat{\zeta})\right]
\, ,
\end{equation}
where $\mathbf{\zeta}\equiv\mathbf{x}/R_\star$ is the coordinate in units of $R_\star$, and we omitted a phase that drops out in the absolute value.\footnote{\label{high_n}As discussed below Eq.\,\eqref{eq:psik_bessel}, the Bessel approximation does not extend over all space: at distances larger than $2\pi/k$, $\psi_\mathbf{k}$ recovers a plane wave behaviour. For the $k$ of interest, these distances are however outside the region $\zeta\lesssim1$, where the integrand of Eq.\,\eqref{eq:M_bessel} is exponentially suppressed anyway, being far outside the radius $R_\star$. Very high $n$ states have an effective integration radius larger than $1/k$, and for these the matrix element will be suppressed; see discussion at the end of this Section.} This quantifies the spatial superposition of three scattering states $\psi_\mathbf{k}$ (of the form in Figure~\ref{fig:psi0} (right), oriented in three generic directions) with the ground state. In this superposition, the dependence on $\alpha$ has completely dropped out. This occurs because the typical region where $\psi_\mathbf{k}$ is of order-one follows the ground-state radius $R_\star$ (as $\alpha/v_{\rm dm}$ changes), as in Figure~\ref{fig:psi0} (right); as a result, there is always overlap of order-one between the two. Additionally, as $\alpha/v_{\rm dm}$ gets larger, $R_\star$ and therefore the volume of integration in $\mathcal{M}$ gets smaller, but at the same time the amplitude of $\psi_{\mathbf{k}}$ increases. These two effects compensate to make $\mathcal{M}$ independent of $\alpha/v_{\rm dm}$. Thus, contrary to the previous case, $\mathcal{M}$ is not suppressed. %
Note that the magnitudes of the momenta $k_i$ only appear as an overall factor. The remaining integral in Eq.\,\eqref{eq:M_bessel} is, by rotational invariance, a calculable (order-one) dimensionless function $F[\hat{k}_i\cdot\hat{k}_j]$ of the three scalar products of $\hat{k}_i$,  %
clearly maximized when all momenta are parallel.\footnote{Although we calculated it numerically, $F$ could be in principle approximated analytically by expanding the $J_0$ factors around the origin, making use of spherical integrals such as $\int d\Omega/(4\pi)\hat{\xi}^i\hat{\xi}^j=\delta_{ij}/3$.%
}
For the 100 level, similarly to Eq.\,\eqref{eq:Gamma_partial}, $\Gamma=\Gamma_1-\Gamma_2$ becomes
\begin{equation}\label{eq:Gamma_partial_1}
\frac{8 g^2 \rho_{\rm dm}^2 }{\pi^2 m^3v_{\rm dm}^2}\int dp_1dp_2\, d\cos\theta_1\,d\cos\theta_2\,d\cos\theta_3\,d\phi_1d\phi_2\,{p_1p_2}F^2[\hat{p}_i\cdot\hat{p}_j]\left[e^{-(\mathbf{p}_1-\hat{z})^2}-2e^{-(\mathbf{p}_3-\hat{z})^2}\right]e^{-(\mathbf{p}_2-\hat{z})^2}\, ,
\end{equation}
where we redefined $\mathbf{p}_i=\mathbf{k}_i/(mv_{\rm dm})$ and it is understood that $p_3=\sqrt{p_1^2+p_2^2+\alpha^2/v_{\rm dm}^2}$ and $\phi_3=0$. The integral in  %
$\Gamma_1$ is independent of $\alpha$ and $v_{\rm dm}$, 
while $\Gamma_2$ depends on $\alpha^2/v_{\rm dm}^2$ via $p_3$. Thus, %
\begin{equation}\label{eq:Gamma_exp}
1/\Gamma= c\,\tau_{\rm rel}/[1-s(\alpha^2/v_{\rm dm}^2)] \, ,
\end{equation}
where a numerical calculation gives $c\simeq0.34$, and $s(a)$ arises from the stripping term: %
as anticipated, it is exponentially suppressed for $a\to \infty$, and of order-one otherwise (see Eq.\,\eqref{eq:Cn00} in Appendix \ref{app:twolevel}). 
The result of Eq.\,\eqref{eq:Gamma_exp} is that, when $\xi_{\rm foc}\gg1$, the timescale $1/\Gamma$ is of order $\tau_{\rm rel}$ %
regardless of  the value of $\alpha$%
. The external body acts as a `catalyzer' for the formation of the gravitational atom, but the rate of accretion does not depend on its mass. We show this characteristic time, including the correction from $s(a)$, in Figure~\ref{fig:Gamma_rho0} (left). This correction is of order one for $2\pi\alpha/v_{\rm dm}\simeq3\div4$ (for smaller values the Bessel approximation is not valid anymore), but still leads to a positive $\Gamma$.

The exponential increase starts from the mass $M_{\star}^{\rm in}\simeq
C/\Gamma$ accumulated by DM capture. However, given that capture is kinematically disallowed for $\xi_{\rm foc}\gg1$, this mass and the corresponding 
bound density are also suppressed in this limit. More precisely%
\begin{equation}\label{eq:Mrho_in} 
M_{\star}^{\rm in}\simeq\frac{8 \pi ^{3/2} \rho_{\rm dm}}{(m v_{\rm dm})^3}h(\alpha^2/v_{\rm dm}^2)  \, , \qquad\quad \frac{\rho_{\rm in}(0)}{\rho_{\rm dm}}=\pi^{-\frac52}\left(\frac{2\pi\alpha}{v_{\rm dm}}\right)^3h(\alpha^2/v_{\rm dm}^2) \, 
\end{equation}
where $h(a)\simeq 0.12\, s(a)/(1-s(a))$. We show the density $\rho_{\rm in}(0)/\rho_{\rm dm}$ in Figure~\ref{fig:Gamma_rho0} (right). %
The function $h(a)$ is exponentially suppressed for $a\to\infty$. Consequently, if $\xi_{\rm foc}\lesssim10$, when the exponential growth starts the gravitational atom has a density comparable to the local DM density, as also shown in Figure~\ref{fig:Gamma_rho0}. However, if $\xi_{\rm foc}\gtrsim10$ the suppression of $C$ is more significant and the initial density is much smaller.\footnote{As mentioned in Section~\ref{sec:basic}, for a gravitational atom on the Sun, the point where $2\pi\alpha/v_{\rm dm}\gtrsim 10$ corresponds to $R_\star \lesssim R_\odot$, so our analysis breaks down around this point anyway. This regime is however relevant for other, smaller astrophysical bodies, such as neutron stars.} 

Despite the suppression of $M_\star^{\rm in}$ from direct capture, an overdense 
bound state will still form. In fact, the $100$ state is irreducibly populated, possibly via the decay of higher levels (see below) or by quantum fluctuations. These will get exponentially enhanced and reach a macroscopic density -- comparable to the local DM density -- at the latest by $t\simeq \mathcal{O}(10)\tau_{\rm rel}$.

Let us finally briefly comment on the higher levels, focusing on $n00$. If the typical radius of the excited state (of order $n^2R_\star$) is smaller than $\lambda_{\rm dB}=2\pi/mv_{\rm dm}$, which occurs for (small) $n\lesssim \xi_{\rm foc}^{1/2}$, the Bessel approximation of $\psi_{\mathbf k}$ is accurate to evaluate $\mathcal{M}$ (see also footnote~\ref{high_n}). In this case the matrix element for the $n00$ states is similar to Eq.\,\eqref{eq:M_bessel}, but with an additional overall factor $n^{-3/2}$, and $e^{-\zeta}\to e^{-\zeta/n}P_n(\zeta)$, where $P_n$ is a polynomial of $\zeta$%
; see Eq.\,\eqref{eq:psi_nlm} in Appendix~\ref{app:bound_states}. Depending on the form of $P_n$, $\mathcal{M}$ could be of the same order as for the ground state.

In fact, from a direct calculation, the matrix element $\mathcal{M}$ for e.g. $nlm=200$ 
is of the same order as that of 100 level in the limit $\xi_{\rm foc}\gg1$, so this level is populated at a similar rate. By studying a system with the 100 and 200 states only, we show in Appendix~\ref{app:twolevel} that for $\xi_{\rm foc} \gtrsim 1$ the higher level in any case quickly decays (in a timescale comparable or shorter than $\tau_{\rm rel}$) into the ground state, via the nonlinear terms omitted from Eq.\,\eqref{eq:Mdot}. Note that direct capture is less exponentially suppressed for these higher $n00$ states %
(because $\alpha/v_{\rm dm}$ in Eq.\,\eqref{eq:Mrho_in} is substituted with $\alpha/(nv_{\rm dm})$ with $n>1$); therefore, the ground state will be also populated by relaxation of higher levels.
 Consequently, the result of this Section should be thought as a lower bound to the captured mass.

On the other hand, for (large) $n\gtrsim \xi_{\rm foc}^{1/2}$, the effective integration region in Eq.\,\eqref{eq:Melement}, set by the radius of the excited state, is larger than $\lambda_{\rm dB}$ and one should use the full scattering states in Eq.\,\eqref{eq:psik}. These resemble plane waves for $r\gtrsim \lambda_{\rm dB}$ and, similarly to Eq.\,\eqref{eq:M_plane_wave}, this behavior suppresses the integral in Eq.\,\eqref{eq:Melement}. %
Consequently, for states with $n\gg1$, $\mathcal{M}$ is expected to be small compared to the first few levels, and these are not efficiently populated to start with. %

\subsubsection*{Intermediate regime: $\xi_{\rm foc} \simeq \mathcal{O}(1)$} %

Due to the complex hypergeometric function in Eq.\,\eqref{eq:psik}, it is not feasible to calculate $C$ and $\Gamma$ analytically %
for intermediate values of $\xi_{\rm foc}\simeq \mathcal{O}(1)$; in Figure~\ref{fig:Gamma_rho0} we show a naive interpolation of this region by the dotted lines. 
Nevertheless, we expect that $\Gamma$ changes sign at some critical value $%
 \xi_{\rm foc}=\mathcal{O}(1)$, such that for larger 
  $\xi_{\rm foc}$ the density increases exponentially%
  , and saturates otherwise. In fact, by simulating the system, in Section~\ref{sec:simulation} we provide evidence that this is the case, and we estimate the critical value to be $\xi_{\rm foc}\simeq 1$. The numerical simulations also suggest that $|\Gamma|^{-1}$ remains of order $\tau_{\rm rel}$ even for $\xi_{\rm foc}\simeq 1$, and as a result the interpolation in Figure~\ref{fig:Gamma_rho0} likely overestimates $|\Gamma|^{-1}$ in the transition region. %

\subsection{The fate of the 
gravitational atom and instability}
\label{sec:fate}

As anticipated in Section~\ref{sec:basic}, the exponential growth for $\xi_{\rm foc} \gtrsim 1$ stops once %
the density grows beyond
\begin{equation}\label{eq:rhocrit}
\rho_{\rm crit}\equiv\frac{2|\Phi_{\rm ex}|m^2}{|g|}\simeq \frac{2\alpha^2m^2}{|g|}\simeq 6\cdot 10^4\rho_{\rm dm}          \left[\frac{f_a}{5\cdot 10^7\,{\rm GeV}}\right]^{2}
                \left[\frac{m}{1.7\cdot 10^{-14}\,{\rm eV}}\right]^4
                \left[\frac{M}{M_\odot}\right]^2
                \left[\frac{0.4\,{\rm GeV/cm}^3}{\rho_{\rm dm}}\right]\,.
\end{equation}
In the second equality of Eq.\,\eqref{eq:rhocrit} we fixed $r=R_\star$. %
Although the sign of $g$ does not play a role in the formation of the gravitational atom, when $\rho\simeq \rho_{\rm crit}$ this 
does in fact affect the bound state evolution and stability. Note that in this regime the self-interactions can no longer be treated perturbatively.
\vspace{-2mm}
\begin{itemize}[leftmargin=0.2in] \setlength\itemsep{0.15em}
\item Once the density reaches $\rho_{\rm crit}$, attractive self-interactions ($g<0$) {cannot be compensated by the gradient pressure} and destabilize the bound state, leading to its collapse
in a manner analogous that of a self-gravitating boson star, widely studied in~\cite{Eby:2016cnq,Levkov:2016rkk,Helfer:2016ljl,Eby:2017xrr,Michel:2018nzt}. During this collapse, the density increases until the point where higher order terms in the potential of Eq.\,\eqref{eq:pot} become relevant in the dynamics, and a full relativistic treatment is needed. In the related case of self-gravitating boson stars for an axion-like potential, in the final moments of the collapse the typical field value is of order $f_a$ and the self-interactions lead a rapid emission of relativistic scalar particles (through e.g. $3\to1$ processes~\cite{Eby:2015hyx,Eby:2017azn}), known as a \emph{Bosenova} explosion~\cite{Eby:2016cnq,Levkov:2016rkk}.\footnote{This Bosenova signal has been analyzed as a potential target for axion direct-detection experiments~\cite{Eby:2021ece}, cosmological searches for decaying dark matter~\cite{Fox:2023aat}, and axion indirect detection from photon emission~\cite{Du:2023jxh,Escudero:2023vgv}.}
The analogous process for a collapsing gravitational atom is expected to release an order-one fraction of the bound-state mass into relativistic scalar radiation, which would present novel opportunities for detection. %
 The collapse of a solar halo is worthy of dedicated study, %
 though we discuss some implications in Section~\ref{sec:solarsoliton}.

\item If instead the self-interactions are repulsive ($g>0$), as soon as the density reaches the critical value%
, the balance of forces supporting the gravitational atom changes, such that the external gravitational potential is balanced by the self-interactions rather than the gradient pressure. The critical mass (corresponding the density $\rho_{\rm crit}$) at which this happens is
\begin{equation} \label{eq:Mstarcrit}
  M_{\rm crit} 
            \equiv
            \frac{\pi}{|g| m\alpha}
            = \frac{8\pi f_a^2}{G M m^2}
        \simeq 3\cdot 10^{-13}M_\odot 
    \left[\frac{f_a}{10^8\,{\rm GeV}}\right]^{2}
    \left[\frac{10^{-14}\,{\rm eV}}{m}\right]^2
                \left[\frac{M_\odot}{M}\right]\, .
\end{equation}
By solving $m\Phi=g\rho/m$ at $r=R_\star^{(g>0)}$, using $\rho(R_\star) \simeq M_\star/(\pi R_\star^3)$, we find that this new bound state has radius
\begin{equation} \label{eq:RstarRep}
    R_\star^{(g>0)} \simeq \frac{1}{m}
            \sqrt{\frac{g M_\star}{\pi G M}}\,,
    \qquad {\rm for }\,\, M_\star \gtrsim M_{\rm crit}\,.
\end{equation}

The radius $R_\star^{(g>0)}$ is larger than the Bohr radius $(m\alpha)^{-1}$ and grows as $\sqrt{M_\star}$.\footnote{With an abuse of notation, we wrote $g=1/(8f_a^2)$ in Eq.\,\eqref{eq:Mstarcrit} for repulsive self-interactions. Note that Eq.\,\eqref{eq:RstarRep} is analogous to the Thomas-Fermi radius \cite{Boehmer:2007um} of repulsively-interacting boson stars, and the mass in Eq.\,\eqref{eq:Mstarcrit} is analogous to the critical mass of a self-gravitating boson star~\cite{Chavanis:2011zi,Chavanis:2011zm,Eby:2014fya}; both are easily obtained in the limit $M \to M_\star$. In the repulsively-interacting case, there is an instability stimulated by general-relativistic effects, though the critical mass is too large to be relevant here; it is also easily obtained using the $M\to M_\star$ limit of the self-gravitating case~\cite{Colpi:1986ye}. 
}    At this point, our analytic analysis of the capture rate no longer directly holds; numerical simulations (see next Section and Appendix\,\ref{app:simulations}) suggest that the density quickly saturates after this time. A detailed analysis is worthy of further study.

\end{itemize}

As mentioned in Section~\ref{sec:basic}, for the values of $f_a$ and $m$ in Eq.\,\eqref{eq:rhocrit} that predict $\tau_{\rm rel}$ of the order of the age of the Solar System, the critical density is orders of magnitude larger than  $\rho_{\rm dm}$.\footnote{Note that we can safely ignore the instability in the regime $\xi_{\rm foc}\ll1$, because the saturated density is much smaller than the critical density, as can be understood by comparing Eqs.~\eqref{eq:Mrhoeq} and~\eqref{eq:rhocrit} for the appropriate values of $m$ and $f_a$.} %

\section{Comparison with numerical simulations}
\label{sec:simulation}

 To check the analytic expectations of Section~\ref{sec:analytic}, we evolve  the EoM in Eq.\,\eqref{eq:EoM} on a discrete lattice. The simulation requires one to resolve the UV scale of the $\alpha/r$ divergent potential, which is the radius $R_s$ of the Sun,\footnote{Formally we distinguish the physical radius of the Sun, $R_\odot\simeq 7\cdot10^5\km$, from the variable $R_s$ used in the simulation.} so the EoM are solved with $\alpha/r \to \alpha (3-(r/R_s)^2)/2R_s$ for $r<R_s$. The spacing between grid points should be small enough to resolve both $R_s$ and the radius $R_\star$, and the box needs to contain enough de Broglie wavelengths to resemble the effectively-infinite volume of our galaxy.

 We set the initial conditions as a random realization of the wave superposition $\psi_w$ in Eq.\,\eqref{eq:psi_w} with $f(\mathbf{k})$ fixed to its Gaussian form in Eq.\,\eqref{eq:rhok} and dispersion $\sigma=v_{\rm dm}/\sqrt{2}$.  As shown in Appendix~\ref{app:simulations}, by rescaling coordinates and time as $\mathbf{x}\to m v_{\rm dm} \mathbf{x}$ and $t\to m v^2_{\rm dm}t$, the EoM and the initial conditions depend only on the dimensionless combinations $\xi_{\rm foc}=2\pi\alpha/v_{\rm dm}$, $\tilde{g}\equiv g\rho_{\rm dm}/(m^2v_{\rm dm}^2)$, and $\tilde{R}_s\equiv mv_{\rm dm}R_s$; in particular, the strength of the self-interactions $g$ always enters together with $\rho_{\rm dm}$. Although $\tilde{g}\simeq \mathcal{O}(10^{-8})$ %
 for the values of $m$ and $g$ for which dense atoms form around the Sun in 5 Gyr, in the following we will fix $|\tilde{g}|= 0.006$ and $\tilde{R}_s= 0.3$. These values do not affect the interpretation of the simulation results, but  allow \emph{(i)} the relaxation time $\tau_{\rm rel}$ to be short enough (and therefore the atom to form within the available simulation time), and \emph{(ii)} the Sun to be easily resolved by the grid. More details on the simulations can be found in Appendix~\ref{app:simulations}.

Figure~\ref{fig:rho_sim} (left) shows the evolution of the overdensity $\rho(0)/\rho_{\rm dm}$ at center of the Sun $r=0$ for a fixed initial condition and different values of $\xi_{\rm foc}=2\pi\alpha/v_{\rm dm}$. All the lines have ${g}<0$ except for $\xi_{\rm foc}=2.5,1.5$ for which ${g}>0$ (see below for discussion). At early times, the density fluctuates over the coherence time of the DM waves, $2\pi/mv^2_{\rm dm}$; therefore, we also plot the average over times longer than this period (solid thick lines). Although the fluctuations depend on the stochastic phase of the waves in the initial conditions, their average should be independent of this phase. To compare with our expectations from Section\,\ref{sec:analytic}, we show time in units of the relaxation time $\tau_{\rm rel}$ in Eq.\,\eqref{eq:tau_rel}. We also indicate with a solid disk the prediction of the exponential timescale $1/\Gamma$ in Eq.\,\eqref{eq:tau_rel} based on the Bessel approximation, valid when $\xi_{\rm foc}\gtrsim 3\div 4$ (see Section~\ref{sec:analytic} and Figure~\ref{fig:psi0}), 
and with an empty disk (placed at the end of each simulated range) the value of the critical density $\rho_{\rm crit}/\rho_{\rm dm}=\xi_{\rm foc}^2/(2\pi^2\tilde{g})$ in Eq.\,\eqref{eq:rhocrit}, multiplied by an order-one factor equal to $0.7$.

\begin{figure}%
\begin{center}
	\includegraphics[width=0.5\textwidth]{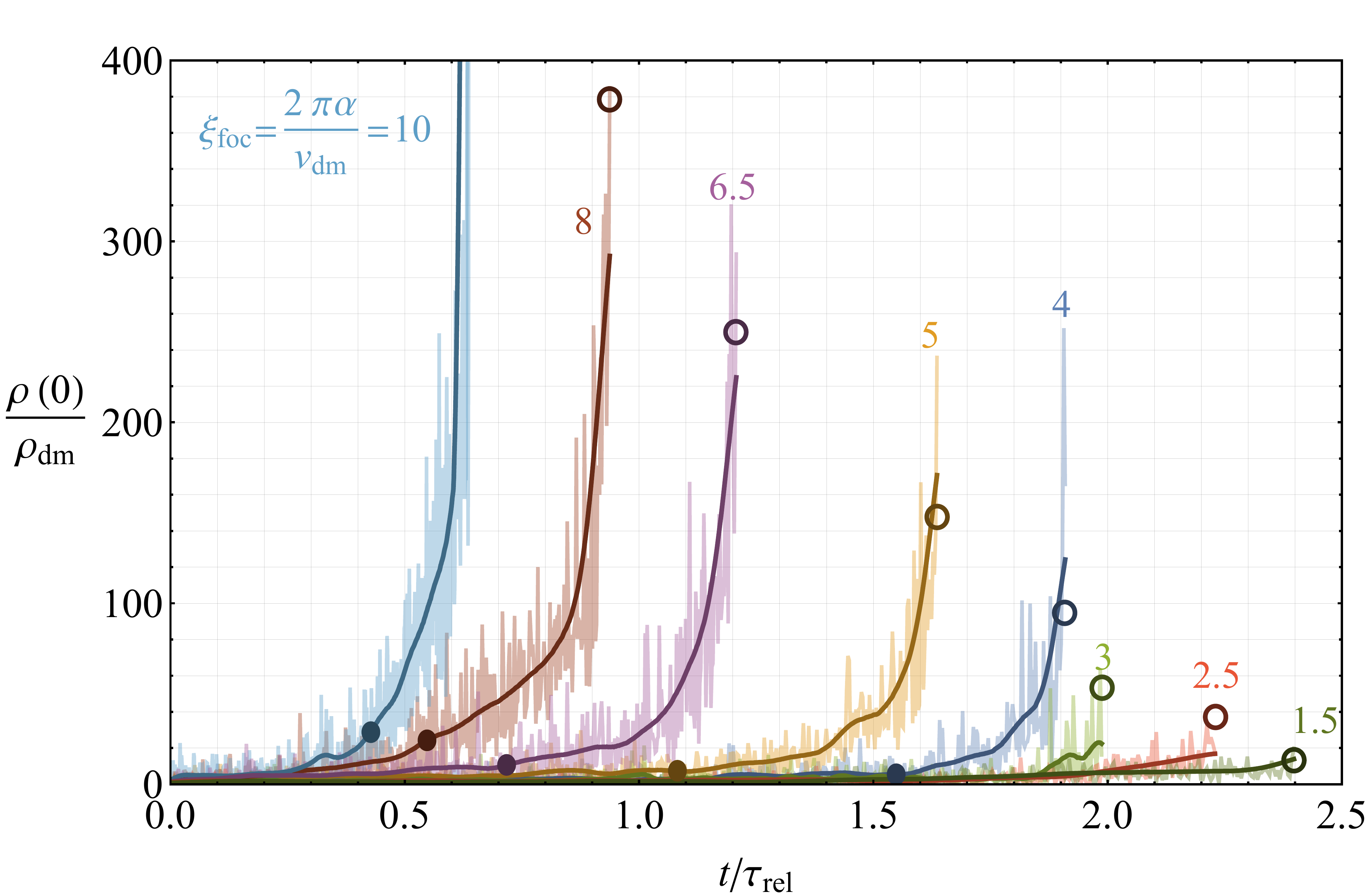}  
	\  \ \  	\includegraphics[width=0.44\textwidth]{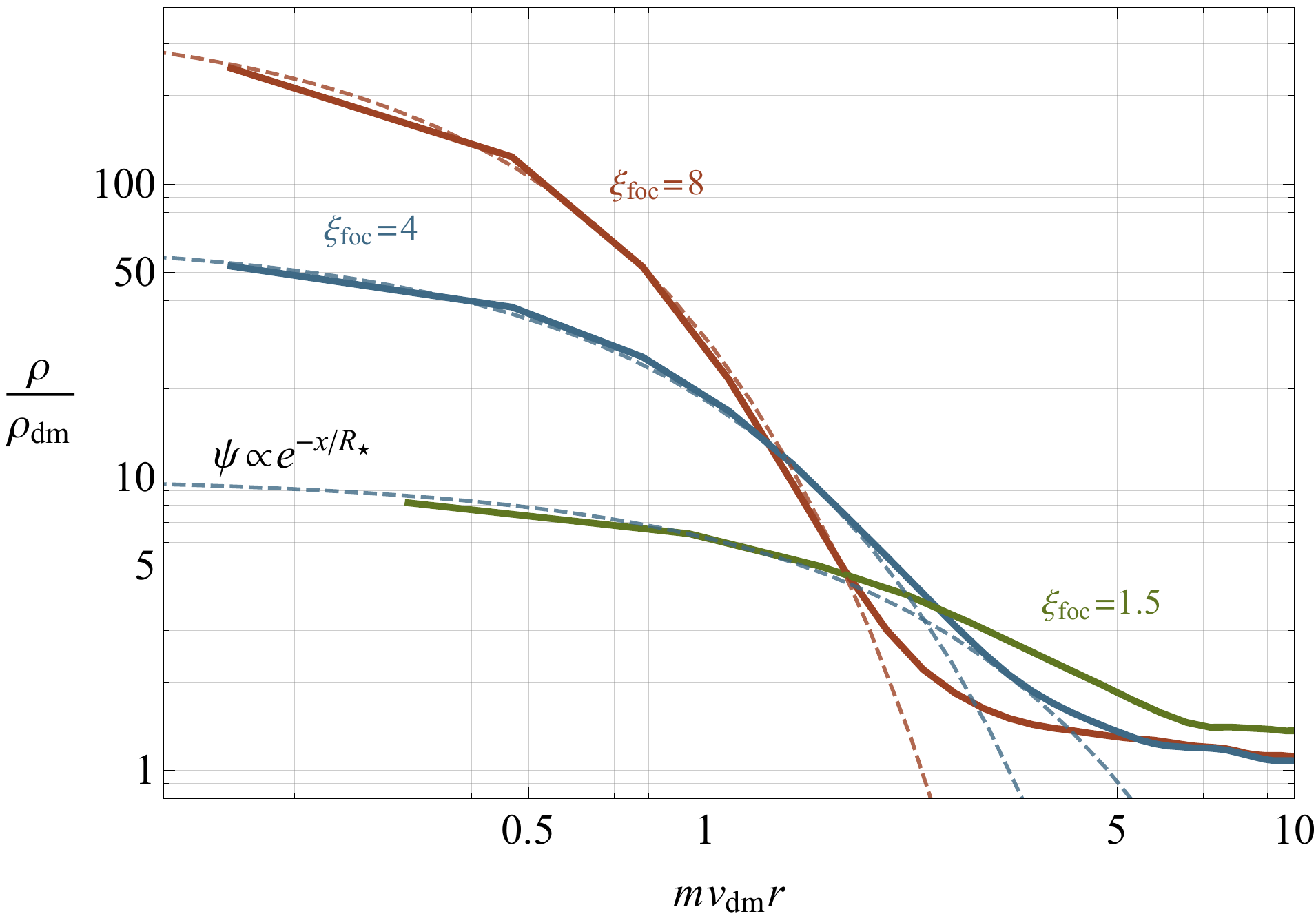}
	\caption{\small{
	\textbf{\emph{Left:}} The time evolution of the density $\rho(0)$ at the position $r=0$ (the center of the Sun) in our numerical simulation, normalized to the average DM density $\rho_{\rm dm}$. Time $t$ is shown in units of the relaxation time via self-interactions $\tau_{\rm rel}$ in Eq.\,\eqref{eq:tau_rel}. Different lines correspond to various choices of $\xi_{\rm foc}=2\pi\alpha/v_{\rm dm}$ as labeled, and we fix $g<0$ (i.e. $\lambda<0$) except for $\xi_{\rm foc}=2.5,1.5$%
 . Empty and solid disks indicate, respectively, the predicted value of the critical density $\rho_{\rm crit}$, and of $1/\Gamma$, which sets the exponential increase timescale. Both of these match well the simulation results. The density fluctuations are related to the stochastic phases of the DM waves and occur over times of order the coherence time of the DM waves. We show with solid lines the average over times larger than this period. \textbf{\emph{Right:}} The density profile of the field obtained by averaging $\rho$ over a spherical volume shell of radius $r$ centered at $r=0$, for $\xi_{\rm foc}=\{8,4,1.5\}$ (solid lines). The profiles are shown during the exponential growth and correspond to the times $t/\tau_{\rm rel}\simeq\{0.9,1.8,2.4\}$. For $r\lesssim R_\star$, the profile is well-fit by the ground state, with the predicted radius $R_\star = (m\alpha)^{-1}$ (dashed), while at larger distances it deviates as the density approaches the DM background.%
	}}\label{fig:rho_sim}
	\end{center}
	\vspace{-3mm}
\end{figure} 

For values of $\xi_{\rm foc}$ definitely larger than 1, the density increases at the predicted timescale. In fact, the growth might be faster than exponential, because also the first few excited levels also experience exponential growth, and then quickly decay to the 100 level, as discussed in Section~\ref{ss:light_vs_heavy} and Appendix~\ref{app:twolevel}. Note that the density continues to oscillate %
also during the exponential increase. These oscillations are not captured by our analytic derivation, which only determines $\dot{N}_\star$ averaged over times much larger than $2\pi/mv_{\rm dm}^2$; see Eq.\,\eqref{eq:Ndotavg}.\footnote{Note that these oscillations might be related to the population of higher levels and could have an impact in the determination of the coherence time of the atom.} %

For $g<0$, once the density is large enough, the halo quickly collapses and the nonrelativistic energy $\int d^3x\,\epsilon$ (see Eq.\,\eqref{eq:energy_density}), is not conserved anymore in the simulation regardless of the time and space-steps, and the simulation is stopped;
to describe the collapse and subsequent evolution of the bound state, the full relativistic EoM are needed. %
 In particular, the density when this occurs in the simulation beautifully matches our predicted critical density, as shown by the empty disks. On the other hand, for $g>0$, energy conservation 
 continues to hold, as the exponential increase stops and the density saturates 
 shortly after the critical density is reached; see also Figure~\ref{fig:rho_sim_small_xi} in Appendix~\ref{app:simulations}.

As expected, as $\xi_{\rm foc}$ decreases, the exponential growth timescale increases, but the latter remains of the order of $\tau_{\rm rel}$ even for $\xi_{\rm foc}\simeq 1$, where the Bessel approximation of Eq.\,\eqref{eq:psik_bessel} has fully broken down.
For values of $\xi_{\rm foc}$ close to 1, %
given the large values of $\tilde{g}$ that are 
feasible for the simulation, the critical density is so small (just a few times the background density%
) that energy non-conservation occurs soon after the increase. Thus, for these values of $\xi_{\rm foc}$ it is more feasible to test the exponential increase for $g>0$. %
Such simulations, shown in Figure~\ref{fig:rho_sim} (left) for $\xi_{\rm foc}=2.5,1.5$, %
 provide evidence that the exponential increase (i.e. $\Gamma>0$) holds for $\xi_{\rm foc}$ close to $1$, %
 or even as small as $\xi_{\rm foc}\simeq1$ as suggested by Figure~\ref{fig:rho_sim_small_xi} in Appendix~\ref{app:simulations}. The exponential timescale continues to be of order the relaxation time, as expected, although we do not attempt to extract it precisely %
 in this work.%

In Figure~\ref{fig:rho_sim} (right) we also show the `spherically averaged' density profile, obtained by averaging $\rho(\mathbf{x})$ over a spherical volume shell of radius $r$ centered at $r=0$ for $\xi_{\rm foc}=1.5,4,8$ during the exponential increase. Also the profile oscillates in time, so we show its time-average over times larger than $2\pi/mv_{\rm dm}^2$. The profile matches that of the ground state with the expected radius given in Eq.\,\eqref{eq:Rstar_sun} (dashed lines). This holds throughout the exponential increase, as shown in Figure~\ref{fig:rho_sim_time} of Appendix \ref{app:simulations}, where the profile is shown at two subsequent times. %
These results nicely confirm our analytic analysis, in particular that for $\xi_{\rm foc}\gtrsim 1$ the mass of the atom increases exponentially over a timescale comparable to the relaxation time.

\section{Solar gravitational atoms}%
\label{sec:solarsoliton}

As anticipated in Section~\ref{sec:basic}, the exponential growth of the gravitational atom around the Sun occurs for $(1\div2)\cdot10^{-14}\eV\lesssim m\lesssim2\cdot10^{-13}\eV$. The upper value corresponds to a radius $R_\star$ smaller than the radius of the Sun, for which the point-like approximation of the Sun is not applicable; see the intersection between the blue and green lines in Figure~\ref{fig:focussing_SS}. In the following we will focus on the mass range above. %

\begin{figure}%
\begin{center}
	\includegraphics[width=0.48\textwidth]{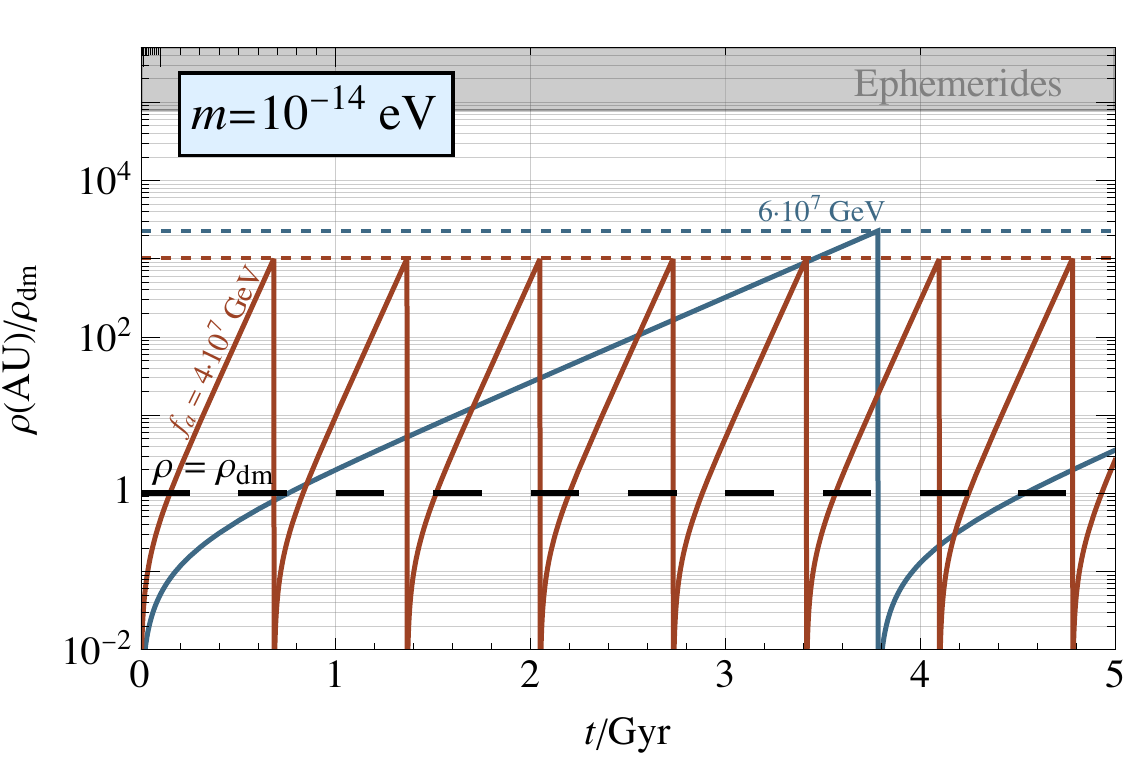}  
	\ \ \ \ 	\includegraphics[width=0.48\textwidth]{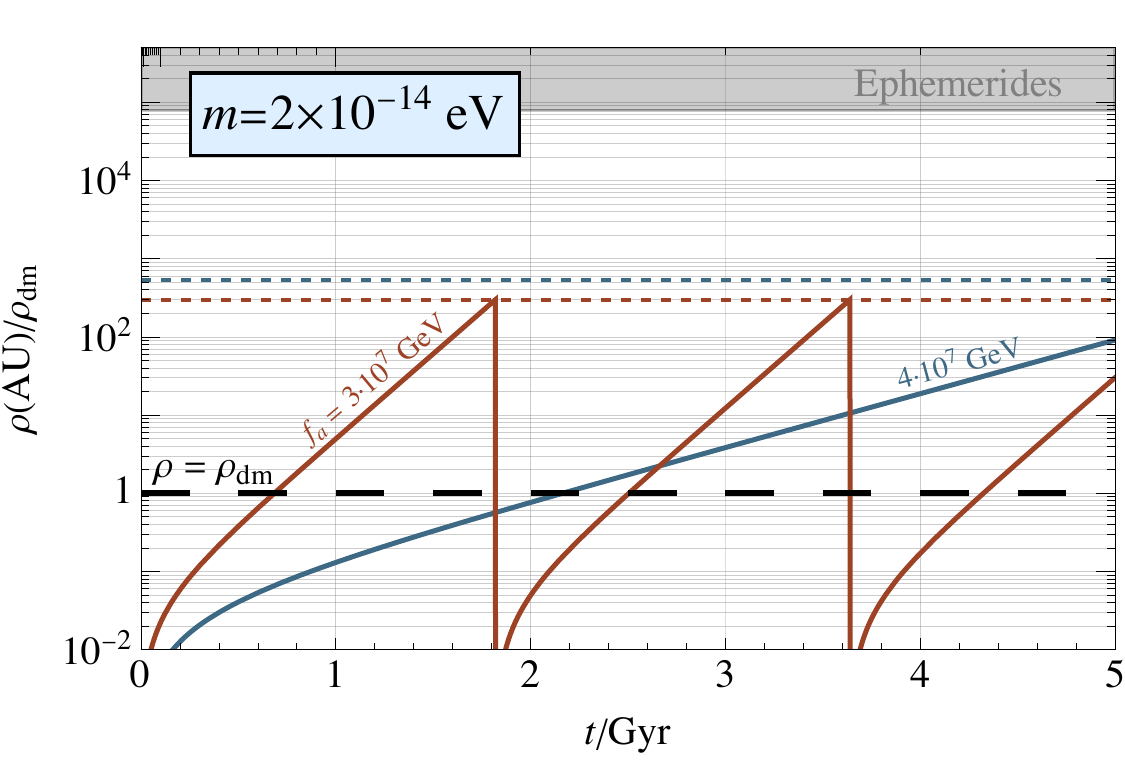}
	\caption{\small{ A sketch of the time-evolution of the density of the gravitational atom (bound to the Sun) at the position of the Earth, $r=1\,{\rm AU}$, for $m=1$ ($2$) $\cdot10^{-14}\eV$ and for $\lambda=-m^2/f_a^2<0$ in the left (right) panel, with $f_a$ fixed as labeled. For the sake of illustration, we assume that the contraction and Bosenova explosion -- expected to occur once the critical density is reached -- are immediate, and that the process then repeats from the beginning. 
	The shaded region represents the direct constraint on the density from Solar System ephemerides \cite{Pitjev:2013sfa}.
	}
	}\label{fig:ExpGrowth}
	\end{center}
	\vspace{-3mm}
\end{figure} 

As an illustrative example, we sketch in Figure \ref{fig:ExpGrowth} the expected time evolution of the overdensity at the position of the Earth, $\rho({\rm AU})/\rho_{\rm dm}$, resulting from the formation of the gravitational atom bound to the Sun, for $m=10^{-14}\eV$ (left) and $m=2\cdot10^{-14}\eV$ (right), and $\lambda=-m^2/f_a^2<0$. In both cases we show the result for two values of $f_a$. Figure \ref{fig:ExpGrowth} assumes that the exponential time scale is of order $\tau_{\rm rel}$ also for $m$ as low as $m\simeq 10^{-14}\eV$ (corresponding to $\xi_{\rm foc}$ slightly smaller than $1$) and that the collapse and Bosenova explosion are immediate.

As can be seen in Figure \ref{fig:ExpGrowth}, the self-interaction strength determines the capture timescale via Eq.\,\eqref{eq:tau_rel} and the maximum density via Eq.\,\eqref{eq:rhocrit}.
For $m\simeq 10^{-14}\eV$, self-interactions with $f_a\simeq 10^7\div10^8\GeV$ are strong enough for the exponential time-scale to be of order the age of the Solar System, while predicting a density $\rho({\rm AU}) > \rho_{\rm dm}$ at the position of Earth. Additionally, larger $m$ requires a smaller $f_a$ to achieve the same density. %
As mentioned in Section~\ref{sec:basic}, the DM overdensity in our Solar System is constrained by direct observation using Solar System ephemerides, at the level of $\rho/\rho_{\rm dm} \lesssim 10^5$ ($10^7$) at a distance of $r \simeq 1$ ($0.3$) AU from the Sun \cite{Pitjev:2013sfa}. This is shown by the gray regions in Figure~\ref{fig:ExpGrowth}.\footnote{Competitive constraints have been set using asteroids as test masses as well \cite{Tsai:2022jnv}.}

\begin{figure}%
\begin{center}
	\includegraphics[width=0.48\textwidth]{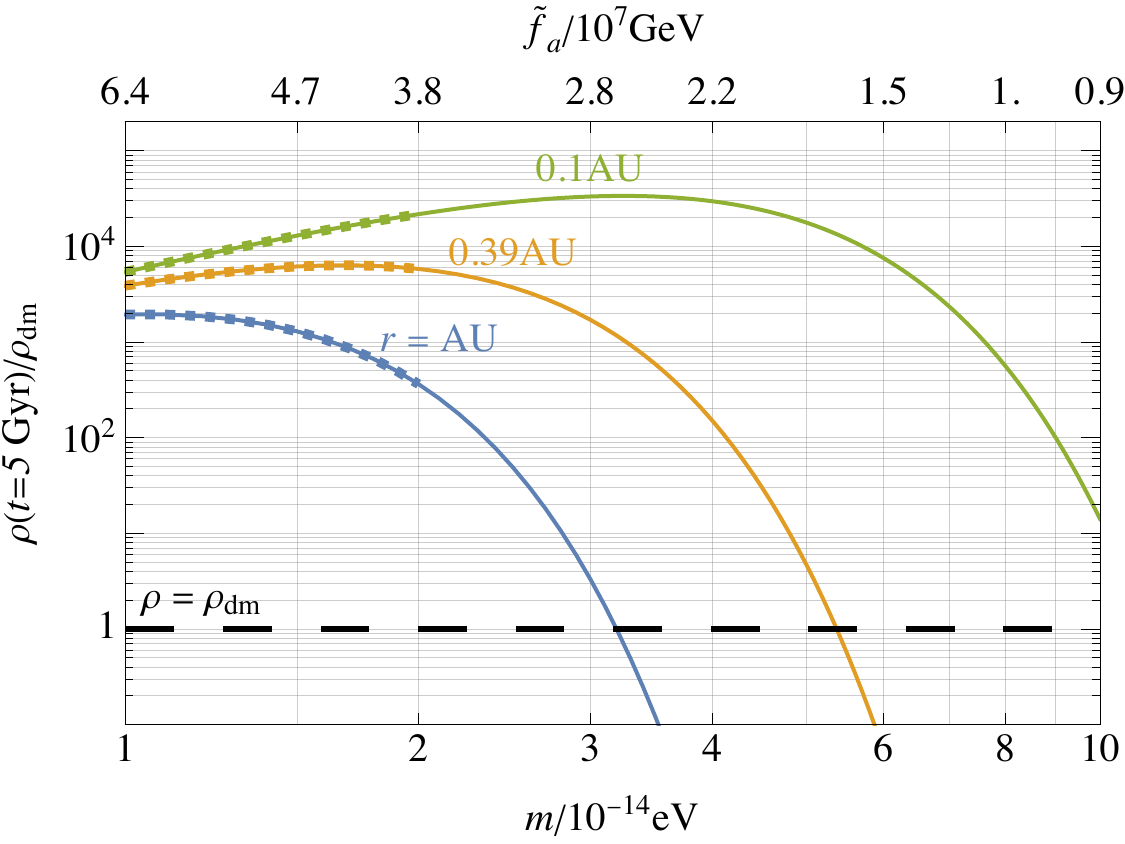}  
 	\includegraphics[width=0.48\textwidth]{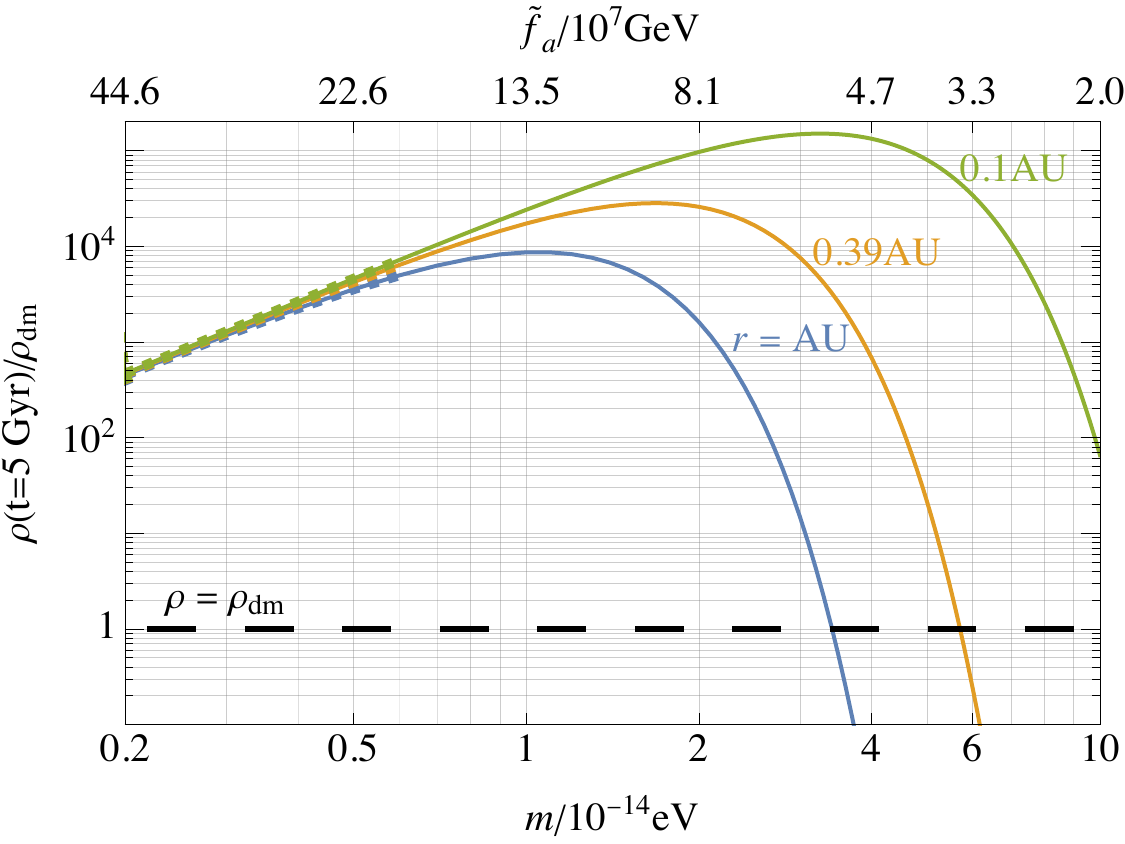}  
	\caption{\small{ The maximum density of the gravitational atom after $5\Gyr$ at $r=1\ {\rm AU}$, $0.39$ AU, and $0.1$ AU (blue, yellow, and green lines, respectively), %
 as a function of $m$. In the upper axis we show the corresponding `optimal' value of $f_a$ for which this maximum density occurs.
 In the left (right) panel, the DM velocity is set to $v_{\rm dm} = 240 \km/\s$ $(50 \km/\s)$. The dashed lines indicate the values of $m$ for which $\xi_{\rm foc}\simeq 1$, which determine the approximate boundary for the exponential growth.
	}
	}\label{fig:MaxRho}
	\end{center}
	\vspace{-3mm}
\end{figure} 

Note that for a fixed value of $m$, there is an `optimal' value of $f_a$, which we denote as $\tilde{f}_a$, for which the density after $t=5\,{\rm Gyr}$ is maximized (this is obtained by requiring $\rho(t=5\,{\rm Gyr}) = \rho_{\rm crit}$). 
For $f_a \gtrsim \tilde{f}_a$, the growth timescale increases as in Eq.\,\eqref{eq:tau_rel}, while for $f_a \lesssim \tilde{f}_a$, the critical density in Eq.\,\eqref{eq:rhocrit} decreases; in both cases, the final density is smaller than that for $f_a=\tilde{f}_a$.  The result for the maximum overdensity as a function of $m$ is shown in Figure~\ref{fig:MaxRho}, where we also report in the upper axis the corresponding optimal value $\tilde{f}_a$. We illustrate the density at a distance of $r=\,\{1,0.39,0.1\}\,$AU (blue, yellow and green lines); the latter two may be relevant for proposed DM searches aboard space missions \cite{Tsai:2021lly}. The dashing between $m=(1\div2)\cdot10^{-14}\eV$ indicates the uncertainty in the point where exponential growth becomes relevant, near $\xi_{\rm foc} \simeq 1$. We use the DM parameters of the standard halo model $v_{\rm dm} = \sigma/\sqrt{2} = 240$ km/s (left panel) and, as an example, also $v_{\rm dm} = \sigma/\sqrt{2} = 50$ km/s (right panel), which could be plausible is the DM resides in a low-dispersion dark disk (see e.g. Refs.~\cite{Read:2008fh,2009ApJ...703.2275P,Kim:2021yyo}).

\subsection*{Solar halos with $\lambda>0$}
For $\lambda>0$, as the density reaches the critical value in Eq.\,\eqref{eq:rhocrit}, %
the (repulsive) self-interaction balance the (attractive) gravitational potential of the Sun, and the radius is given in Eq.\,\eqref{eq:RstarRep}. While no Bosenova occurs in this case, our prediction of a large, stable overdensity is striking and worthy of further study.%

\subsection{Effect of Standard Model couplings}

\label{sec:SMcouplings}
We have so far ignored possible direct couplings of $\phi$ to the SM. These, however, determine the DM direct detection prospects%
. In this Section we provide evidence that such couplings do not significantly alter the formation of the gravitational atom in the Solar System. We will consider the case where $\phi$ %
is a pseudo-scalar field, so that these couplings take the form
\begin{equation} \label{eq:SMcouplings}
    \mathcal{L} \supset 
            \frac14g_{\phi\gamma\gamma}\phi F_{\mu\nu}\tilde{F}^{\mu\nu} 
            + g_\psi \partial_\mu\phi \, \bar{\psi} \gamma^\mu\gamma_5 \psi \, ,
\end{equation}
where $\psi$ is a SM fermion field and $F^{\mu\nu}$ is the photon field strength tensor, with $\tilde{F}^{\mu\nu} = \epsilon^{\mu\nu\alpha\beta}F_{\alpha\beta}/2$. %
The coupling constants $g_{\phi\gamma\gamma}$ and $g_\psi$ in Eq.\,\eqref{eq:SMcouplings} are model-dependent and not necessarily related to the self-interactions, although in many plausible axion theories they are fixed in terms of $f_a$. 

The interaction with the photon in Eq.\,\eqref{eq:SMcouplings} induces the decay $\phi\to\gamma\gamma$, which could deplete the gravitational atom.\footnote{In principle, the scalar field could decay into neutrinos through a $g_\psi$ coupling, if at least one neutrino mass eigenstate has $2m_\nu\lesssim m$. However, the rate is strictly smaller than those considered for photons due to phase-space suppression. Decay to other SM fermions is forbidden for the range of $m$ we consider here.}  The corresponding rate in vacuum, %
\begin{equation}\label{eq:Gamma_a}
   \Gamma_a  \simeq \frac{m^3 g_{\phi\gamma\gamma}^2}{64\pi}
   \simeq 2\cdot 10^{-53}\,{\rm Gyr}^{-1}\left[\frac{m}{10^{-14}\,{\rm eV}}\right]^3\left[\frac{g_{\phi\gamma\gamma}}{10^{-11}\,{\rm GeV}^{-1}}\right]^2\,,
\end{equation} 
is far too slow to be relevant on Solar-System time scales (the benchmark value $g_{\phi\gamma\gamma} \simeq 10^{-11}$ GeV$^{-1}$ in Eq.\,\eqref{eq:Gamma_a} is close to the upper limit for this coupling, see e.g.~\cite{FermiLAT:2016nkz,Reynolds:2019uqt,Dolan:2022kul}). On the other hand, 
`stimulated' decay of the scalars can be induced in the background of a radiation field, e.g. by the large number of photons around the Sun. This enhances the decay rate by a factor of the occupation number of (soft) photons with frequency $\omega \simeq  m/2$; see e.g.~\cite{Caputo:2018vmy}. 
Since the temperature of the Sun and therefore the typical photon energy are $\gg \eV$,  this process is not enhanced in our case.

In the background of a large scalar density $\rho$, also parametric resonance enhances the decay  $\phi\to\gamma\gamma$%
. This effect has been analyzed for both homogeneous and inhomogeneous field configurations and applied to boson stars \cite{Hertzberg:2018zte,Hertzberg:2020dbk,Levkov:2020txo}, and is negligible if
\begin{equation}
    g_{\phi\gamma\gamma}R \sqrt{\rho} 
    \simeq 2\cdot 10^{-5} 
        \left[\frac{g_{\phi\gamma\gamma}}{10^{-11}\,{\rm GeV}^{-1}}\right]
        \left[\frac{10^{-14}\,{\rm eV}}{m}\right]^2
        \sqrt{\frac{\rho}{\rho_{\rm dm}}}
            \lesssim 1\,,
\end{equation}
where $R\simeq R_\star=1/(m\alpha)$ is the typical size of the region where the density is $\rho$. %
This condition is satisfied for the values of $m$ and $\rho$ for the solar halo, %
as shown for instance in Figure~\ref{fig:rhovsr}.\footnote{The analogous scalar couplings, $g_{\gamma}\phi F^{\mu\nu}F_{\mu\nu}$, could be treated similarly, but the constraints on $g_{\gamma}$ are even more restrictive in this case; see \cite{Antypas:2022asj} for a recent review.}

SM couplings could also disrupt the bound state by inducing $2\to2$ scattering processes that eject bound axions out of the gravitational atom. One example is $X+100\to X+{\bf k}$, with $X$ either the photon or the electron. Because the typical energy of photons and electrons from the Sun (of order keV) is larger than the binding energy of the atom, the amplitude for this process is suppressed compared to that of the capture processes in %
Figure \ref{fig:processes}. Additionally, unless the emitted axion remains nonrelativistic, the final state will see negligible Bose enhancement, resulting in further suppression of the scattering rate from SM particles.

A second possibility is that the scalars in the bound state get absorbed as in a process of the form $e+100\to e+\gamma$ (where $e$ is the electron field),  studied for example in \cite{Hochberg:2016sqx}. We merely note that absorption is the inverse of a Compton-like production process, $e+\gamma \to a+e$, which was studied in e.g. \cite{VanTilburg:2020jvl,Lasenby:2020goo}; its effect was found to be negligible for $m\ll \keV$ due to a very large absorption timescale $\tau_{\rm abs} \propto (e^{m/T}-1)^{-1}$, where $T$ is the temperature in the Sun. In our case, there will be further suppression due to the fact that the overlap region between the gravitational atom and the Sun is a small fraction of the total volume of the bound state (as long as $R_\star \gg R_\odot$, or equivalently $m\lesssim 10^{-13}\eV$). We therefore expect axion absorption to be too slow to significantly impact the formation of the gravitational atom.

The discussion above suggests that, for the allowed values of the SM couplings in Eq.\,\eqref{eq:SMcouplings}, the formation of the solar halo is unlikely to be affected by solar dynamics of e.g. photons, electrons, or other SM particles.

\section{Bounds and impact on structure formation}
\label{sec:lss}
 
Sufficiently strong DM self-interactions alter the early-universe evolution of DM perturbations, and are constrained by measurements e.g. of the matter power spectrum, as discussed in~\cite{Arvanitaki:2014faa,Fan:2016rda,Cembranos:2018ulm}.  %
Specifically, from Eq.\,\eqref{eq:pot}, the Fourier transform of the DM overdensity field $\delta\equiv (\rho-\bar{\rho})/\bar{\rho}$ follows~\cite{Chavanis:2011uv,Suarez:2015fga,Hwang:2009js,Chavanis:2016dab,Hu:2000ke,Arvanitaki:2019rax}
\begin{equation}\label{eq:deltak}
\ddot{\delta}_\mathbf{k}+2H\dot{\delta}_\mathbf{k}-4\pi G\,\bar{\rho}\left[1\pm\left(\frac{k}{k_\lambda}\right)^2-\left(\frac{k}{k_J}\right)^4\right]\delta_\mathbf{k}=0 \, ,
\end{equation}
where $\bar{\rho}\propto a^{-3}$ is the average DM energy density, $k$ is the comoving momentum, $H\equiv \dot{a}/a$ is the Hubble parameter and $a$ is the scale factor. The second and third terms in the square bracket of Eq.\,\eqref{eq:deltak} modify the standard Meszaros equation and are a consequence of the self-interactions and the small ULDM mass, and can be interpreted as arising from an effective sound speed (in the second term, $\pm$ refers to attractive/repulsive self-interactions). %
The two comoving momentum scales $k_\lambda(a)$ and $k_J(a)$ in Eq.\,\eqref{eq:deltak} depend on time and are associated to the self-interactions and to the so-called `quantum pressure' of the ULDM field. They are defined by
\begin{equation}\label{eq:kl}
    k_\lambda\equiv a\left(\frac{16\pi G m^4}{|\lambda|}\right)^\frac12 \simeq 2.7 \Mpc^{-1}\left[\frac{a}{a_{\rm eq}}\right]\left[\frac{m}{10^{-14}\eV}\right]\left[\frac{f_a}{10^7\GeV}\right] \,, 
\end{equation}
and 
\begin{equation}\label{eq:kJ}
    k_J\equiv a\left(16\pi G\bar{\rho}m^2\right)^\frac14  \simeq \, 10 \Mpc^{-1}\left[\frac{a}{a_{\rm eq}}\right]^\frac14\left[\frac{m}{10^{-22}\eV}\right]^\frac12 \, ,
\end{equation}
where $a_{\rm eq}$ is the scale factor at the time of matter-radiation equality. 
Note that Eq.\,\eqref{eq:deltak} holds only when the mode $\delta_\mathbf{k}$ is subhorizon, i.e. $k/a>H$.%

The self-interaction and quantum pressure terms become dominant and change the standard growth of the primordial perturbations (which is $\delta_\mathbf{k}\propto 1+(3/2)a/a_{\rm eq}$) for $k$ larger than the critical momenta $k_\lambda$ and $k_J$ respectively, while smaller momentum scales evolve as for standard cold DM. In particular, attractive self-interactions lead to an exponential increase of $\delta_\mathbf{k}$ (until $\delta_\mathbf{k}\simeq 1$, at which point non-linear collapse into bound objects is expected to occur), while repulsive self-interactions and quantum pressure prevent their growth by making $\delta_{\mathbf{k}}$ oscillate.%

\begin{figure}%
\begin{center}
 \includegraphics[width=0.7\textwidth]{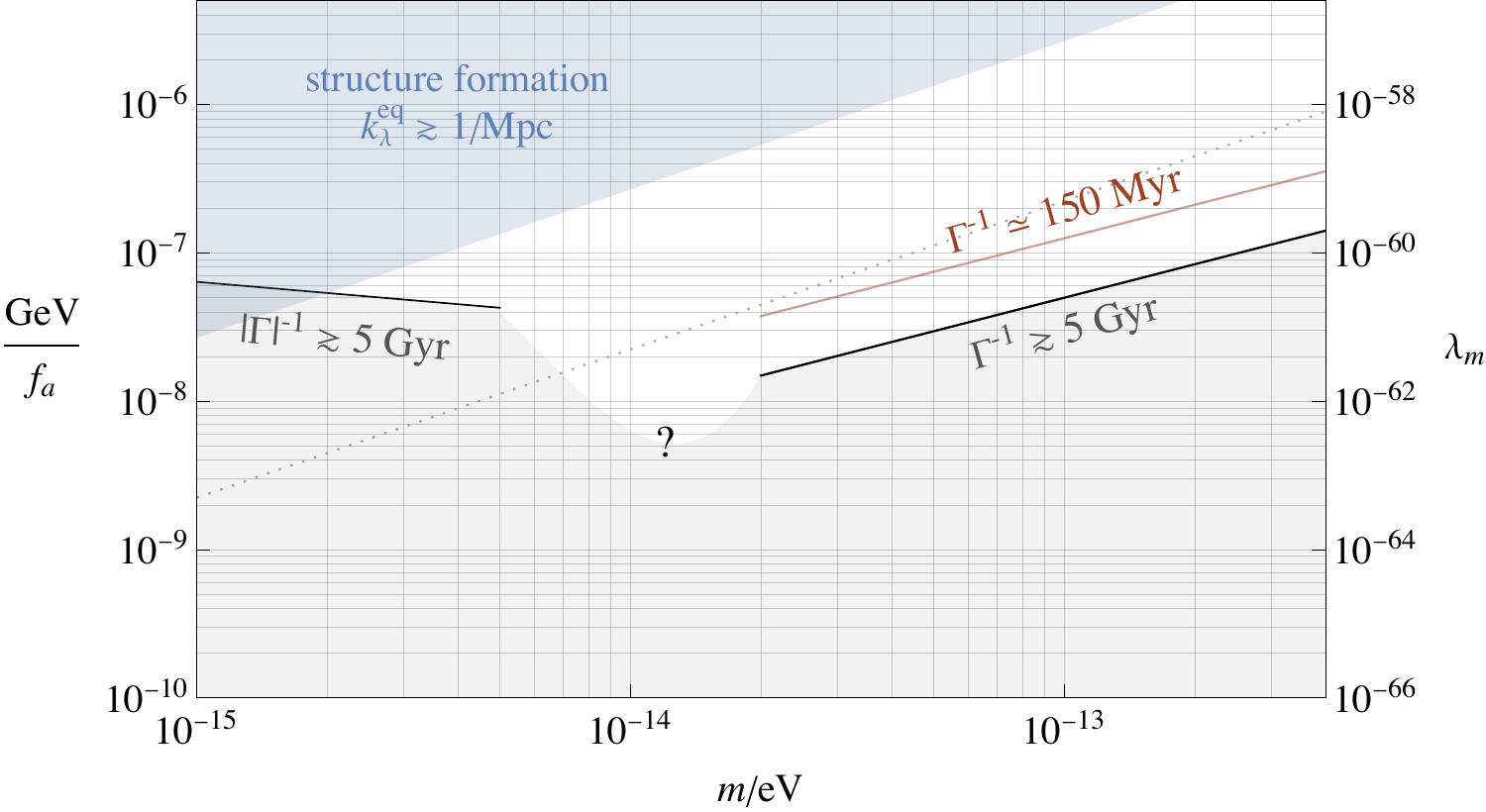} 
	\caption{\small{ Blue region: Approximate constraint on the size of the self-coupling $\lambda$ (parameterized  for convenience by $f_a$) from the matter power spectrum at largest measured momenta, $k\simeq 1\Mpc^{-1}$. This bound applies both for $\lambda>0$ and $\lambda<0$, and considers the effect of the self-interactions only after matter-radiation equality. %
The dotted line is an indication of an upper bound that could be set on $\lambda$ (for attractive self-interactions) from the matter power spectrum due to the exponential enhancement of the perturbations during radiation domination, see Eq.\,\eqref{eq:delta_rad}. Black lines show the values of $m$ and $\lambda$ for which the typical formation time $|\Gamma|^{-1}$ of the atom in the Solar System equals $5\Gyr$ or $150\,{\rm Myr}$. Over the gray region, the self-coupling is too weak for the formation to take place within $5\Gyr$. We show in the axis on the right the corresponding value of $\lambda$ parameterized by $\lambda_m\equiv|\lambda|(10^{-14}\eV/m)^{2}$. {Although the misalignment mechanism for a conventional axion potential underproduces the DM abundance for the plotted values of $m$ and $f_a$, other scalar potentials or different production mechanisms could provide the correct abundance.}%
	}
	}
	\label{fig:structure_form}
\end{center}
	\vspace{-4mm}
\end{figure}

As described in Refs.~\cite{Arvanitaki:2014faa,Fan:2016rda}, a rough conservative bound is obtained by requiring that both of these effects are negligible at the largest comoving momentum at which the matter power spectrum has been measured precisely (from CMB, large scale structure or Lyman-$\alpha$ data)%
, i.e. approximately $k\simeq 1\Mpc^{-1}$ (see e.g.~\cite{SDSS:2003eyi,Ivanov:2019pdj,Appleby:2020pem}). %
Given that $k_\lambda\propto a$ and $k_J\propto a^{1/4}$, see Eqs.~(\ref{eq:kl}-\ref{eq:kJ}), the corresponding bound on $m$ and $\lambda$ is the strongest at the earliest times at which Eq.\,\eqref{eq:deltak} is valid (i.e. the field behaves as DM), which we assume to be at matter-radiation equality. The blue region of Figure~\ref{fig:structure_form} shows the constraint $k_\lambda^{\rm eq}\equiv k_\lambda|_{a=a_{\rm eq}}\gtrsim 1\Mpc^{-1}$ %
for ULDM mass and self-coupling in the range of interest for the formation of the solar halo, while the equivalent constraint from $k_J$ (not shown) only bounds much smaller $m$ (around $10^{-22}\eV$, see Eq.\,\eqref{eq:kJ}).  %
The self-interactions also affect larger momenta, though these are poorly constrained by direct observations; should such measurements improve, the corresponding lower bound on $f_a$ in Figure~\ref{fig:structure_form} would be stronger and scale proportionally to $k$. %
Note that ref.~\cite{Cembranos:2018ulm} carried out a much more detailed analysis for repulsive self-interactions, including the precise evolution of perturbations via a Boltzmann code and comparison with CMB measurements and large scale structure data. The resulting bound matches (within $10\%$) with that obtained in our rough estimate.%

The self-interactions could in principle modify the evolution of %
the perturbations also during radiation domination (this is the case for the so-called large-misalignment mechanism~\cite{Arvanitaki:2019rax}), potentially strengthening the bound on $\lambda$. It is known that the gravitational potential induced by the perturbations in the radiation bath during radiation domination introduces a source term into Eq.\,\eqref{eq:deltak}, which provides a (small) non-zero value for the time-derivative of $\delta_\mathbf{k}$ once the modes enter the horizon,  so that $\delta_\mathbf{k}$ increases logarithmically from the time when the mode enters the horizon until matter-radiation equality. In particular, the mode at $k=1\Mpc^{-1}$ enters the horizon at redshift $z_i\simeq 5\cdot 10^5$, corresponding to the temperature $T_i\simeq 100 \eV$. In the following we briefly discuss how the self-interactions could affect this behavior, and give a crude estimate of the corresponding bound from the matter power spectrum. %

For attractive self-interactions, %
Eq.\,\eqref{eq:deltak} can be rewritten as
\begin{equation}\label{eq:deltakp}
    \delta_\mathbf{k}''+\frac{3y+2}{2y(y+1)}\delta_\mathbf{k}'-\frac{3}{2y(y+1)}\left[1+\left(\frac{k}{k_\lambda^{\rm eq}}\right)^2\frac{1}{y^2}\right]\delta_\mathbf{k}=0\, , 
\end{equation}
where $y\equiv a/a_{\rm eq}$, $\delta'\equiv d\delta/dy$ and we neglected the quantum pressure term, which is an excellent approximation for $m\gtrsim 10^{-21}\eV$. The solution of Eq.\,\eqref{eq:deltakp} deep into radiation domination %
starting from $a=a_i$ (when the perturbation enters the horizon) with initial conditions $\delta_\mathbf{k}(a_i)=1$ and $\dot{\delta}_\mathbf{k}(a_i)=0$ is
\begin{equation}\label{eq:delta_rad}
    \delta_\mathbf{k}(a)= \left[\frac{a}{a_i}\right]^\frac14 \cosh\left[\sqrt{6\frac{a_{\rm eq}}{a_i}}\frac{k}{k_\lambda^{\rm eq}}\left(1-\sqrt{\frac{a_i}{a}}\right)\right] \,.
\end{equation}
Eq.\,\eqref{eq:delta_rad} shows that, soon after the time when $a=a_i$, perturbations with momentum $k\gtrsim k_\lambda^{\rm eq} \sqrt{a_i/a_{\rm eq}}$ increase substantially during radiation domination, receiving an enhancement factor much larger than one, in particular 
of order $(a_{\rm eq}/a_i)^{1/4}\exp{[\sqrt{6a_{\rm eq}/a_i}k/k_\lambda^{\rm eq}]}$ by the time of matter-radiation equality.\footnote{The expression in Eq.\,\eqref{eq:delta_rad} is a solution only in the limit $k\gtrsim k_\lambda^{\rm eq} \sqrt{a/a_{\rm eq}}$, and overestimates the true solution otherwise. Eq.\,\eqref{eq:delta_rad} has been obtained by considering the leading $1/y$ terms in Eq.\,\eqref{eq:deltakp} and, in particular, only the second term in the square bracket, which is a good approximation for $k\gtrsim k_\lambda^{\rm eq} \sqrt{a_i/a_{\rm eq}}$. }
Measurements of the matter power spectrum can rule out such an increase and provide a (stronger) bound on the attractive self-coupling. This can be estimated by requiring $k_\lambda^{\rm eq} \sqrt{a_i/a_{\rm eq}}\gtrsim 1\Mpc^{-1}$, with $a_i=1/(1+z_i)$. The corresponding upper limit on $f_a$, shown with dotted lines in Figure~\ref{fig:structure_form}, is stronger by a factor $1/\sqrt{a_i/a_{\rm eq}}\simeq 10$ than the one discussed previously. Given the roughness of this analysis, we only take this bound as indicative, and leave a full analysis (which could differ from this estimate) for future work.

On the other hand, if the self-interactions are repulsive, the solution of Eq.\,\eqref{eq:deltak} does not increase, but oscillates at momenta $k\gtrsim k_\lambda^{\rm eq} \sqrt{a_i/a_{\rm eq}}$. Therefore, in this case, the effect of the self-interactions before matter-radiation equality is much less dramatic. Indeed %
the analysis of ref.~\cite{Cembranos:2018ulm}, carried out from very high redshift, finds a bound comparable to that at matter-radiation equality, as discussed above.

Finally, black and red lines in Figure~\ref{fig:structure_form} represent the values of $m$ and $f_a$ for which the formation time scale $|\Gamma|^{-1}$ is $5\Gyr$ and $150\,{\rm Myr}$ respectively: For $m\lesssim 10^{-14}\eV$, $\Gamma$ is given by Eq.\,\eqref{rel2}, otherwise by Eq.\,\eqref{rel1}; the transition region is uncertain%
. In particular, in the gray region the self-coupling is too small for the formation to occur within $5\Gyr$. %
We conclude that gravitational atoms can form within the age of the Solar System consistently with direct bounds from structure formation, for about one or two orders of magnitude in $f_a$, in the range $10^6\div10^8\GeV$.

\vspace{-1mm}
\subsubsection*{Other limits on ULDM self-interactions}
\vspace{-1mm}
Reference \cite{Li:2013nal} obtained a bound on the self-interactions that is stronger than that of Figure~\ref{fig:structure_form} by requiring that the ULDM equation of state is $w=p/\rho< 10^{-3}$ at matter-radiation equality, where $p$ is the pressure, though the choice of this threshold is artificial. 
Additionally, any effectively massless scalar field, with mass $m<H$, gets isocurvature perturbations during inflation, which are strongly constrained by cosmic microwave background (CMB) observations. However, these depend on the initial field value during inflation, and are evaded e.g. if the initial field value is large (e.g. of order $\Mpl$), if the theory is a different phase or the scalar field gets a large effective mass via couplings to the inflation field; see also~\cite{Arvanitaki:2014faa}.

Finally, the existence of ultralight particles, regardless of them being DM, induces the spinning down of rotating black holes  -- a phenomenon known as superradiance -- and is constrained by the observation of such spinning objects \cite{Arvanitaki:2010sy,Arvanitaki:2016qwi}. However, the self-interactions suppress the superradiance process for mass and couplings in Figure~\ref{fig:structure_form} which are relevant for the formation of the gravitational atom \cite{Baryakhtar:2020gao,Branco:2023frw}.%

\section{Conclusions and outlook}
\label{sec:conclusion}

We have shown that a halo bound to massive astrophysical bodies, resembling a `gravitational atom', forms from the capture of ultralight (pseudo-)scalar dark matter (DM) particles, as a consequence of their (weak) quartic self-interactions.
This formation takes place also around our Sun, leading to a \emph{solar halo} whose density can be orders of magnitude larger than the background DM density at the position of the Earth.
More generally, the formation of gravitational atoms  around astrophysical objects in our Galaxy could provide novel ULDM signatures.

{The efficiency of the capture process is controlled by the amount of gravitational focusing of the galactic DM waves around the body, which is measured by the ratio $\xi_{\rm foc} \equiv \lambda_{\rm dB}/R_\star = 2\pi\alpha/v_{\rm dm}$ between their de Broglie wavelength, $\lambda_{\rm dB}$, and the gravitational Bohr radius, $R_\star$ in Eq.\,\eqref{eq:Rstar_sun}.} In particular, the bound state either saturates to an (underdense) steady-state configuration (when $\xi_{\rm foc}\lesssim 1$, i.e. $m\lesssim 10^{-14}\eV$ for the Sun) or it exponentially grows in density (when $\xi_{\rm foc}\gtrsim 1$). We have confirmed this behavior through numerical simulations. On the other hand, the size of the self-interaction coupling $\lambda$ only dictates the formation time scale, which in the regime   $\xi_{\rm foc}\gtrsim 1$ is of the order of the relaxation time via self-interactions, defined in Eq.\,\eqref{eq:tau_rel}.  The macroscopic properties of the resulting 
bound state, e.g. its radius and maximum density after the exponential increase (in Eq.\,\eqref{eq:Mrhoeq}), depend on the particle mass $m$, the central (for example, solar) mass $M$, the properties of the galactic DM near the system (through $\rho_{\rm dm}$ and $v_{\rm dm}$), and the 
magnitude of $\lambda$. Note that the basic mechanism depends neither on the sign of $\lambda$ nor the parity of the field. 

The most striking implications occur in the exponential-growth regime, when $\xi_{\rm foc}\gtrsim 1$. For instance, for axion DM with mass  $m\simeq {\rm few}\cdot 10^{-14}$\,eV and decay constants $f_a\simeq 10^7\div 10^8$\,GeV, on Gyr timescales the solar halo density grows to be larger than the background DM density by a factor as large as $10^4$ at a distance $r=1 \,{\rm AU}$ from the Sun; see Figures~\ref{fig:rhovsr} and \ref{fig:boson_stars}. The density at $r\ll $\,AU can be even larger for smaller masses, $m\simeq 10^{-13}$\,eV, (see Figure\,\ref{fig:MaxRho}) 
providing a clear target for proposed searches in space~\cite{Tsai:2021lly}. In the presence of low-velocity and low-dispersion DM substructure, like a dark disk, the density could be further enhanced.

The exponential growth stops once the density of the bound states approaches the critical density in Eq.\,\eqref{eq:rhocrit}. For repulsive self-interactions ($\lambda>0$), at this point the gravitational atom becomes a stable bound object, whose density can be large compared to that of the background DM. 
Instead, for attractive self-interactions ($\lambda<0$), the atom is unstable and collapses, possibly triggering a Bosenova emission of relativistic particles. Both the large density enhancements and a possible nearby Bosenova can provide novel signals in ULDM detection experiments.

In this work we assumed that the gravitational potential is proportional to $1/r$, which implies that $R_\star$ is larger than the radius $R$ of the astrophysical body that captures the ULDM. For large enough axion masses ($m\gtrsim2\cdot10^{-13}\eV$ for the solar halo, see Figure~\ref{fig:focussing_SS}), this assumption is no longer satisfied. In this case, assuming a uniform density inside the body, the potential is proportional to $r^2$ rather than $1/r$%
, and, if captured, the particles occupy bound states  \emph{inside} the bulk of the astrophysical body. A full analysis of the formation of bound states in this case is worthy of further study.

We emphasize that the formation mechanism introduced in this paper is generic, requiring only a light DM (pseudo-)scalar field, with sufficiently strong quartic self-interactions for the formation to 
occur within Gyr time scales. In particular, this leads to conclude that gravitational atoms form %
around any sufficiently compact object (i.e. with radius 
smaller than $R_\star$). Here we merely note that when $\xi_{\rm foc}\gtrsim 1$, the Bohr radius $R_\star$ of the atom bound to a few different astrophysical systems is given by:
\begin{equation}
    R_\star \lesssim
    \begin{cases}
        \displaystyle
            \,\,\,1.3\,R_J\left[\frac{1.7\cdot10^{-11}\eV}{m}\right]^2\left[\frac{10^{-3}M_\odot}{M}\right] 
            & \text{(Jupiter-like\,\,planet)} \\
        
        \displaystyle
            \,\,\,\,1\,\AU \,\left[\frac{1.7\cdot 10^{-14}\eV}{m}\right]^2\left[\frac{M_\odot}{M}\right]
            & {\rm (Star\,\,or\,\,neutron\,\,star)} \\
            
        \displaystyle
            \,\,\,30\,\,{\rm pc}\, \left[\frac{1.7\cdot10^{-21}\eV}{m}\right]^2\left[\frac{10^7 M_\odot}{M}\right]
            & \text{(Supermassive black hole)}\,
    \end{cases}
    \,
\end{equation}
where for reference $R_J = 7\cdot 10^4\km$ is the radius of Jupiter. Therefore, neutron stars, black holes (both solar-mass and supermassive), and planets are among promising targets for a dedicated study. Note that the gravitational atom bound to the Earth would have a radius larger than that of the Earth only for $m \lesssim 10^{-9}$ eV~\cite{Banerjee:2019epw}. However, for these masses and $M=M_\oplus$ (the mass of the Earth), $\xi_{\rm foc} \lesssim 0.1$ and therefore the exponential growth does not occur. Using Eq.\,\eqref{eq:Mrhoeq} one finds that $\rho_{\rm eq} \ll \rho_{\rm dm}$ for all relevant masses $m$.%
It is worthwhile to compare our results to other known configurations of (light) particles bound to  astrophysical objects. The superradiant instability of ultralight fields around rapidly rotating black holes can populate $l> 0$ modes by extracting the angular momentum of the host (see, for example, \cite{Arvanitaki:2010sy,Arvanitaki:2016qwi,Baryakhtar:2020gao,Branco:2023frw}). Contrary to the mechanism discussed in this paper, superradiance does not require the field to be the DM
. Additionally, the process is effective when $m$ is of the order of the black hole radius $1/(2GM)$, and therefore, for fixed $M$, is active for larger boson masses (by a factor $2\pi/v_{\rm dm}$) than those for which {DM capture starts to be effective}, $\xi_{\rm foc}\simeq1$.\footnote{The analysis of solar-mass and supermassive spinning black holes led to constraints in the mass range $m\simeq 10^{-13}\div 10^{-11}$\,eV \cite{Arvanitaki:2014wva,Baryakhtar:2020gao} and $m\simeq 10^{-21}\div 10^{-17}\eV$ \cite{Unal:2020jiy}, respectively.} Similar processes can occur in dense stellar objects like neutron stars, but require additional SM couplings beyond gravity and are not relevant for less dense objects like our Sun \cite{Chadha-Day:2022inf}. Finally, the particles produced by the Sun that lie in the low-velocity tail of the distribution can get captured into bound orbits around the Sun, giving rise to yet-another bound configuration, a \emph{solar basin} \cite{VanTilburg:2020jvl,Lasenby:2020goo,DeRocco:2022jyq}. This is however most relevant for particle masses $m\sim \mathcal{O}(\keV)$, much larger than those considered here. Note that as with superradiance, the solar basin will form even if the field constitutes a negligible fraction of the DM.

Self-gravitating bound configurations of ultralight bosons, known as boson stars \cite{Kaup:1968zz,Ruffini:1969qy,Colpi:1986ye}, can form %
on astrophysical timescales (see e.g. \cite{Schive:2014dra,Levkov:2018kau}). After formation, their mass $M_{\rm bs}$ changes in time through processes mediated by gravitational interactions (see e.g.~\cite{Chen:2020cef,Dmitriev:2023ipv}), similar in form to those that occur in the formation of the gravitational atom discussed in this paper. The authors of ref.\,\cite{Chan:2022bkz,Chan:2023crj} analyzed the mass growth of boson stars defining the characteristic parameter $\nu \equiv v_{\rm dm}/v_b$ ($\nu$ is analogous to $2\pi/\xi_{\rm foc}$): Boson stars with $\nu \gg 1$ %
decrease in mass
as a result of processes analogous to \emph{stripping}, see Figure\,\ref{fig:processes}(c). In the other limit, $\nu\ll 1$, their mass %
increases, but at a rate that decreases in time, as $M_{\rm bs}$ becomes larger. %
While there is a close connection between the conditions for the growth of the gravitational atom and the boson stars%
, the dynamical evolution of the system is different in the two cases. Vector boson stars, similar configurations for spin-$1$ ULDM particles, can also form during the evolution of the Universe~\cite{Brito:2015pxa,Minamitsuji:2018kof,Amin:2019ums,Jain:2021pnk,Gorghetto:2022sue}. 
Finally, we observe that the gravitational atoms could form also from the capture of such spin-1 DM, although the capture process could be different given the derivative form of the self-interactions.
\section*{Acknowledgements}
The authors are grateful to Hyungjin Kim and Christopher McKee for crucial contributions in the early stages of this work. The authors thank Kfir Blum, Elisa Ferreira, Ricardo Ferreira, Ed Hardy, Oleksii Matsedonskyi, Mehrdad Mirbabayi, Wolfram Ratzinger, Giovanni Villadoro, and Wei Xue for valuable discussions. The authors also thank Kfir Blum and Ed Hardy for insightful feedbacks on a draft. MG is grateful to Ed Hardy for %
substantial contributions to the Schr\"odinger--Poisson code used for the numerical results of this paper. The work of DB was supported in part by the Cluster of Excellence ``Precision Physics, Fundamental Interactions, and Structure of Matter'' (PRISMA+ EXC 2118/1) and the German-Israeli Foundation (GIF). The work of JE was supported by the World Premier International Research Center Initiative (WPI), MEXT, Japan and by the JSPS KAKENHI Grant Numbers 21H05451 and 21K20366. The work of MJ is supported by a research grant from the Musk Foundation.
The work of GP is supported by grants from BSF-NSF, Friedrich Wilhelm Bessel research award of the Alexander von Humboldt Foundation, GIF, ISF, Minerva,
SABRA - Yeda-Sela - WRC Program, the Estate of Emile Mimran, and the Maurice and Vivienne Wohl
Endowment.

\appendix

\section{Energy density and bound states}
\label{app:bound_states}
In this Appendix we derive the energy density $\epsilon$ in Eq.\,\eqref{eq:energy_density} and provide the expression of excited states $\psi_{nlm}$.%

\vspace{-3mm}
\paragraph{Energy density.} As mentioned in Section~\ref{ss:quantum_scattering}, in the nonrelativistic limit the action of the scalar field reduces to $S\equiv\int dt d^3x\mathcal{L}$ in Eq.\,\eqref{eq:S_nonrel}. If the self-gravitational potential $\Phi_{\rm se}$ (neglected in the main text) is relevant, $\mathcal{L}$ contains the additional terms $-(8\pi G)^{-1}|\nabla\Phi_{\rm se}|^2-m\Phi_{\rm se}|\psi|^2$. The invariance of $S$ under time translations implies that the time-like component `00' of the (nonrelativistic) energy momentum tensor, which is given by 
\begin{equation}\label{eq:Econs}
    \frac{\partial\mathcal{L}}{\partial \dot{\psi}}\dot{\psi}+\frac{\partial\mathcal{L}}{\partial \dot{\psi}^*}\dot{\psi}^*+\frac{\partial\mathcal{L}}{\partial \dot{\Phi}}\dot{\Phi}-\mathcal{L}=  \frac{1}{2m}|\nabla\psi|^2+\frac{1}{8\pi G}|\nabla\Phi_{\rm se}|^2\, +m(\Phi_{\rm ex}+\Phi_{\rm se})|\psi|^2+\frac12g|\psi|^4 \,,
\end{equation}
(with $\Phi=\Phi_{\rm se}+\Phi_{\rm ex}$) is conserved on the EoM, when integrated over the whole space. Eq.\,\eqref{eq:Econs} can be simplified using $|\nabla\Phi_{\rm se}|^2=\nabla\cdot(\Phi_{\rm se}\nabla\Phi_{\rm se})-\Phi_{\rm se}\nabla^2\Phi_{\rm se}$: The first term vanishes upon spatial integration%
, while the second can be rewritten using the Poisson equation (see footnote~\ref{footnotePoisson}). This leads to the conclusion that the energy density 
\begin{equation}\label{eq:epsilon_tot}
    \epsilon=\frac{|\nabla\psi|^2}{2m}+m\left(\Phi_{\rm ex}+\frac{\Phi_{\rm se}}{2}\right)|\psi|^2+\frac12g|\psi|^4\, 
     \,,
\end{equation}
is globally conserved.%

\vspace{-3mm}
\paragraph{Bound states.} The quasi-stationary solutions $\psi=e^{-i\omega t}\Psi(\mathbf{x})$ of the EoM in Eq.\,\eqref{eq:EoM} satisfy the eigenvalue equation $(-\nabla^2/2m-\alpha/r)\Psi=\omega\Psi$, whose most general solutions with negative energy ($\omega<0$) are the well-known hydrogen-like bound states%
\begin{equation}\label{eq:psi_nlm}
\Psi=\psi_{nlm}=R_{nl}(r)Y_l^m(\theta,\varphi) \, , \quad R_{nl}(r)\equiv\left[\frac{2}{nR_\star}\right]^\frac32 \left[\frac{2 r}{nR_\star}\right]^l%
\sqrt{\frac{ (n-l-1)!}{2 n (l+n)!}}  L_{n-l-1}^{2 l+1}\left[\frac{2 r}{n R_\star}\right]e^{-\frac{r}{nR_\star}} \, ,
\end{equation}
with eigenvalues $\omega=\omega_n=- m\alpha^2/(2n^2)$. In Eq.\,\eqref{eq:psi_nlm}, $Y_l^m(\theta,\varphi)$ are the spherical harmonics and $L_a^b[x]$ the generalized Laguerre polynomials of degree $a$ and order $b$. The bound states $\psi_{nlm}$ and the scattering states $\psi_\mathbf{k}$ in Eq.\,\eqref{eq:psik} are orthonormalized, and are all orthogonal to one another, as %
\begin{align}\label{eq:orthonormality}
        \int d^3x \,\psi^*_{nlm}\psi_{n'l'm'}&=\delta_{nn'}\delta_{ll'}\delta_{mm'}\, , \quad \int d^3x \psi_{{\bf k}'}^*\psi^{}_{\bf k} = (2\pi)^3\delta^3({\bf k}-{\bf k}')\, , \quad     \int d^3x \psi^*_{nlm}\psi_\mathbf{k}=0\, .
\end{align}

\section{Perturbative solution of the Gross--Pitaevskii equation}%
\label{app:onelevel}
In this Appendix we provide a detailed derivation of the change in the number of bound particles $\langle \dot{N}_{nlm} \rangle$ in Eq.\,\eqref{eq:Ndotavg}. This is obtained by solving the equation of motion
\begin{equation}
\label{eq:eom}
\left(i\partial_t+\frac{\nabla^2}{2m}+\frac{\alpha}{r}\right)\psi=g|\psi|^2\psi \, ,
\end{equation}
with initial condition  $\psi(t=0)=(\psi_w+\psi_b)_{t=0}$ in Eqs.~\eqref{eq:psi_w} and~\eqref{eq:psi_s_ini}%
,  perturbatively in the (local) effective %
self-interaction parameter $g_{\rm eff}=g\rho/(m\omega)$, where $\rho$ and $\omega$ are the typical values of the local density and frequency of the field. %
As mentioned in Section~\ref{ss:classical_p}, in the regime $\rho<\rho_{\rm crit}$, the self-interactions are small enough that it is possible to approximate the generic solution of Eq.\,\eqref{eq:eom} as a superposition of solutions of the EoM for $\lambda=g=0$, i.e.
\begin{equation}\label{eq:psi_sol}
\psi= \sum_{nlm}c_{nlm}(t)e^{-i \tilde{\omega}_{nlm} t} \psi_{nlm}+\int [dk]c_{\mathbf{k}}(t) e^{-i \tilde{\omega}_\mathbf{k} t} \psi_{\mathbf{k}} \,,
\end{equation}
where $\psi_{nlm}$ is given in Eq.\,\eqref{eq:psi_nlm} and the coefficients $c_{nlm}$ and $c_\mathbf{k}$ depend on $\lambda$ (or $g$) and vary in time more slowly than the typical field oscillation periods, given by $1/\tilde{\omega}_{nlm}$ and $1/\tilde{\omega}_{\mathbf k}$ for bound and scattering states respectively. %
We can expand  $c_{nlm}$ and $c_\mathbf{k}$ in a series in $g_{\rm eff}$ as  %
\begin{equation}\label{eq:c_exp}
c_{nlm}(t)=c_{nlm}^{(0)}(t)+c_{nlm}^{(1)}(t)+c_{nlm}^{(2)}(t)+\ldots, \qquad 
c_{\mathbf{k}}(t)=c_{\mathbf{k}}^{(0)}(t)+c_{\mathbf{k}}^{(1)}(t)+c_{\mathbf{k}}^{(2)}(t)+\ldots\, ,
\end{equation}
where the superscripts indicate the order in $g_{\rm eff}$ of each term. In general $\tilde{\omega}_{nlm}$ and $\tilde{\omega}_\mathbf{k}$ in Eq.\,\eqref{eq:psi_sol} are different from the original eigenfrequencies $\omega_n$ and $\omega_k$, which indeed receive time-independent corrections (or `shifts') when $\lambda\neq0$; %
 see for instance~\cite{Baryakhtar:2020gao}. These corrections may be also expanded as
\begin{equation}\label{eq:omega_exp}
\tilde{\omega}_{nlm}=\omega_{n}+\omega_{nlm}^{(1)}+\omega_{nlm}^{(2)}+\ldots, \qquad 
\tilde{\omega}_{\mathbf{k}}=\omega_{k}+\omega_{\mathbf{k}}^{(1)}+\omega_{\mathbf{k}}^{(2)}+\ldots,
\end{equation}
where $\omega_{n}=-m\alpha^2/2n^2$ and $\omega_{k}=k^2/2m$ are the original $\lambda=0$ frequencies, and $|\omega_{nlm}^{(1)}|\ll |\omega_n|$, etc.. %
As a result, we can organize the full solution $\psi$ in Eq.\,\eqref{eq:psi_sol} in the perturbative series%
$$
\psi=\psi^{(0)}+\psi^{(1)}+\psi^{(2)}+\ldots \, ,
$$
where, for instance,
\begin{align}\label{eq:psi_00}
\psi^{(0)}&= \sum_{nlm}c_{nlm}^{(0)}(t)e^{-i \omega_{n} t} \psi_{nlm}+\int [dk]c_{\mathbf{k}}^{(0)}(t) e^{-i \omega_k t} \psi_{\mathbf{k}}\ , \\
\psi^{(1)}&= \sum_{nlm}\left[c_{nlm}^{(1)}(t)-i \omega_{nlm}^{(1)} t c_{nlm}^{(0)}(t) \right]e^{-i \omega_{n} t} \psi_{nlm}+\int [dk]\left[c_{\mathbf{k}}^{(1)}(t)-i\omega_{\mathbf{k}}^{(1)} t c_{\mathbf{k}}^{(0)}(t) \right]e^{-i \omega_{k} t} \psi_{\mathbf{k}} \ .\label{eq:psi_11}
\end{align}
The expansions above allow us to easily solve Eq.\,\eqref{eq:eom} at every order in the perturbative expansion. First, at the zeroth order Eq.\,\eqref{eq:eom} simply reads %
\begin{equation}
\left(i\partial_t+\frac{\nabla^2}{2m}+\frac{\alpha}{r}\right)\psi^{(0)}=0\, ,
\end{equation}
and implies that %
$c_{nlm}^{(0)}(t)$ and $c_{\mathbf{k}}^{(0)}(t)$ are in fact time-independent. Imposing that $\psi$ equals the initial condition $\psi_w+\psi_b$ at $t=0$, we find $\psi^{(0)}=\psi_w+\psi_b$, and $c_{nlm}^{(0)}=\sqrt{N_{nlm}^{(0)}}$ and $c_{\mathbf{k}}^{(0)}=a(\mathbf{k})$. The zeroth-order solution in fact coincides with the initial conditions, and 
the full nontrivial time evolution is determined by the time-dependent coefficients $c_{nlm}^{(i)}$ and $c_{\mathbf{k}}^{(i)}$.

As discussed in Section\,\ref{ss:classical_p}, the number of particles bound to the $nlm$ level follows Eq.\,\eqref{eq:Nnlm}, reported here for simplicity: %
\begin{equation}\label{eq:Nnlm_full}
     N_{nlm}(t)=|c_{nlm}(t)|^2=N_{nlm}^{(0)}+2\sqrt{N_{nlm}^{(0)}}{\rm Re}[c_{nlm}^{(1)}(t)] +\left\{|c_{nlm}^{(1)}(t)|^2+2\sqrt{N_{nlm}^{(0)}}{\rm Re}[c_{nlm}^{(2)}(t)]\right\} +\dots \, .
\end{equation}
In the following we will describe the solution of the EoM~\eqref{eq:eom} up to the second order, which will be sufficient to calculate the first non-trivial part of $\langle \dot{N}_{nlm} \rangle$.

\subsubsection*{First-order solution}
The first order EoM is
\begin{equation}
\left(i\partial_t+\frac{\nabla^2}{2m}+\frac{\alpha}{r}\right) \psi^{(1)}= g|\psi^{(0)}|^2 \psi^{(0)} \, ,
\label{eq:1order}
\end{equation}
which is a source equation for $\psi^{(1)}$. We solve Eq.\,\eqref{eq:1order} with initial conditions containing a generic configuration of scattering states $a(\mathbf{k})$, but only particles in the ground state, i.e. $N_{nlm}^{(0)}=0$ unless $nlm=100$, or equivalently
\begin{equation}
\label{eq:psi0}
\psi^{(0)}= \sqrt{N_{100}^{(0)}}e^{-i \omega_{1} t} \psi_{100}+\int [dk]a({\mathbf{k}}) e^{-i \omega_k t} \psi_{\mathbf{k}} \, .
\end{equation}
With these initial conditions, $\langle \dot{N}_{100}\rangle$ %
will describe the evolution of the 
bound-state occupation number both before the ground state is populated (upon taking $N_{100}^{(0)}=0$), and after a 
ground-state 
population has been captured.
Plugging the expression of $\psi^{(1)}$ of Eq.\,\eqref{eq:psi_11} into %
Eq.\,\eqref{eq:1order}, we obtain the time evolution of the first-order coefficients:
\begin{equation}
\sum_{nlm}\left[i\dot c_{nlm}^{(1)}(t)+ \sqrt{N_{nlm}^{(0)}}\omega_{nlm}^{(1)}\right]e^{-i \omega_{n} t} \psi_{nlm}+\int [dk]\left[c_{\mathbf{k}}^{(1)}(t)+a(\mathbf{k}) \omega_{\mathbf{k}}^{(1)}\right]e^{-i \omega_\mathbf{k} t} \psi_{\mathbf{k}}=g|\psi^{(0)}|^2\psi^{(0)} \, .
\end{equation}
Using the orthonormality relations of Eq.\,\eqref{eq:orthonormality} we can single out the contribution of the individual bound-state levels and waves:%
\begin{align}\label{eq:cdot100}
i\dot c_{nlm}^{(1)}(t)+ \sqrt{N_{nlm}^{(0)}}\omega_{nlm}^{(1)}  \ &= g e^{i \omega_{n} t}\int d^3x |\psi^{(0)}|^2 \psi^{(0)} \psi_{nlm}^* \,, \\
i \dot c_{\mathbf{k}}^{(1)} +a(\mathbf{k})\omega_{\mathbf{k}}^{(1)}&= g e^{i \omega_{k} t}\int d^3x |\psi^{(0)}|^2 \psi^{(0)} \psi_{\mathbf{k}}^* \, .  \label{eq:cdotk}
\end{align}
Substituting the expression of $\psi^{(0)}$ of Eq.\,\eqref{eq:psi0} into Eqs.~\eqref{eq:cdot100} and~\eqref{eq:cdotk} and expanding the products, we obtain multiple terms on the right hand side (RHS) of these equations with different physical meaning. Crucially, a generic such `source' term $S(\mathbf{x}) e^{-i\omega_st}\subset g|\psi^{(0)}|^2 \psi^{(0)}$ that oscillates with frequency $\omega_s$ contributes to a significant change in $c_{nlm}$ or $c_\mathbf{k}$ only if the frequencies in the phases in the RHS of Eqs.~\eqref{eq:cdot100} and~\eqref{eq:cdotk} sum up to zero, in such a way that $c_{nlm}$ or $c_\mathbf{k}$ have a chance of growing with $t$ instead of oscillating. In particular (see also~\cite{Baryakhtar:2020gao}):
\vspace{-1mm}
\begin{enumerate}[leftmargin=0.2in%
] \setlength\itemsep{0.15em}
	\item In Eq.\,\eqref{eq:cdot100}, terms with $\omega_s<0$ and  $\omega_s=\omega_n$ for some $n$ correspond to a `resonant' increase of the bound modes labeled by $n$ (i.e. production of particles in the state $n$), or, as we will see later, to a correction of the eigenfrequency $\omega_{nlm}^{(1)}$. If instead %
	$\omega_s\neq\omega_n$ for all $n$, the term is a driven off-resonance oscillator, and represents a (suppressed) off-resonant production of particles.%
	\item In Eq.\,\eqref{eq:cdotk}, terms with $\omega_s>0$ correspond to a resonant production of scattering states or a correction to their frequency $\omega_\mathbf{k}^{(1)}$.
\end{enumerate}

As we will see, only resonant terms will effectively contribute to $c_{nlm}$ and $c_\mathbf{k}$. Let us first focus on the ground state, $nlm=100$. 
There are five types of terms %
on the RHS of Eq.\,\eqref{eq:cdot100}. As Eq.~\eqref{eq:diagrams1} in the main text, these can be pictured diagrammatically as in Figure~\ref{fig:diagrams}, where the states `100' and $\mathbf{k}$ correspond to the factors of $\psi_{100}$ or $\psi_\mathbf{k}$ appearing in the integrand of each term, and are placed on the right or left depending on whether the factor appears with or without the complex conjugate. The dashed line represents the $\psi_{100}^*$ (or in general $\psi_{nlm}^*$) factor common in all the terms. This representation makes it clear that the 100 level is populated, together with the other state that enters with the conjugate. We will now consider each diagram in turn:

\begin{figure}
\begin{center}
 \begin{tikzpicture}
[decoration={markings, 
    mark= at position 0.5 with {\arrow{stealth}}}
]

  \node at (-1.,1.) {\small{$100$}};
  \node at (-1.,-1.) {\small{$100$}};
  \node at (1.,1.) {\small{$100$}};
  \node at (1.,-1.) {\small{$100$}};
  \node at (0.,-1.4) {\small{$(1)$}};
\draw[thick,postaction={decorate}] (0,0) to (0.8,0.8);
\draw[thick,dotted,postaction={decorate}] (0,0) to (0.8,-0.8);
\draw[thick,postaction={decorate}] (-0.8,-0.8) to (0,0) ;
\draw[thick,postaction={decorate}] (-0.8,0.8) to (0,0);
  
  \begin{scope}
    [xshift=3.5cm]
\node at (-1.,1.) {\small{$\mathbf{k}_1$}};
  \node at (-1.,-1.) {\small{$100$}};
  \node at (1.,1.) {\small{$100$}};
  \node at (1.,-1.) {\small{$100$}};
  \node at (0.,-1.4) {\small{$(2)$}};
\draw[thick,postaction={decorate}] (0,0) to (0.8,0.8);
\draw[thick,dotted,postaction={decorate}] (0,0) to (0.8,-0.8);
\draw[thick,postaction={decorate}] (-0.8,-0.8) to (0,0) ;
\draw[thick,postaction={decorate}] (-0.8,0.8) to (0,0); 
   \end{scope}
   
   \begin{scope}
    [xshift=7cm]
\node at (-1.,1.) {\small{$\mathbf{k}_1$}};
  \node at (-1.,-1.) {\small{$100$}};
  \node at (1.,1.) {\small{$\mathbf{k}_2$}};
  \node at (1.,-1.) {\small{$100$}};
  \node at (0.,-1.4) {\small{$(3)$}};
\draw[thick,postaction={decorate}] (0,0) to (0.8,0.8);
\draw[thick,dotted,postaction={decorate}] (0,0) to (0.8,-0.8);
\draw[thick,postaction={decorate}] (-0.8,-0.8) to (0,0) ;
\draw[thick,postaction={decorate}] (-0.8,0.8) to (0,0); 
   \end{scope}

    \begin{scope}
    [xshift=10.5cm]
\node at (-1.,1.) {\small{$\mathbf{k}_1$}};
  \node at (-1.,-1.) {\small{$\mathbf{k}_2$}};
  \node at (1.,1.) {\small{$100$}};
  \node at (1.,-1.) {\small{$100$}};
  \node at (0.,-1.4) {\small{$(4)$}};
\draw[thick,postaction={decorate}] (0,0) to (0.8,0.8);
\draw[thick,dotted,postaction={decorate}] (0,0) to (0.8,-0.8);
\draw[thick,postaction={decorate}] (-0.8,-0.8) to (0,0) ;
\draw[thick,postaction={decorate}] (-0.8,0.8) to (0,0); 
   \end{scope}
     \begin{scope}
    [xshift=14cm]
\node at (-1.,1.) {\small{$\mathbf{k}_1$}};
  \node at (-1.,-1.) {\small{$\mathbf{k}_2$}};
  \node at (1.,1.) {\small{$\mathbf{k}_3$}};
  \node at (1.,-1.) {\small{$100$}};
  \node at (0.,-1.4) {\small{$(5)$}};
\draw[thick,postaction={decorate}] (0,0) to (0.8,0.8);
\draw[thick,dotted,postaction={decorate}] (0,0) to (0.8,-0.8);
\draw[thick,postaction={decorate}] (-0.8,-0.8) to (0,0) ;
\draw[thick,postaction={decorate}] (-0.8,0.8) to (0,0); 
   \end{scope}
\end{tikzpicture}
	\caption{\small{Diagrammatic representation of the terms in the right hand side of Eq.\,\eqref{eq:cdot100} that could contribute to the time-evolution of the coefficient $c_{100}^{(1)}$. Eventually only diagram (5) leads to a significant change in $c_{100}^{(1)}$, as all other diagrams do not conserve energy or do not change the particle number in the 100 state. %
	}
	}\label{fig:diagrams}
	\end{center}
	\vspace{-3mm}
\end{figure}
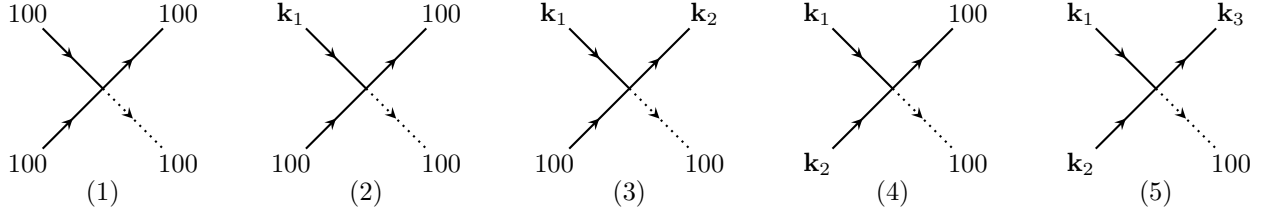

\vspace{-2mm}
\begin{itemize}[leftmargin=0.2in] \setlength\itemsep{0.15em}
	\item Diagram (4) in Figure~\ref{fig:diagrams} corresponds to
	$$
  g \sqrt{N_{100}^{(0)}}  \int [dk_1][dk_2] e^{i (2\omega_{1}-\omega_{k_1}-\omega_{k_2}) t}a(\mathbf{k}_1)a(\mathbf{k}_2)\int d^3x \psi_{\mathbf{k}_1}\psi_{\mathbf{k}_2} {\psi_{100}^{*2}} \, 
$$
	 and leads to an off-resonance oscillation since the argument in the exponential is always negative. This also happens for diagram (2), whose integrand is proportional to $e^{i(\omega_1-\omega_{k_1})t}$%
	 .  Intuitively, for these terms the energy in the corresponding processes is not conserved for all values of $k_1$ and $k_2$, and the production of bound states is not effective. Instead, all the other diagrams have a chance of obtaining a vanishing phase.
	 \item Diagram (1) corresponds to $g N_{100}^{(0)3/2}  \int d^3x  |\psi_{100}|^{4}$ and is time-independent, and would in principle lead to a linear increase of $c^{(1)}_{100}$. In reality, this term is cancelled by the term $\sqrt{N_{nlm}^{(0)}}\omega_{nlm}^{(1)}$ on the left hand side of Eq.\,\eqref{eq:cdot100}, because the frequency shift $\omega_{100}^{(1)}$ contains the contribution $g N_{100}^{(0)}  \int d^3x  |\psi_{100}|^{4}=g N_{100}^{(0)}/8\pi R_\star^3$. Consequently, diagram (1) represents the change in energy of the $100$ state due to self-interactions among the particles present in the ground state, and does not actually change their occupation number. Note that the condition $|\omega_{100}^{(1)}|<|\omega_1|$ for our perturbative method to work can be translated to $N_{100}^{(0)}< 4\pi/g m^2 \alpha$. Equivalently, this can be expressed in terms of $\rho_{\rm crit}$ defined in Eq.\,(\ref{eq:rhocrit}) as $\rho%
  (r=0)<2\rho_{\rm crit}$. 
	 
	 \item Diagram (3) corresponds to
	 $$
 2 g \sqrt{N_{100}^{(0)}}  \int [dk_1][dk_2] e^{-i (\omega_{k_1}-\omega_{k_2}) t}a^*(\mathbf{k}_2)a(\mathbf{k}_1)\int d^3x |\psi_{100}|^2\psi_{\mathbf{k}_1}\psi^*_{\mathbf{k}_2} \, .
$$
    Although the phase could vanish when $k_1=k_2$, this term also does not lead to a change in the number of particles $N_{100}$.  To understand its physical interpretation, it is convenient to write $a^*(\mathbf{k}_2)a(\mathbf{k}_1)$ as  $\langle a^*(\mathbf{k}_2) a(\mathbf{k}_1)\rangle+[a^*(\mathbf{k}_2)a(\mathbf{k}_1)-\langle a^*(\mathbf{k}_2)a(\mathbf{k}_1)\rangle]$. Using Eq.\,\eqref{eq:ff}, the first part becomes $$2 g \sqrt{N_{100}^{(0)}} \int [dk] f(\mathbf{k})\int d^3x |\psi_{100}|^2|\psi_\mathbf{k}|^2,$$ and thus represents a shift of the energy of the ground state due to the dark matter waves, $\omega_{100}^{(1)}\supset 2 g \int [dk] f(\mathbf{k})\int d^3x |\psi_{100}|^2|\psi_\mathbf{k}|^2$. (For instance, in the plane wave limit of $\psi_\mathbf{k}$, the frequency correction is $\omega_{100}^{(1)}\supset 2 g n_{\rm dm}$, with $n_{\rm dm}=\rho_{\rm dm}/m$ being the number density of the waves. Thus the perturbative condition $|\omega_{100}^{(1)}|<|\omega_1|$ is equivalent $
    \rho_{\rm dm}<\rho_{\rm crit}/8$.) The second part is proportional to $a^*(\mathbf{k}_2)a(\mathbf{k}_1)-\langle a^*(\mathbf{k}_2)a(\mathbf{k}_1) \rangle$, and describes the elastic scattering of the waves with the %
    bound state, and vanishes upon taking the average over times much longer than the coherence time, so does not contribute to $\langle \dot{N}_{100}\rangle$ at this order. %
    Although it actually affects $\dot{c}_{100}$, it can be verified that its contribution to $N_{100}(t)$ at second order in $g_{\rm eff}$ is canceled between  $\langle |c_{100}^{(1)}|^2 \rangle$ and $2\sqrt{N_{100}^{(0)} }{\rm Re}[\langle c_{100}^{(2)} \rangle]$ in Eq.\,\eqref{eq:Navg2}. We thus ignore this term in the following.

    \item Finally, diagram (5) corresponds to a resonant production of the $100$ state via the `capture' process ${k}_1+{k}_2\rightarrow {k}_3+100$, and is the only diagram  in Figure~\ref{fig:diagrams} that leads to an effective change in ${c}_{100}^{(1)}$.  %
    \end{itemize}
\noindent As a result %
the part of the solution of Eq.\,\eqref{eq:cdot100} that grows in time is, for $nlm=100$, 
\begin{align}
c_{100}^{(1)}(t)&=-i g \int [dk_1][dk_2][dk_3]a^*(\mathbf{k}_3)a(\mathbf{k}_1)a(\mathbf{k}_2) \mathcal{M}_{{k}_1+{k}_2\rightarrow {k}_3+100}\int_0^te^{i(
\omega_{1}+\omega_{k_3}-\omega_{k_1}-\omega_{k_2})t'}dt' \, ,
\label{eq:c1001}
\end{align}
where $$\mathcal{M}_{a+b\rightarrow c+d} \equiv \int \psi_{a}\psi_{b}\psi_{c}^*\psi_{d}^* d^3x\,.$$ This reproduces Eq.\,\eqref{eq:cnlm1} in the main text, and, correspondingly, diagram (5) reproduces $c_{nlm}^{(1)}$ in Eq.\,\eqref{eq:diagrams1}.  %

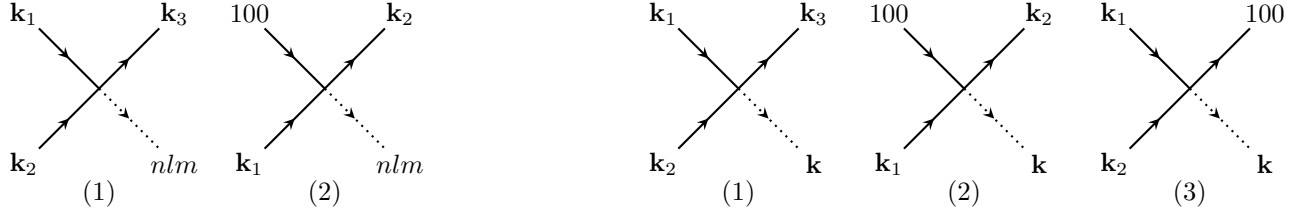
\begin{figure}
\begin{center}
  \begin{tikzpicture}
[decoration={markings, 
    mark= at position 0.5 with {\arrow{stealth}}}
]

  \node at (-1.,1.) {\small{$\mathbf{k}_1$}};
  \node at (-1.,-1.) {\small{$\mathbf{k}_2$}};
  \node at (1.,1.) {\small{$\mathbf{k}_3$}};
  \node at (1.,-1.) {\small{$nlm$}};
  \node at (0.,-1.4) {\small{$(1)$}};
\draw[thick,postaction={decorate}] (0,0) to (0.8,0.8);
\draw[thick,dotted,postaction={decorate}] (0,0) to (0.8,-0.8);
\draw[thick,postaction={decorate}] (-0.8,-0.8) to (0,0) ;
\draw[thick,postaction={decorate}] (-0.8,0.8) to (0,0);
  
  \begin{scope}
    [xshift=3.cm]
\node at (-1.,1.) {\small{$100$}};
  \node at (-1.,-1.) {\small{$\mathbf{k}_1$}};
  \node at (1.,1.) {\small{$\mathbf{k}_2$}};
  \node at (1.,-1.) {\small{$nlm$}};
  \node at (0.,-1.4) {\small{$(2)$}};
\draw[thick,postaction={decorate}] (0,0) to (0.8,0.8);
\draw[thick,dotted,postaction={decorate}] (0,0) to (0.8,-0.8);
\draw[thick,postaction={decorate}] (-0.8,-0.8) to (0,0) ;
\draw[thick,postaction={decorate}] (-0.8,0.8) to (0,0); 
   \end{scope}
   
   \begin{scope}
    [xshift=8.5cm]
\node at (-1.,1.) {\small{$\mathbf{k}_1$}};
  \node at (-1.,-1.) {\small{$\mathbf{k}_2$}};
  \node at (1.,1.) {\small{$\mathbf{k}_3$}};
  \node at (1.,-1.) {\small{$\mathbf{k}$}};
  \node at (0.,-1.4) {\small{$(1)$}};
\draw[thick,postaction={decorate}] (0,0) to (0.8,0.8);
\draw[thick,dotted,postaction={decorate}] (0,0) to (0.8,-0.8);
\draw[thick,postaction={decorate}] (-0.8,-0.8) to (0,0) ;
\draw[thick,postaction={decorate}] (-0.8,0.8) to (0,0); 
   \end{scope}

    \begin{scope}
    [xshift=11.5cm]
\node at (-1.,1.) {\small{$100$}};
  \node at (-1.,-1.) {\small{$\mathbf{k}_1$}};
  \node at (1.,1.) {\small{$\mathbf{k}_2$}};
  \node at (1.,-1.) {\small{$\mathbf{k}$}};
  \node at (0.,-1.4) {\small{$(2)$}};
\draw[thick,postaction={decorate}] (0,0) to (0.8,0.8);
\draw[thick,dotted,postaction={decorate}] (0,0) to (0.8,-0.8);
\draw[thick,postaction={decorate}] (-0.8,-0.8) to (0,0) ;
\draw[thick,postaction={decorate}] (-0.8,0.8) to (0,0); 
   \end{scope}
     \begin{scope}
    [xshift=14.5cm]
\node at (-1.,1.) {\small{$\mathbf{k}_1$}};
  \node at (-1.,-1.) {\small{$\mathbf{k}_2$}};
  \node at (1.,1.) {\small{$100$}};
  \node at (1.,-1.) {\small{$\mathbf{k}$}};
  \node at (0.,-1.4) {\small{$(3)$}};
\draw[thick,postaction={decorate}] (0,0) to (0.8,0.8);
\draw[thick,dotted,postaction={decorate}] (0,0) to (0.8,-0.8);
\draw[thick,postaction={decorate}] (-0.8,-0.8) to (0,0) ;
\draw[thick,postaction={decorate}] (-0.8,0.8) to (0,0); 
   \end{scope}
\end{tikzpicture}
	\caption{\small{ Diagrammatic representation of the terms in the right hand side of Eqs.~\eqref{eq:cdot100} and~\eqref{eq:cdotk} that are relevant for the change in $c_{nlm}^{(1)}$ (left) and $c_{\mathbf{k}}^{(1)}$ (right). Only diagram (2, left) and diagrams (2-3, right) end up contributing to $\langle \dot{N}_{100}\rangle$, and give rise to the processes of excitation, stripping and stimulated capture respectively.
	}\label{fig:diagrams1}}
	\end{center}
	\vspace{-3mm}
\end{figure} 

We observe that, in practice, the relevant terms to consider in the RHS of Eqs.~\eqref{eq:cdot100} and \eqref{eq:cdotk} are those that \emph{(i)} conserve energy and \emph{(ii)} change the number of particles in the state (in this case 100) in the corresponding diagram. We can use this information to determine $c^{(1)}_{nlm}$ for $nlm\neq100$ and $c^{(1)}_\mathbf{k}$, whose relevant diagrams are shown in Figure~\ref{fig:diagrams1} (left and right, respectively). The result for $c^{(1)}_{nlm}$ is
\begin{align}
c_{nlm}^{(1)}(t)&=-i g \int\nonumber [dk_1][dk_2][dk_3]a^*(\mathbf{k}_3)a(\mathbf{k}_1)a(\mathbf{k}_2) \mathcal{M}_{{k}_1+{k}_2\rightarrow {k}_3+nlm}\int_0^te^{i(
\omega_{n}+\omega_{k_3}-\omega_{k_1}-\omega_{k_2})t'}dt' \, \\
&-2ig\sqrt{N_{100}^{(0)}} \int [dk_1][dk_2]a(\mathbf{k}_1)a^*(\mathbf{k}_2)\mathcal{M}_{{100}+{k}_1\rightarrow {k}_2+nlm}\int_0^t e^{i(
\omega_{n}+\omega_{k_2}-\omega_{k_1}-\omega_{1})t'} \, ,
\label{eq:c1nlmt}
\end{align}
where the first term represents the `capture' process in diagram (1) of Figure~\ref{fig:diagrams1} (left) and is completely analogous to that of the $nlm=100$ case. The second represents the `excitation' process in diagram (2) where a particle in the $100$ state gets excited to a higher state ($n>1$) because of the interactions with the waves. This is absent for $nlm=100$ and can occur because the $100$ state is initially populated and $|\omega_{n}|<|\omega_{1}|$. 

The result for $c^{(1)}_\mathbf{k}$ is
\begin{align}
c_{\mathbf{k}}^{(1)}(t)&=-i g \int\nonumber [dk_1][dk_2][dk_3]a^*(\mathbf{k}_3)a(\mathbf{k}_1)a(\mathbf{k}_2) \mathcal{M}_{{k}_1+{k}_2\rightarrow {k}_3+k}\int_0^te^{i(
\omega_{k}+\omega_{k_3}-\omega_{k_1}-\omega_{k_2})t'}dt' \, \\
&-i g \sqrt{N_{100}^{(0)}}\int [dk_1][dk_2]\big\{ a(\mathbf{k}_1)a(\mathbf{k}_2) \mathcal{M}_{{k}_1+{k}_2\rightarrow {k}+100}\int_0^te^{i(
\omega_{1}+\omega_{k}-\omega_{k_1}-\omega_{k_2})t'}dt'\nonumber\\
&+2a(\mathbf{k}_1)a^*(\mathbf{k}_2) \mathcal{M}_{{k}_1+100\rightarrow {k}+{k}_2}\int_0^t e^{i(
\omega_{k}+\omega_{k_2}-\omega_{k_1}-\omega_{1})t'}dt'\big\} \, , \label{eq:ck1sol}
\end{align}
where the first term represents the pure scattering of waves in diagram (1) of Figure~\ref{fig:diagrams1} (right), while the second and third term represent the diagrams (3) and (2). These -- as we will see next -- are at the origin of the stimulated capture and stripping processes discussed in the main text.

Eqs.\,\eqref{eq:c1001},~\eqref{eq:c1nlmt} and~\eqref{eq:ck1sol} provide %
solution of the EoM at the first order in $g_{\rm eff}$ for the initial condition in Eq.\,\eqref{eq:psi0}, containing only particles in the scattering states and in the ground state. Note that the results above are also valid if the initial conditions contain particles \emph{only} at one particular level $\tilde{n}\tilde{l}\tilde{m}$, i.e. %
\begin{equation}
\label{eq:psi0000}
\psi^{(0)}= \sqrt{N_{\tilde{n}\tilde{l}\tilde{m}}^{(0)}}e^{-i \omega_{{\tilde{n}}} t} \psi_{\tilde{n}\tilde{l}\tilde{m}}+\int [dk]a({\mathbf{k}}) e^{-i \omega_k t} \psi_{\mathbf{k}} \, ,
\end{equation}
in which case it is sufficient to substitute $100\to\tilde{n}\tilde{l}\tilde{m}$ in Eqs.~\eqref{eq:c1001},~\eqref{eq:c1nlmt} and~\eqref{eq:ck1sol}. %
In Appendix~\ref{app:twolevel} we will further generalize this derivation for a system with \emph{both} $100$ and $200$ levels populated, i.e. $N_{100}^{(0)}\neq0$ and $N_{200}^{(0)}\neq0$. Although in principle doable, we do not attempt to write the general solution for $N_{nlm}^{(0)}\neq0$ in this work.

\subsubsection*{Occupation number and second-order solution}

As mentioned in Section~\ref{ss:classical_p}, the contribution at order $g_{\rm eff}$ to $\langle{N}_{100}\rangle$ (the second term in Eq.\,\eqref{eq:Nnlm_full}) vanishes,  since %
$\langle c_{100}^{(1)}\rangle \propto \langle a(\mathbf{k}_1)a(\mathbf{k}_2) a^*(\mathbf{k}_3)\rangle=0$, and the first nontrivial correction to $\langle N_{100}\rangle$  comes at order $g_{\rm eff}^2$ 
and reads
\begin{equation}\label{eq:Navg2}
    \langle |c_{100}^{(1)}|^2 \rangle +2 \sqrt{N_{100}^{(0)} }{\rm Re}[\langle c_{100}^{(2)} \rangle] \, .
\end{equation}

As discussed in Section~\ref{ss:classical_p}, the first term of Eq.\,\eqref{eq:Navg2} is simply the square of  Eq.\,\eqref{eq:c1001} and can be readily calculated using the expression of $\langle a^*(\mathbf{k}_1)a(\mathbf{k}_2)\rangle$ in Eq.\,\eqref{eq:ff}, along with the identity $\lim\limits_{\omega t \to \infty}|\int_0^tdt'e^{i\omega t'}|^2=\lim\limits_{\omega t \to \infty} 4\omega^{-2}\sin^2(\frac{\omega t}{2})=2\pi\delta(\omega)t$. We obtain
\begin{equation}
\label{eq:c1}
\langle |c_{100}^{(1)}|^2 \rangle =2 g^2  \int [d k_1] [d k_2][d k_3]|\mathcal{M}_{{k}_1+{k}_2\rightarrow 100+{k}_3}|^2 f(\mathbf{k}_1)f(\mathbf{k}_2)f(\mathbf{k}_3) (2\pi) \delta(\Delta \omega)t \, ,
\end{equation}
with $\Delta\omega\equiv\omega_{k_1}+\omega_{k_2}-\omega_{k_3}-\omega_{1}$. 
On the other hand, the second term in Eq.\,\eqref{eq:Navg2} %
requires the solution of the EoM at the second order, which we report here for simplicity:
\begin{equation} 
\left(i\partial_t+\frac{\nabla^2}{2m}+\frac{\alpha}{r}\right) \psi^{(2)}= g\left(\psi^{(0)2} \psi^{(1)*}+2|\psi^{(0)}|^2 \psi^{(1)}\right) \,.
\label{eq:2order}
\end{equation}
Similarly to Eq.\,\eqref{eq:1order}, this is a source equation for $\psi^{(2)}$, generated by a combination of $\psi^{(0)}$ and $\psi^{(1)}$. By plugging the explicit expression of $\psi^{(2)}$ in terms of $c_{nlm}^{(2)}$ and $c_\mathbf{k}^{(2)}$ and using again the orthonormality relations, we obtain (as in Eq.\,\eqref{eq:c1001})
\begin{equation}
\label{eq:c2}
i\dot{c}_{100}^{(2)}(t)+\dots = g  e^{i\omega_{1}t}\int d^3x \left(\psi^{(0)2} \psi^{(1)*}+2|\psi^{(0)}|^2 \psi^{(1)}\right) \psi_{100}^*  \, ,
\end{equation}
where the dots stand for frequency corrections analogous to those in Eq.\,\eqref{eq:c1001}. It is via $\psi^{(1)}$ in the RHS of Eq.\,\eqref{eq:c2}  that the stimulated capture and stripping diagrams (2) and (3) of Figure~\ref{fig:diagrams1} (right), and the excitation process in diagram (2) of Figure~\ref{fig:diagrams1} (left), affect $\langle \dot{N}_{100}\rangle$.%

We can substitute into Eq.\,\eqref{eq:c2} the expressions of $\psi^{(0)}$ in Eq.\,\eqref{eq:psi0} and the solution of $\psi^{(1)}$ computed in Eqs.~\eqref{eq:c1001},~\eqref{eq:c1nlmt} and~\eqref{eq:ck1sol}. However, as for the first-order coefficients, only the terms corresponding to processes that conserve energy and change the particle number in the 100 state contribute to a change in $N_{100}$. In particular:

\vspace{-2mm}
\begin{itemize}[leftmargin=0.2in] \setlength\itemsep{0.15em}
	\item Only the unbound component of $\psi^{(0)}$, i.e. $ \int [dk]a(\mathbf{k}) e^{-i \omega_\mathbf{k} t} \psi_{\mathbf{k}}\subset\psi^{(0)}$, gives rise to processes that modify the particle number. 
 \item Additionally, only the unbound component of $\psi^{(1)}$, i.e. $\int [dk]c_{\mathbf{k}}^{(1)}(t) e^{-i \omega_{k} t} \psi_{\mathbf{k}}\subset\psi^{(1)}$, can conserve energy in the first term in the RHS of Eq.\,\eqref{eq:c2} (this is the reason why we considered only the unbound component of the first order field in the diagrams of Eq.\,\eqref{eq:diagrams2} in the main text).
In the second term of Eq.\,\eqref{eq:c2}  there is also a contribution from its bound component with $nlm\neq100$, i.e. Eq.\,\eqref{eq:c1nlmt}, which gives rise to an excitation term not reported in the main text (see discussion below).
\end{itemize}

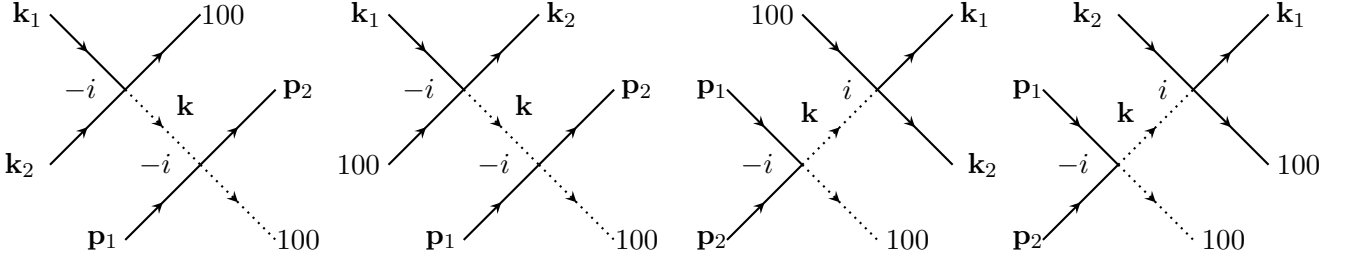
\begin{figure}
    \centering
\begin{tikzpicture}
[decoration={markings, 
    mark= at position 0.5 with {\arrow{stealth}}}
]

  \node at (-1.3,1.) {$\mathbf{k}_1$};
  \node at (-1.4,-1.) {$\mathbf{k}_2$};
  \node at (1.3,1.) {$100$};
  \node at (0.8,-0.2) {$\mathbf{k}$};
    \node at (2.3,0) {$\mathbf{p}_2$};
    \node at (2.3,-2.) {$100$};
   \node at (-0.3,-2.) {$\mathbf{p}_1$};
     \node at (-.6,0)  {$-i$};
    \node at (0.4,-1.)  {$-i$};
  
\draw[thick,postaction={decorate}] (0,0) to  (1,1);
\draw[thick,dotted,postaction={decorate}] (0,0) to  (1,-1);
\draw[thick,postaction={decorate}] (-1,-1) to (0,0) (-1,-1);
\draw[thick,postaction={decorate}] (-1,1) to (0,0);
\draw[thick,postaction={decorate}] (1,-1) to (2,0) ;
\draw[thick,dotted,postaction={decorate}] (1,-1) to (2,-2) ;
\draw[thick,postaction={decorate}] (0,-2) to (1,-1);
 \begin{scope}[xshift=4.5cm]
  
  \node at (-1.3,1.) {$\mathbf{k}_1$};
  \node at (-1.4,-1.) {$100$};
  \node at (1.3,1.) {$\mathbf{k}_2$};
  \node at (0.8,-0.2) {$\mathbf{k}$};
    \node at (2.3,0) {$\mathbf{p}_2$};
    \node at (2.3,-2.) {$100$};
   \node at (-0.3,-2.) {$\mathbf{p}_1$};
        \node at (-.6,0)  {$-i$};
    \node at (0.4,-1.)  {$-i$};
\draw[thick,postaction={decorate}] (0,0) to  (1,1);
\draw[thick,dotted,postaction={decorate}] (0,0) to  (1,-1);
\draw[thick,postaction={decorate}] (-1,-1) to (0,0) (-1,-1);
\draw[thick,postaction={decorate}] (-1,1) to (0,0);
\draw[thick,postaction={decorate}] (1,-1) to (2,0) ;
\draw[thick,dotted,postaction={decorate}] (1,-1) to (2,-2) ;
\draw[thick,postaction={decorate}] (0,-2) to (1,-1);
  \end{scope}
  \begin{scope}[xshift=10.cm]
  \node at (-2.2,0) {$\mathbf{p}_1$};
   \node at (-2.2,-2) {$\mathbf{p}_2$};
 \node at (-1.4,1.) {$100$};
  \node at (-0.9,-0.3) {$\mathbf{k}$};
  \node at (1.3,1.) {$\mathbf{k}_1$};
  \node at (1.4,-1) {$\mathbf{k}_2$};
   \node at (0.4,-2) {$100$};
\node at (-1.6,-1.) {$-i$};
\node at (-.4,0.) {$i$};

\draw[thick,postaction={decorate}] (0,0) to (1,1);
\draw[thick,postaction={decorate}] (0,0) to (1,-1);
\draw[thick,dotted,postaction={decorate}] (-1,-1) to (0,0) ;
\draw[thick,postaction={decorate}] (-1,1) to (0,0);
\draw[thick,dotted,postaction={decorate}] (-1,-1) to [out=-45,in=135] (0,-2) ;
\draw[thick,postaction={decorate}] (-2,-2) to (-1,-1);
\draw[thick,postaction={decorate}] (-2,0) to (-1,-1);
  \end{scope}
  \begin{scope}[xshift=14.2cm]

\node at (-2.2,0) {$\mathbf{p}_1$};
   \node at (-2.2,-2) {$\mathbf{p}_2$};
 \node at (-1.4,1.) {$\mathbf{k}_2$};
  \node at (-0.9,-0.3) {$\mathbf{k}$};
  \node at (1.3,1.) {$\mathbf{k}_1$};
  \node at (1.4,-1) {$100$};
   \node at (0.4,-2) {$100$};
\node at (-1.6,-1.) {$-i$};
\node at (-.4,0.) {$i$};

\draw[thick,postaction={decorate}] (0,0) to (1,1);
\draw[thick,postaction={decorate}] (0,0) to (1,-1);
\draw[thick,dotted,postaction={decorate}] (-1,-1) to (0,0) ;
\draw[thick,postaction={decorate}] (-1,1) to (0,0);
\draw[thick,dotted,postaction={decorate}] (-1,-1) to [out=-45,in=135] (0,-2) ;
\draw[thick,postaction={decorate}] (-2,-2) to (-1,-1);
\draw[thick,postaction={decorate}] (-2,0) to (-1,-1);
\end{scope}
\end{tikzpicture}
   \caption{\small{ The diagrams representing each term of $c_{100}^{(2)}$ in Eq.\,\eqref{eq:c21002}. Only (2) and  (3) contribute to  $\langle\dot{N}_{100}\rangle$.
	}\label{fig:diagrams2order}}
\end{figure}

\noindent
Let us first consider the contributions arising from the unbound component of $\psi^{(1)}$. The first term of $c_\mathbf{k}^{(1)}$ in Eq.\,\eqref{eq:ck1sol} does not contribute to $\langle\dot{c}_{100}^{(2)}\rangle$, because this term contains three powers of $a$ and always appears in the RHS of Eq.\,\eqref{eq:c2} in combination with two additional powers of $a$ from $\psi^{(0)}$, and $\langle f^5\rangle=0$ as discussed in Section~\ref{ss:ourDM}. As a result, the process of pure wave scattering in diagram (1) of Figure~\ref{fig:diagrams1} (right) does not contribute at the leading order to the increase of bound particles. On the other hand, the other two terms in $c_\mathbf{k}^{(1)}$ (diagrams (2) and (3) of Figure~\ref{fig:diagrams1} (right)), corresponding to the stripping-like and stimulated capture-like diagrams, are relevant. Once substituted into Eq.\,\eqref{eq:c2} %
they lead to
\begin{align}
c_{100}^{(2)}(t)&\supset - g^2 \sqrt{N_{100}^{(0)}} \bigg\{2 \int [dk][dp_1][dp_2]a(\mathbf{p}_1)a^*(\mathbf{p}_2) \mathcal{M}_{{p}_1+{k}\rightarrow 100+{p}_2} \int_0^t dt' e^{i(\omega_{1}+\omega_{p_2}-\omega_{p_1}-\omega_k)t'} 
\nonumber\\
&\int[dk_1][dk_2] \big(a(\mathbf{k}_1)a(\mathbf{k}_2)\mathcal{M}_{{k}_1+{k}_2\rightarrow 100+{k}}\int_0^{t'} e^{i(
\omega_{1}+\omega_{k}-\omega_{k_1}-\omega_{k_2})t''}dt''\nonumber\\
&+2a(\mathbf{k}_1)a^*(\mathbf{k}_2)\mathcal{M}_{{k}_1+100\rightarrow {k}+{k}_2}\int_0^{t'} e^{i(
\omega_{k}+\omega_{k_2}-\omega_{1}-\omega_{k_1})t''}dt''\big)\nonumber\\
&-\int [dk][dp_1][dp_2] a(\mathbf{p}_1)a(\mathbf{p}_2)\mathcal{M}_{{p}_1+{p}_2\rightarrow 100+{k}} \int_0^t dt' e^{i(\omega_{1}+\omega_{k}-\omega_{p_1}-\omega_{p_2})t'}
\nonumber\\
&\int [dk_1][dk_2] \big(a^*(\mathbf{k}_1)a^*(\mathbf{k}_2)\mathcal{M}_{100+{k}\rightarrow {k}_1+{k}_2}\int_0^{t'} e^{-i(
\omega_{1}+\omega_{k}-\omega_{k_1}-\omega_{k_2})t''}dt''\nonumber\\
&+2a^*(\mathbf{k}_1)a(\mathbf{k}_2)\mathcal{M}_{{k}+{k}_2\rightarrow 100+{k}_1}\int_0^{t'} e^{-i(
\omega_{k}+\omega_{k_2}-\omega_{1}-\omega_{k_1})t''}dt''\big)\bigg\} \, .\label{eq:c21002}
 \end{align}
The first (last) three lines of Eq.\,\eqref{eq:c21002} arise from the second (first) term in the RHS of Eq.\,\eqref{eq:c2}, where the two contributions in the first order field $\psi^{(1)}$ come from the second and third terms of Eq.\,\eqref{eq:ck1sol} respectively. We can represent the four addends of Eq.\,\eqref{eq:c21002} via the diagrams (1-4) in Figure~\ref{fig:diagrams2order}. As in the main text, one of the incoming/outgoing legs is substituted with the first order field corresponding to the diagrams~(2-3) of Figure~\ref{fig:diagrams1} (right), with momenta labelled by $\mathbf{k}_1,\mathbf{k}_2,\mathbf{k}$. Note that in Eq.\,\eqref{eq:diagrams2} we only reported the diagrams (2) and (3), which are the only that do not vanish in the determination of $\langle\dot{N}_{100}\rangle$, see below.  %

Using the identity $$\langle a(\mathbf{k}_1)a(\mathbf{k}_2) a^*(\mathbf{k}_3)a^*(\mathbf{k}_4)\rangle=(2\pi)^6f(\mathbf{k}_1)f(\mathbf{k}_2)\left[\delta(\mathbf{k}_1-\mathbf{k}_3)\delta(\mathbf{k}_2-\mathbf{k}_4)+\delta(\mathbf{k}_1-\mathbf{k}_4)\delta(\mathbf{k}_2-\mathbf{k}_3)\right]\, ,$$
which follows from Eq.\,\eqref{eq:ff} and  Wick's theorem, Eq.\,\eqref{eq:c21002} simplifies to
\begin{align}
\langle c_{100}^{(2)}(t)\rangle
&\supset- 4 g^2 \sqrt{N_{100}^{(0)}}   \int [d k] [d k_1][d k_2]f(\mathbf{k}_1)f(\mathbf{k}_2)|\mathcal{M}_{{k}_1+{k}_2\rightarrow 100+{k}}|^2 \int_0^t dt' \int_0^{t'} dt'' e^{i(\omega_{1}+\omega_{k_1}-\omega_{k_2}-\omega_k)(t'-t'')} \nonumber\\
&+2 g^2 \sqrt{N_{100}^{(0)}}   \int [d k] [d k_1][d k_2]f(\mathbf{k}_1)f(\mathbf{k}_2)|\mathcal{M}_{100+{k}\rightarrow {k}_1+{k}_2}|^2 \int_0^t d t'\int_0^{t'} d t''e^{i(\omega_{1}+\omega_{}-\omega_{k_1}-\omega_{k_2})(t'-t'')} \, .
\end{align}
The time integration can be performed explicitly using $\int_0^t dt' \int_0^{t'} dt'' e^{ i\omega (t'-t'')}+c.c=4\omega^{-2}\sin^2 (\omega t)\rightarrow 2\pi \delta( \omega)t$, and yields %
\begin{align}
\langle c_{100}^{(2)}+c_{100}^{(2)*}\rangle
& \supset 2 g^2 \sqrt{N_{100}^{(0)}}   \int [d k_1] [d k_2][d k_3]|\mathcal{M}_{{k}_1+{k}_2\rightarrow 100+{k}_3}|^2 [f(\mathbf{k}_1)f(\mathbf{k}_2)-2f(\mathbf{k}_1)f(\mathbf{k}_3)] (2\pi) \delta (\Delta\omega) t \, . \label{eq:c1002_1}
\end{align}
This equation provides the capture and stripping terms in Eq.\,\eqref{eq:Ndotavg}. As mentioned in Section~\ref{ss:classical_p}, this corresponds to the diagrams in Eq.\,\eqref{eq:Ndot2contr} of the main text, which can be thought as arising from the `contraction' of the diagrams~(2) and (3) in Figure~\ref{fig:diagrams2order}.

Similarly, only the second term in the bound component $c^{(1)}_{nlm}$ in Eq.\,\eqref{eq:c1nlmt}, representing the excitation process, contributes to $c_{100}^{(2)}$; this term %
gives 
\begin{align}
\langle c_{100}^{(2)}+c_{100}^{(2)*}\rangle
& \supset 2 g^2 \sqrt{N_{100}^{(0)}}   \int [d k_1] [d k_2]\sum_{nlm}|\mathcal{M}_{nlm+{k}_1\rightarrow 100+{k}_2}|^2 [-2f(\mathbf{k}_1)f(\mathbf{k}_2)] 2\pi \delta (\omega_n+\omega_{k_1}-\omega_{1}-\omega_{k_2}) t \, . \label{eq:c1002_2}
\end{align}
Plugging the last two equations into Eq.\,\eqref{eq:Navg2} we obtain the full result: 
\begin{align}\label{eq:ndot100full}
\langle  \dot{N}_{100}&\rangle = \frac{d}{dt} \left\langle |c_{100}^{(1)}|^2+\sqrt{N_{100}^{(0)}}(c_{100}^{(2)}+c_{100}^{(2)*})\right\rangle \nonumber\\
&=2g^2 \int [dk_1][dk_2][d k_3] |\mathcal{M}_{{k}_1+{k}_2\rightarrow {k}_3+100}|^2 (2\pi) \delta(
\Delta \omega)\{f(\mathbf{k}_1)f(\mathbf{k}_2)f(\mathbf{k}_3)+N_{100}^{(0)}[f(\mathbf{k}_1)f(\mathbf{k}_2)-2f(\mathbf{k}_2)f(\mathbf{k}_3)]\} \,  \nonumber\\
&+2g^2N_{100}^{(0)}\int[dk_1][dk_2]\sum_{nlm}|\mathcal{M}_{{nlm}+k_1\to 100+k_2}|^2(2\pi)\delta(\omega_{n}+{\omega}_{k_1}-{\omega}_{k_2}-{\omega}_{1})\left[-2f(\mathbf{k}_1)f(\mathbf{k}_2)\right] \, .
\end{align}
This equation coincides with Eq.\,\eqref{eq:Ndotavg} up to the term in the last line%
, neglected in the main text. This describes the depletion of particles from the 100 level because of their excitation to all higher levels via scatterings with the DM waves, as depicted in diagram (2) in Figure~\ref{fig:diagrams1} (left). Note the similarity between the stripping and excitation terms, which have an analogous form provided $\int[dk]$ is substituted with $\sum_{nlm}$ (this is because the diagrams (2, left) and (2, right) in Figure~\ref{fig:diagrams1}  are identical upon substituting $\mathbf{k}\leftrightarrow nlm$).

The excitation term could in principle reduce the number of particles in the $100$ state (when the captured mass is large enough, since it is proportional to $N_{100}^{(0)}$). However, at the same time, the occupation number of excited states %
would grow. %
As we will show in Appendix~\ref{app:twolevel}, for a system consisting of $100$ and $200$ states, the excited state (if populated) %
decays (`relaxes' or `de-excites') to the ground state, at least in the regime $\xi_{\rm foc
}=2\pi\alpha/v_{\rm dm}\gg1$. Consequently, $N_{100}$ increases. As a result, the excitation terms are not expected to affect the discussion in the main text. %

\section{Role of excited states and two-level system} \label{app:twolevel}

In the main text we studied the capture of ULDM to the ground state, 
but neglected excited states in most of our discussions. In this Appendix, we show that these are likely to only 
enhance the accretion of DM onto the ground state%
, and not to change the overall conclusions of Section~\ref{sec:analytic}. %

As argued in Section~\ref{ss:light_vs_heavy}, the matrix element $\mathcal{M}\equiv \mathcal{M}_{k_1+k_2\to k_3+nlm}$ in Eq.\,\eqref{eq:Melement} %
is 
suppressed for $n\gg1$, in both limits of $\xi_{\rm foc} \gg 1$ and $\xi_{\rm foc} \ll 1$,
which inhibits the population of very high-$n$ states. However, we also argued that for $\xi_{\rm foc} = 2\pi\alpha/v_{\rm dm}\gg1$ the magnitude of $\mathcal{M}$ for the first few levels (those satisfying $n\lesssim \xi_{\rm foc}^{1/2}$) could be comparable to that of $nlm=100$ (this can be seen, for instance for $nlm=200$, by using the explicit expression of $\psi_{nlm}$ in Eq.\,\eqref{eq:psi_nlm}). Such levels are therefore populated with a similar rate as the 100 state, and can affect the density of the
gravitational atom.
We now study the evolution of a system where both the 100 and 200 levels are included (and populated by capture from the DM waves) and show that: 
For $\xi_{\rm foc}\ll 1$ the population of the excited state $200$ is comparable to $100$, but with significantly reduced density due to its large radius; and
in the limit 
$\xi_{\rm foc}\gg 1$, the excited state is rapidly populated but transitions to the ground state over %
a timescale similar to the exponential increase time, $\tau_{\rm rel}$. %

Let us first derive a generalized version of Eq.\,\eqref{eq:Ndotavg} that describes the simultaneous evolution of the occupation number of both $100$ and $200$ states, and comment later about the inclusion of all other levels. To do so, we solve the EoM in Eq.\,\eqref{eq:eom} perturbatively, starting from the initial condition:
\begin{equation}\label{eq:psi0_2}
\psi^{(0)}= \sqrt{N_{100}^{(0)}}e^{-i \omega_{1} t} \psi_{100}+\sqrt{N_{200}^{(0)}}e^{-i \omega_{2} t} \psi_{200}+\int [dk]a({\mathbf{k}}) e^{-i \omega_k t} \psi_{\mathbf{k}} \, .
\end{equation}
Contrary to Eq.\,\eqref{eq:psi0}, this has both $N_{100}^{(0)}\neq0$ and $N_{200}^{(0)}\neq0$. Of course, $N_{100}^{(0)}=N_{200}^{(0)}=0$ at $t=0$ in our Solar System, but, as in Section~\ref{ss:classical_p}, keeping them generic will allow us to study the rates $\langle \dot{N}_{100}\rangle$ and $\langle \dot{N}_{200}\rangle$ at an arbitrary time after such levels get populated from the DM capture. %

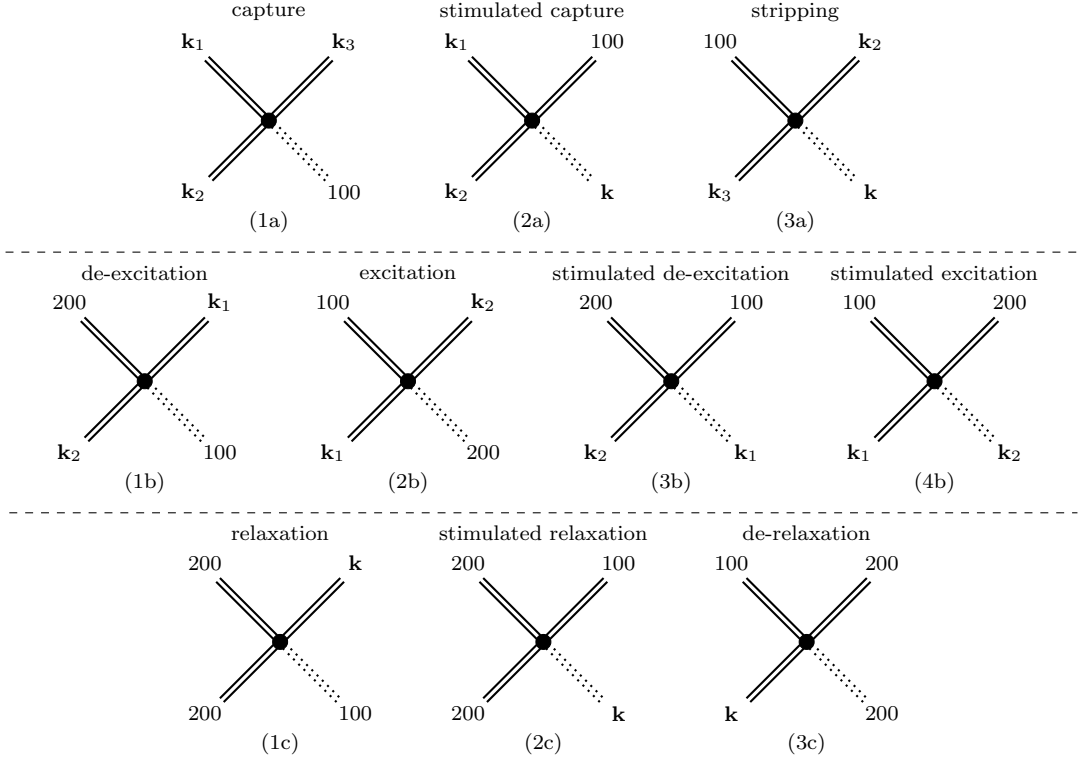
\begin{figure}
\begin{center}
 \begin{tikzpicture}
 \draw[dashed] (-3.5,-1.8) to (10.8,-1.8);
 \node at (-1.,1.) {\scriptsize{$\mathbf{k}_1$}};
  \node at (-1.,-1.) {\scriptsize{$\mathbf{k}_2$}};
  \node at (1.,1.) {\scriptsize{$\mathbf{k}_3$}};
  \node at (1.,-1.) {\scriptsize{$100$}};
 \node at (0,1.4) {\scriptsize{capture}};
  \node at (0,-1.4) {\scriptsize{(1a)}};
  \node[draw,circle,fill=black,inner sep=2pt] at (0,-0.05) {};

\draw[thick] (0,0) to (0.8,0.8);
\draw[thick] (0.85,0.75) to (0.05,-0.05) ;
\draw[thick,dotted] (0,0) to  (0.8,-0.8) ;
\draw[thick,dotted] (0.75,-0.85) to (-0.05,-0.05) ;
\draw[thick] (0,0) to (-0.8,-0.8) ;
\draw[thick] (-0.75,-0.85) to (0.05,-0.05) ;
\draw[thick] (0,0) to  (-0.8,0.8) ;
\draw[thick] (-0.85,0.75) to (-0.05,-0.05) ;
  \begin{scope}[xshift=3.5cm]
    
 \node at (-1.,1.) {\scriptsize{$\mathbf{k}_1$}};
  \node at (-1.,-1.) {\scriptsize{$\mathbf{k}_2$}};
  \node at (1.,1.) {\scriptsize{$100$}};
  \node at (1.,-1.) {\scriptsize{$\mathbf{k}$}};
 \node at (0,1.4) {\scriptsize{stimulated capture}};
  \node at (0,-1.4) {\scriptsize{(2a)}};
  \node[draw,circle,fill=black,inner sep=2pt] at (0,-0.05) {};

\draw[thick] (0,0) to (0.8,0.8);
\draw[thick] (0.85,0.75) to (0.05,-0.05) ;
\draw[thick,dotted] (0,0) to  (0.8,-0.8) ;
\draw[thick,dotted] (0.75,-0.85) to (-0.05,-0.05) ;
\draw[thick] (0,0) to (-0.8,-0.8) ;
\draw[thick] (-0.75,-0.85) to (0.05,-0.05) ;
\draw[thick] (0,0) to  (-0.8,0.8) ;
\draw[thick] (-0.85,0.75) to (-0.05,-0.05) ;
\end{scope}

\begin{scope}[xshift=7cm]
    
 \node at (-1.,1.) {\scriptsize{$100$}};
  \node at (-1.,-1.) {\scriptsize{$\mathbf{k}_3$}};
  \node at (1.,1.) {\scriptsize{$\mathbf{k}_2$}};
  \node at (1.,-1.) {\scriptsize{$\mathbf{k}$}};
 \node at (0,1.4) {\scriptsize{stripping}};
  \node at (0,-1.4) {\scriptsize{(3a)}};
  \node[draw,circle,fill=black,inner sep=2pt] at (0,-0.05) {};

\draw[thick] (0,0) to (0.8,0.8);
\draw[thick] (0.85,0.75) to (0.05,-0.05) ;
\draw[thick,dotted] (0,0) to  (0.8,-0.8) ;
\draw[thick,dotted] (0.75,-0.85) to (-0.05,-0.05) ;
\draw[thick] (0,0) to (-0.8,-0.8) ;
\draw[thick] (-0.75,-0.85) to (0.05,-0.05) ;
\draw[thick] (0,0) to  (-0.8,0.8) ;
\draw[thick] (-0.85,0.75) to (-0.05,-0.05) ;
\end{scope}
\end{tikzpicture}\\
\begin{tikzpicture}
 \draw[dashed] (-1.8,-1.8) to (12.4,-1.8);
 \node at (-1.,1.) {\scriptsize{$200$}};
  \node at (-1.,-1.) {\scriptsize{$\mathbf{k}_2$}};
  \node at (1.,1.) {\scriptsize{$\mathbf{k}_1$}};
  \node at (1.,-1.) {\scriptsize{$100$}};
 \node at (0,1.4) {\scriptsize{de-excitation}};
  \node at (0,-1.4) {\scriptsize{(1b)}};
  \node[draw,circle,fill=black,inner sep=2pt] at (0,-0.05) {};

\draw[thick] (0,0) to (0.8,0.8);
\draw[thick] (0.85,0.75) to (0.05,-0.05) ;
\draw[thick,dotted] (0,0) to  (0.8,-0.8) ;
\draw[thick,dotted] (0.75,-0.85) to (-0.05,-0.05) ;
\draw[thick] (0,0) to (-0.8,-0.8) ;
\draw[thick] (-0.75,-0.85) to (0.05,-0.05) ;
\draw[thick] (0,0) to  (-0.8,0.8) ;
\draw[thick] (-0.85,0.75) to (-0.05,-0.05) ;
  \begin{scope}[xshift=3.5cm]
    
 \node at (-1.,1.) {\scriptsize{$100$}};
  \node at (-1.,-1.) {\scriptsize{$\mathbf{k}_1$}};
  \node at (1.,1.) {\scriptsize{$\mathbf{k}_2$}};
  \node at (1.,-1.) {\scriptsize{$200$}};
 \node at (0,1.4) {\scriptsize{excitation}};
  \node at (0,-1.4) {\scriptsize{(2b)}};
  \node[draw,circle,fill=black,inner sep=2pt] at (0,-0.05) {};

\draw[thick] (0,0) to (0.8,0.8);
\draw[thick] (0.85,0.75) to (0.05,-0.05) ;
\draw[thick,dotted] (0,0) to  (0.8,-0.8) ;
\draw[thick,dotted] (0.75,-0.85) to (-0.05,-0.05) ;
\draw[thick] (0,0) to (-0.8,-0.8) ;
\draw[thick] (-0.75,-0.85) to (0.05,-0.05) ;
\draw[thick] (0,0) to  (-0.8,0.8) ;
\draw[thick] (-0.85,0.75) to (-0.05,-0.05) ;
\end{scope}

\begin{scope}[xshift=7cm]
    
 \node at (-1.,1.) {\scriptsize{$200$}};
  \node at (-1.,-1.) {\scriptsize{$\mathbf{k}_2$}};
  \node at (1.,1.) {\scriptsize{$100$}};
  \node at (1.,-1.) {\scriptsize{$\mathbf{k}_1$}};
 \node at (0,1.4) {\scriptsize{stimulated de-excitation}};
  \node at (0,-1.4) {\scriptsize{(3b)}};
  \node[draw,circle,fill=black,inner sep=2pt] at (0,-0.05) {};

\draw[thick] (0,0) to (0.8,0.8);
\draw[thick] (0.85,0.75) to (0.05,-0.05) ;
\draw[thick,dotted] (0,0) to  (0.8,-0.8) ;
\draw[thick,dotted] (0.75,-0.85) to (-0.05,-0.05) ;
\draw[thick] (0,0) to (-0.8,-0.8) ;
\draw[thick] (-0.75,-0.85) to (0.05,-0.05) ;
\draw[thick] (0,0) to  (-0.8,0.8) ;
\draw[thick] (-0.85,0.75) to (-0.05,-0.05) ;
\end{scope} 
\begin{scope}[xshift=10.5cm]
    
 \node at (-1.,1.) {\scriptsize{$100$}};
  \node at (-1.,-1.) {\scriptsize{$\mathbf{k}_1$}};
  \node at (1.,1.) {\scriptsize{$200$}};
  \node at (1.,-1.) {\scriptsize{$\mathbf{k}_2$}};
 \node at (0,1.4) {\scriptsize{stimulated excitation}};
  \node at (0,-1.4) {\scriptsize{(4b)}};
  \node[draw,circle,fill=black,inner sep=2pt] at (0,-0.05) {};

\draw[thick] (0,0) to (0.8,0.8);
\draw[thick] (0.85,0.75) to (0.05,-0.05) ;
\draw[thick,dotted] (0,0) to  (0.8,-0.8) ;
\draw[thick,dotted] (0.75,-0.85) to (-0.05,-0.05) ;
\draw[thick] (0,0) to (-0.8,-0.8) ;
\draw[thick] (-0.75,-0.85) to (0.05,-0.05) ;
\draw[thick] (0,0) to  (-0.8,0.8) ;
\draw[thick] (-0.85,0.75) to (-0.05,-0.05) ;
\end{scope} 
\end{tikzpicture}\\
\begin{tikzpicture}
 \node at (-1.,1.) {\scriptsize{$200$}};
  \node at (-1.,-1.) {\scriptsize{$200$}};
  \node at (1.,1.) {\scriptsize{$\mathbf{k}$}};
  \node at (1.,-1.) {\scriptsize{$100$}};
 \node at (0,1.4) {\scriptsize{relaxation}};
  \node at (0,-1.4) {\scriptsize{(1c)}};
  \node[draw,circle,fill=black,inner sep=2pt] at (0,-0.05) {};

\draw[thick] (0,0) to (0.8,0.8);
\draw[thick] (0.85,0.75) to (0.05,-0.05) ;
\draw[thick,dotted] (0,0) to  (0.8,-0.8) ;
\draw[thick,dotted] (0.75,-0.85) to (-0.05,-0.05) ;
\draw[thick] (0,0) to (-0.8,-0.8) ;
\draw[thick] (-0.75,-0.85) to (0.05,-0.05) ;
\draw[thick] (0,0) to  (-0.8,0.8) ;
\draw[thick] (-0.85,0.75) to (-0.05,-0.05) ;
  \begin{scope}[xshift=3.5cm]
    
 \node at (-1.,1.) {\scriptsize{$200$}};
  \node at (-1.,-1.) {\scriptsize{$200$}};
  \node at (1.,1.) {\scriptsize{$100$}};
  \node at (1.,-1.) {\scriptsize{$\mathbf{k}$}};
 \node at (0,1.4) {\scriptsize{stimulated relaxation}};
  \node at (0,-1.4) {\scriptsize{(2c)}};
  \node[draw,circle,fill=black,inner sep=2pt] at (0,-0.05) {};

\draw[thick] (0,0) to (0.8,0.8);
\draw[thick] (0.85,0.75) to (0.05,-0.05) ;
\draw[thick,dotted] (0,0) to  (0.8,-0.8) ;
\draw[thick,dotted] (0.75,-0.85) to (-0.05,-0.05) ;
\draw[thick] (0,0) to (-0.8,-0.8) ;
\draw[thick] (-0.75,-0.85) to (0.05,-0.05) ;
\draw[thick] (0,0) to  (-0.8,0.8) ;
\draw[thick] (-0.85,0.75) to (-0.05,-0.05) ;
\end{scope}

\begin{scope}[xshift=7cm]
    
 \node at (-1.,1.) {\scriptsize{$100$}};
  \node at (-1.,-1.) {\scriptsize{$\mathbf{k}$}};
  \node at (1.,1.) {\scriptsize{$200$}};
  \node at (1.,-1.) {\scriptsize{$200$}};
 \node at (0,1.4) {\scriptsize{de-relaxation}};
  \node at (0,-1.4) {\scriptsize{(3c)}};
  \node[draw,circle,fill=black,inner sep=2pt] at (0,-0.05) {};

\draw[thick] (0,0) to (0.8,0.8);
\draw[thick] (0.85,0.75) to (0.05,-0.05) ;
\draw[thick,dotted] (0,0) to  (0.8,-0.8) ;
\draw[thick,dotted] (0.75,-0.85) to (-0.05,-0.05) ;
\draw[thick] (0,0) to (-0.8,-0.8) ;
\draw[thick] (-0.75,-0.85) to (0.05,-0.05) ;
\draw[thick] (0,0) to  (-0.8,0.8) ;
\draw[thick] (-0.85,0.75) to (-0.05,-0.05) ;
\end{scope}
\end{tikzpicture}
	\caption{A schematic representation of the processes that contribute to the change in the number of bound particles $\langle \dot{N}_{100}\rangle$, see Eq.\,\eqref{eq:2leveln1}, in a system where both the 100 and 200 levels are present. As in Figure~\ref{fig:processes}, double solid lines represents the occupation number factors in the different terms of Eq.\,\eqref{eq:2leveln1}.
	}\label{fig:diagrams2level}
	\end{center}
	\vspace{-3mm}
\end{figure}

As in Appendix~\ref{app:onelevel}, the occupation number $\langle {N}_{100}\rangle$ is given in Eq.\,\eqref{eq:Navg2}, and requires the calculation of $c_{100}^{(1)}$ and $c_{100}^{(2)}$. By plugging Eq.\,\eqref{eq:psi0_2} into Eq.\,\eqref{eq:cdot100} %
and selecting only the terms that conserve energy and change the particle number in the 100 state, we find %
\begin{align}
c_{100}^{(1)}(t)&=-i g \bigg[\int [dk_1][dk_2][dk_3]a(\mathbf{k}_1)a(\mathbf{k}_2)a^*(\mathbf{k}_3) \mathcal{M}_{{k}_1+{k}_2\rightarrow {k}_3+100}\int_0^te^{i(
\omega_{1}+\omega_{k_3}-\omega_{k_1}-\omega_{k_2})t'}dt'\nonumber\\
&+2\sqrt{N_{200}^{(0)}}\int [dk_1][dk_2]a(\mathbf{k}_1)a^*(\mathbf{k}_2) \mathcal{M}_{{k}_1+200\rightarrow {k}_2+100}\int_0^te^{i(
\omega_{1}+\omega_{k_2}-\omega_{k_1}-\omega_{2})t'}dt'\nonumber\\
&+N_{200}^{(0)}\int [dk]a(\mathbf{k}) \mathcal{M}_{200+200\rightarrow {k}+100}\int_0^te^{i(
\omega_{1}+\omega_{k}-2\omega_{2})t'}dt' \bigg] \, . \label{eq:c100_2level}
\end{align}
The $\langle |c_{100}^{(1)}|^2\rangle$ part of $\langle N_{100}\rangle$ is simply given by the square of the amplitude of each term in Eq.\,\eqref{eq:c100_2level} (the cross terms vanish upon average) and reads%
\begin{align}
\langle |c_{100}^{(1)}|^2 \rangle = & \, g^2 t \Big[2 \int [d k_1] [d k_2][d k_3]|\mathcal{M}_{{k}_1+{k}_2\rightarrow 100+{k}_3}|^2 (2\pi) \delta(\omega_{1}+\omega_{k_3}-\omega_{k_1}-\omega_{k_2})f(\mathbf{k}_1)f(\mathbf{k}_2)f(\mathbf{k}_3) \nonumber\\
&+4N_{200}^{(0)}\int [dk_1][dk_2] |\mathcal{M}_{{k}_1+200\rightarrow {k}_2+100}|^2 (2\pi)\delta(
\omega_{1}+\omega_{k_2}-\omega_{k_1}-\omega_{2})f(\mathbf{k}_1)f(\mathbf{k}_2)\nonumber\\
&+2N_{200}^{(0)2}\int [dk] |\mathcal{M}_{200+200\rightarrow {k}+100}|^2 (2\pi)\delta (\omega_{1}+\omega_{k}-2\omega_{2})f(\mathbf{k})\Big] \, . \label{eq:c100sq_2level}
\end{align}
The first term in Eqs.~\eqref{eq:c100_2level} and~\eqref{eq:c100sq_2level} describes the (already discussed) direct capture from the DM waves, and is shown in diagram (1a) in Figure~\ref{fig:diagrams2level}, where the conventions for the lines are as in Section~\ref{ss:classical_p}. In addition to this, two more terms appeared: They describe the increase in the occupation number from the processes ${k}_1+200 \rightarrow 100+{k}_2$ and $200+200\rightarrow {k}+100$. The former, dubbed `de-excitation', is shown in diagram (1b); the latter, dubbed `relaxation', in diagram (1c). These are only active when $N_{200}^{(0)}\neq0$, and increase the occupation number in the ground state at the expenses of the populated 200 state.

As in Appendix~\ref{app:onelevel}, one could proceed to calculate the second order coefficient $c_{100}^{(2)}$, and thus the second part of $\langle N_{100}\rangle$%
. The only difference is that in $\psi^{(0)}$ on the RHS of the second order EoM, Eq.\,\eqref{eq:2order}, both the 200 state and the unbound component %
contribute. As before, the latter provides the stimulated capture and stripping terms in Eqs.~(\ref{eq:c1002_1}-\ref{eq:c1002_2}), represented in the diagrams (2a-3a) in Figure~\ref{fig:diagrams2level}, while the former provides new terms that involve $N_{200}^{(0)}$.   %

Rather than going through this lengthy derivation, we can obtain (heuristically) the form of the terms in $\langle\dot{N}_{100}\rangle$ that depend on $N_{200}^{(0)}$ as we did in Section~\ref{ss:quantum_scattering}, i.e. by demanding that the dependence on the occupation number of the two new processes ${k}_1+200 \leftrightarrow
 100+{k}_2$ and $200+200\leftrightarrow {k}+100$ (which include direct and inverse processes) is consistent with Bose enhancement. In other words, the contribution to $\langle \dot{N}_{100}\rangle$ from the former %
 should be proportional to
\begin{align}
&f(\mathbf{k}_1)N_{200}[f(\mathbf{k}_2)+1][N_{100}+1]-[f(\mathbf{k}_1)+1][N_{200}+1]f(\mathbf{k}_2)N_{100} \nonumber\\
\simeq \  &f(\mathbf{k}_1)f(\mathbf{k}_2)(N_{200}-N_{100})+N_{100}N_{200}[f(\mathbf{k}_2)-f(\mathbf{k}_1)] \, .
\end{align}
while for the latter to
\begin{align}
&N_{200}^2[f(\mathbf{k})+1][N_{100}+1]-[N_{200}+1]^2f(\mathbf{k})N_{100} \nonumber\\
\simeq \ &f(\mathbf{k})N_{200}^2+N_{100}N_{200}^2-2N_{100}N_{200}f(\mathbf{k}) \, ,
\end{align}
where the approximations are valid for $N_{nlm}\gg1$ (for simplicity of notation, here and in the following we omit the superscript `$(0)$'). Then, the evolution of the occupation number $\langle \dot{N}_{100}\rangle$ in the ground state can be obtained by starting from Eq.\,\eqref{eq:c100sq_2level} and adding more terms in the RHS in such a way that the dependence on the occupation numbers $N_{100}$ and $N_{200}$ matches that of the two previous equations. These terms are simply the permutations (2b-4b) and (2c-3c) of the diagrams (1b) and (1c) respectively, weighted with the appropriate factors. The full result is
\begin{align}
\label{eq:2leveln1}
\dot{N}_{100}&=2g^2 \int [dk_1][dk_2][dk_3] |\mathcal{M}_{{k}_1+{k}_2\rightarrow {k}_3+100}|^2 (2\pi) \delta(
\omega_{k_1}+\omega_{k_2}-\omega_{1}-\omega_{k_3})\nonumber\\
& \left\{f(\mathbf{k}_1)f(\mathbf{k}_2)f(\mathbf{k}_3)+N_{100}[f(\mathbf{k}_1)f(\mathbf{k}_2)-2f(\mathbf{k}_2)f(\mathbf{k}_3)]\right\}\nonumber\\
&+ 4g^2 \int [dk_1][dk_2]|\mathcal{M}_{{k}_1+100\rightarrow {k}_2+200}|^2 (2\pi) \delta(
\omega_{k_1}+\omega_{1}-\omega_{2}-\omega_{k_2})\nonumber\\
& \{N_{100}N_{200}[f(\mathbf{k}_2)-f(\mathbf{k}_1)]-f(\mathbf{k}_1)f(\mathbf{k}_2)[N_{100}-N_{200}]\}\nonumber\\
&+2g^2 \int [dk]|\mathcal{M}_{200+200\rightarrow 100+{k}}|^2 (2\pi) \delta(
2\omega_{2}-\omega_{1}-\omega_{k}) \{f(\mathbf{k})N_{200}^2+N_{100}N_{200}^2-2N_{200}N_{100}f(\mathbf{k})\} \, .
\end{align}
One can similarly obtain the evolution of $\langle N_{200}\rangle$, which reads
\begin{align}
\label{eq:2leveln2}
\dot{N}_{200}&=2g^2 \int [dk_1][dk_2][dk_3] |\mathcal{M}_{{k}_1+{k}_2\rightarrow {k}_3+200}|^2 (2\pi) \delta(
\omega_{k_1}+\omega_{k_2}-\omega_{2}-\omega_{k_3})\nonumber\\
& \{f(\mathbf{k}_1)f(\mathbf{k}_2)f(\mathbf{k}_3)+N_{200}[f(\mathbf{k}_1)f(\mathbf{k}_2)-2f(\mathbf{k}_2)f(\mathbf{k}_3)]\}\nonumber\\
&- 4g^2 \int [dk_1][dk_2]|\mathcal{M}_{{k}_1+100\rightarrow {k}_2+200}|^2 (2\pi) \delta(
\omega_{k_1}+\omega_{1}-\omega_{2}-\omega_{k_2})\nonumber\\
& \{N_{100}N_{200}[f(\mathbf{k}_2)-f(\mathbf{k}_1)]-f(\mathbf{k}_1)f(\mathbf{k}_2)[N_{100}-N_{200}]\}\nonumber\\
&-4g^2 \int [dk]|\mathcal{M}_{200+200\rightarrow 100+{k}}|^2 (2\pi) \delta(
2\omega_{2}-\omega_{1}-\omega_{k}) \{f(\mathbf{k})N_{200}^2+N_{100}N_{200}^2-2N_{200}N_{100}f(\mathbf{k})\} \, ,
\end{align}
and is analogous to Eq.\,\eqref{eq:2leveln1}, provided that $N_{100}$ and $N_{200}$ are exchanged, appropriate minus signs added and a factor of 2 in the last line of Eq.\,\eqref{eq:2leveln2}. Note that:

\vspace{-2mm}
\begin{itemize}[leftmargin=0.2in] \setlength\itemsep{0.15em}
	\item The first two lines of Eq.\,\eqref{eq:2leveln1} reproduce the capture, stimulated capture and stripping processes, diagrams (1a-3a) of Figure~\ref{fig:diagrams2level}.
    \item The third and fourth lines include the de-excitation process (1b) and its inverse  (`excitation', (2b)), as well as the same processes that are `stimulated' by the 200 state being already populated (3b-4b); these appear thanks to Bose enhancement. Excitations deplete the 100 state given their minus sign. Observe that the excitation term comes together with a tower of additional excitations to higher and higher states, similar to the last term in Eq.\,\eqref{eq:ndot100full}. We omitted these because not crucial for what follows.
    \item Finally, the last line includes the relaxation process (1c), as well as its stimulated version (2c) and its inverse process (`de-relaxation', 3c).
\end{itemize}

Eqs.~(\ref{eq:2leveln1}-\ref{eq:2leveln2}) constitute a closed system that determines the evolution of $N_{100}(t)$ and $N_{200}(t)$%
. However, as discussed at the beginning of this Section, higher levels (or levels with nontrivial angular momentum) could be also populated from DM capture. Thus, in principle, one should derive a more general set of coupled equations describing also the evolution e.g. of the $300$ %
level, by including $N_{300}^{(0)}\neq0$ in the initial conditions of Eq.\,\eqref{eq:psi0_2}. In this way one could %
study also the population of these levels, and track their evolution afterwards. %
Eventually this leads to an infinite (and intractable) set of  coupled equations (similar in form to Eqs.~(\ref{eq:2leveln1}-\ref{eq:2leveln2})) describing the simultaneous evolution of the occupation number of all levels.   However, in the following we will prove that: %

\vspace{-2mm}
\begin{itemize}[leftmargin=0.2in] \setlength\itemsep{0.15em}
	\item In the limit $2\pi\alpha/v_{\rm dm}\ll1$, the two-level system in Eqs.~(\ref{eq:2leveln1}-\ref{eq:2leveln2}) %
	tends to an equilibrium configuration where $N_{100}$ and (later) $N_{200}$ have a common value, which turns out to coincide with that obtained from the single-level equation, Eq.\,\eqref{eq:Ndotavg} in the main text. Thus, the density at the center is dominated by the ground state. %
    \item In the limit $2\pi\alpha/v_{\rm dm}\gg1$ %
	the excited state, initially exponentially enhanced, quickly decays to the ground state.
\end{itemize}
We expect this dynamics to be prototypical of the all-level system: Excited states quickly decay to lower-level states in the second limit, and the density is well reproduced by the single level system in the first; as a result, it is likely that the full system behaves, to a good approximation, as the ground state.%

\subsubsection*{Dilute gravitational atoms: $\xi_{\rm foc}=2\pi\alpha/v_{\rm dm}\ll1$}

We first consider the limit $2\pi \alpha /v_{\rm dm} \ll 1$. As mentioned in Section~\ref{ss:light_vs_heavy}, $\mathcal{M}_{{k}_1+{k}_2\rightarrow {k}_3+n00} \propto n^{-3/2}$; thus, the capture and stripping rates for the $200$ state are approximately $1/8$ times those for $100$. Energy conservation,  $\omega_{k_1}+\omega_{1}-\omega_{2}-\omega_{k_2}=0$, forces ${k}_1\simeq {k}_2$ in the $2\pi \alpha /v_{\rm dm} \ll 1$ limit. As a result, since the sum of the excitation and de-excitation terms is proportional to $f(\mathbf{k}_2)-f(\mathbf{k}_1)$, these cancel between each other. Consequently, Eqs.~\eqref{eq:2leveln1} and~\eqref{eq:2leveln2} can be approximated as
\begin{align}
\dot{N}_{100}&=a_1-a_2 N_{100}-a_3 (N_{100}-N_{200})+a_4 (N_{200}^2-2N_{200}N_{100})+a_5 N_{100}N_{200}^2 \, , \nonumber\\
\dot{N}_{200}&=a_1/8-(a_2/8) N_{200}+a_3 (N_{100}-N_{200})-2a_4 (N_{200}^2-2N_{200}N_{100})-2a_5 N_{100}N_{200}^2\, .
\end{align}
As in Section~\ref{ss:light_vs_heavy}, the coefficients $a_1$ and $a_2$ are
\begin{align}
a_1= \frac{16384 \sqrt{\pi} B g^2\rho_{\rm dm}^3 \alpha^4}{m^7 v_{\rm dm}^{9}}~, \qquad a_2 =\frac{A}{2.4}\frac{(2\pi \alpha/v_{\rm dm})^4}{\tau_{\rm rel}}~,
\end{align}
with $A\simeq 0.21$ and $B\simeq 0.08$. The other coefficients may be evaluated in a similar way:%
\begin{align}
a_3 \simeq 0.04\frac{(2\pi \alpha/v_{\rm dm})^4}{\tau_{\rm rel}}~, \qquad a_4= \frac{10816 g^2 \rho_{\rm dm} \alpha^3 m }{ 4782969 \sqrt{\pi} v_{\rm dm}^2 }f_4(\alpha/v_{\rm dm})~ ,\qquad a_5 =\frac{2704\sqrt{2} g^2 m^5 \alpha^4 }{ 4782969 \pi} ~,
\end{align}
where $f_4(a)=2\sqrt{2}a/e$ for $a\ll1$%
.\footnote{The exact form is $f_4(a)\equiv e^{-(1+a/\sqrt{2})^2}(e^{2\sqrt{2}a}-1)$.} 
These equations are greatly simplified when rewritten in terms of $t'=t /\tau_{\rm rel}$ and of the normalized occupation number %
\begin{equation}\label{eq:}
    \tilde{N}_{n00}\equiv  \frac{m^4 \alpha^3 }{\rho_{\rm dm}} N_{n00}=\pi n^3\frac{\rho_{n00}}{\rho_{\rm dm}} ~, \qquad  \rho_{n00}\equiv m N_{n00}|\psi_{n00}(0)|^2=\frac{m N_{n00}}{\pi (nR_\star)^3} \, .
\end{equation}
Note that $\tilde{N}_{n00}$ represents (modulo a $\pi n^3$ factor) the contribution $\rho_{n00}$ from the state $n00$  to the density of the bound state %
at the center $r=0$, normalized to $\rho_{\rm dm}$ (i.e. $\rho(0)/\rho_{\rm dm}$). In particular, %
for the single-level system in Eq.\,\eqref{eq:Ndotavg} at equilibrium we have $\tilde{N}_{100}\sim \rho_{100}/\rho_{\rm dm}\sim (2\pi\alpha/v_{\rm dm})^3\ll1$, see Eq.\,\eqref{eq:Mrhoeq}. Using these variables, we obtain 
\begin{align}
\frac{d \tilde{N}_{100}}{d t'}&=2 \cdot 10^3 \left[\frac{\alpha}{v_{\rm dm}}\right]^7- 10^2 \left[\frac{\alpha}{v_{\rm dm}}\right]^4 \tilde{N}_{100}-60 \left[\frac{\alpha}{v_{\rm dm}}\right]^4 \left[\tilde{N}_{100}-\tilde{N}_{200}\right]\nonumber\\
&+10^{-3} f_4\left(\frac{\alpha}{v_{\rm dm}}\right)\left[\tilde{N}_{200}^2-2\tilde{N}_{100}\tilde{N}_{200}\right]+2.5 \cdot 10^{-4 }\left[\frac{v_{\rm dm}}{\alpha}\right]^2\tilde{N}_{100}\tilde{N}_{200}^2 \, ,  \label{eq:ntildeund1}%
\end{align}
\vspace{-.3cm}
\begin{align}
\frac{d \tilde{N}_{200}}{d t'}&= 3\cdot10^2 \left[\frac{\alpha}{v_{\rm dm}}\right]^7-20 \left[\frac{\alpha}{v_{\rm dm}}\right]^4 \tilde{N}_{200}+60 \left[\frac{\alpha}{v_{\rm dm}}\right]^4 \left[\tilde{N}_{100}-\tilde{N}_{200}\right]\nonumber\\
&-2\cdot 10^{-3} f_4\left(\frac{\alpha}{v_{\rm dm}}\right)\left[\tilde{N}_{200}^2-2\tilde{N}_{100}\tilde{N}_{200 }\right]-5 \cdot 10^{-4 }\left[\frac{v_{\rm dm}}{\alpha}\right]^2\tilde{N}_{100}\tilde{N}_{200}^2 \, .\label{eq:ntildeund2}
\end{align}

For $\xi_{\rm foc}=2\pi\alpha/v_{\rm dm}\ll1$ the equilibrium solution (i.e. $\dot{N}_{nlm}=0$) of this system occurs at $N_{100}=N_{200}\equiv N$, with equilibrium value for $N_{100}$ coinciding with that of the single-level system in Eq.\,\eqref{eq:Ndotavg}. This is because for such a configuration the stimulated excitation and de-excitation terms (proportional to $\tilde{N}_{100}-\tilde{N}_{200}$ in the previous equations) cancel between each other, while all the relaxation terms (in last line) are much smaller than the capture and stimulated capture ones, which therefore dominate. Indeed, the relaxation terms are proportional to $f_4\tilde{N}^2\propto (\alpha/v_{\rm dm})^7$ and $(v_{\rm dm}/\alpha)^2\tilde{N}^3\propto(\alpha/v_{\rm dm})^7$, but their coefficient is suppressed by at least four orders of magnitudes compared to that of the capture and stimulated capture, ultimately because $\tilde{N}\ll1$.

\begin{figure}%
\begin{center}
	\includegraphics[width=0.4\textwidth]{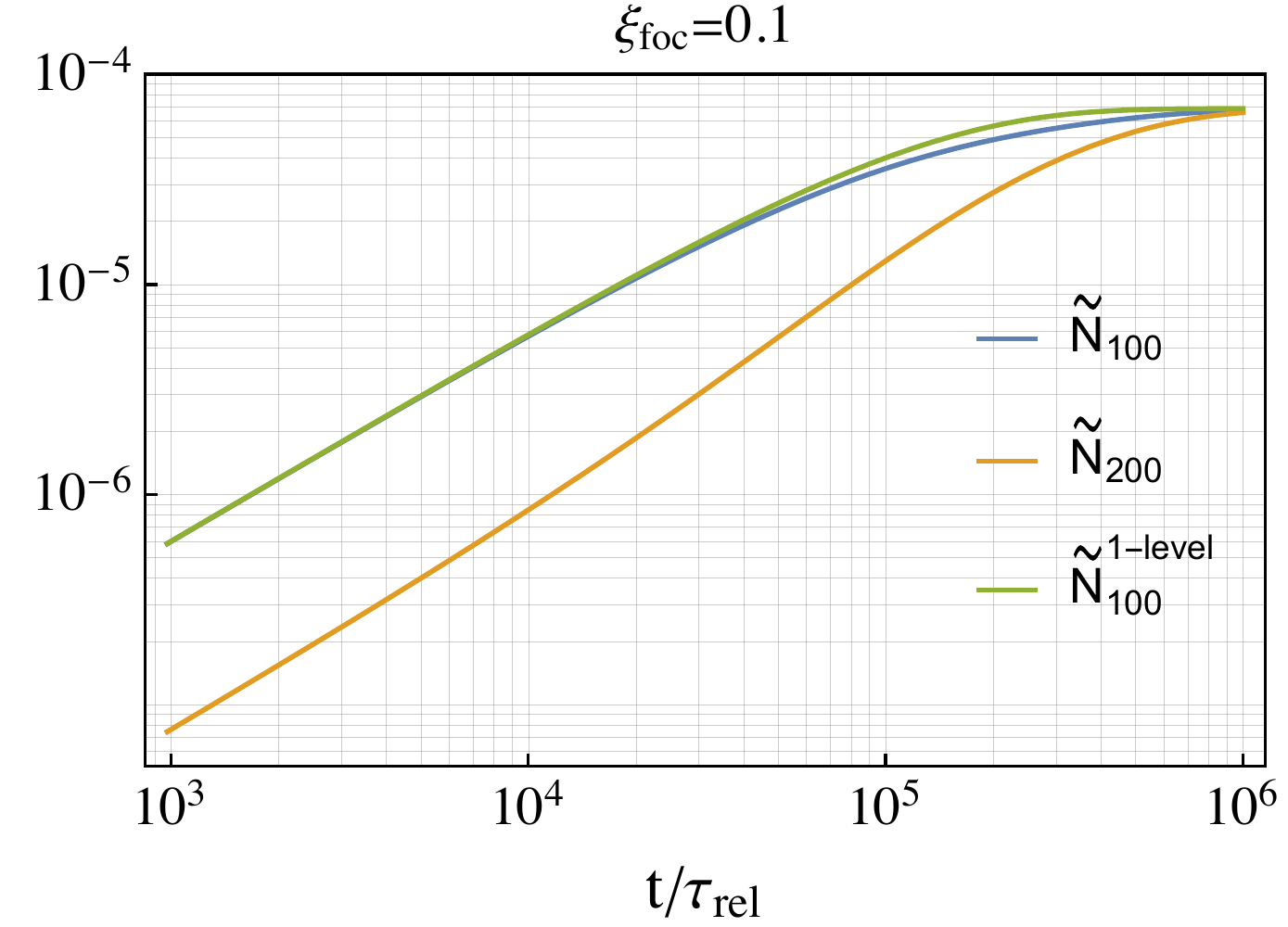}
 \includegraphics[width=0.4\textwidth]{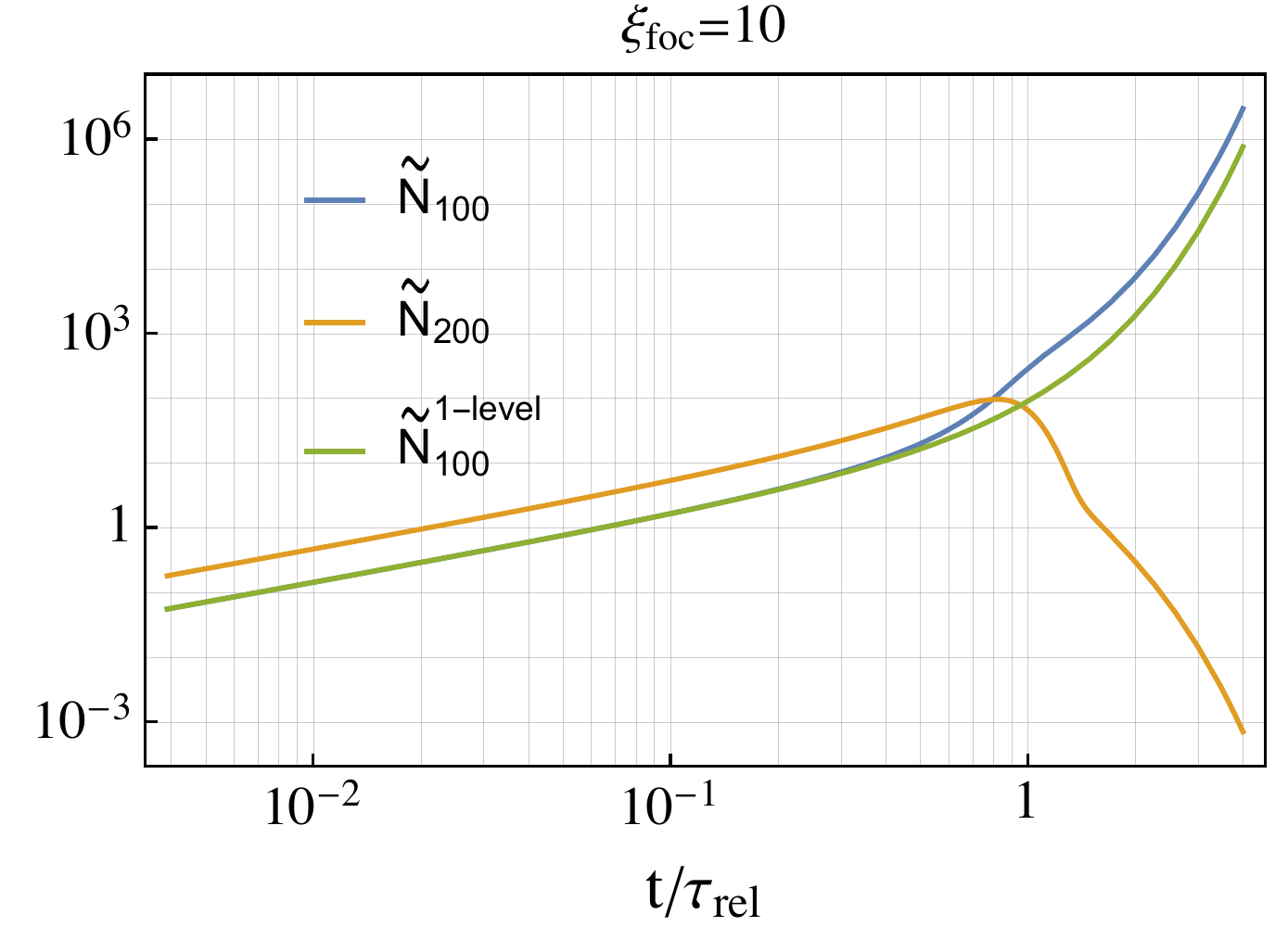}
	\caption{The time evolution of the normalized occupation numbers $\tilde{N}_{100}$  (blue) and $\tilde{N}_{200}$ (orange), by numerically solving the two-level system, Eqs.~(\ref{eq:ntildeund1}-\ref{eq:ntildeund2}), with $\xi_{\rm foc}=2\pi \alpha/v_{\rm dm}=0.1$ (left), and Eqs.~(\ref{eq:Ndot2red1}-\ref{eq:Ndot2red2}) with $\xi_{\rm foc}=10$ (right).  $\tilde{N}_{n00}/(\pi n^3)$ represents the contribution from the state $n00$ to the central %
 density, normalized to $\rho_{\rm dm}$. For comparison, we also show the evolution of the occupation number of the one-level system studied in the main text (green). In the second limit, %
 both the states increases exponentially, but the excited state quickly transitions to the ground state soon after $t\simeq\tau_{\rm rel}$.%
 }\label{fig:2level}
	\end{center}
	\vspace{-3mm}
\end{figure}

Figure~\ref{fig:2level} (left) shows the numerical evolution of the system in Eqs.~(\ref{eq:ntildeund1}-\ref{eq:ntildeund2}) starting from $N_{100}=N_{200}=0$ for $\xi_{\rm foc}=%
0.1$. Initially the 200 level grows slower than the ground state due to the $1/8$ factor. However, when the occupation numbers become large enough, the excitation terms become important and quickly bring the occupation numbers close to each other. In the end, both states reach the equilibrium, with occupation numbers similar to the value in the one level system. As mentioned, since the $200$ level has a radius twice as larger as that of the ground state, the density near of center of the gravitational atom is dominated by the ground state. %

\subsubsection*{Dense gravitational atoms: $\xi_{\rm foc}= 2\pi\alpha/v_{\rm dm}\gg1$}

The only processes in Figure~\ref{fig:diagrams2level} that are not suppressed in the regime $2\pi\alpha/v_{\rm dm}\gg1$ are \emph{(i)} stimulated capture (diagram (2a) of Figure~\ref{fig:diagrams2level}, discussed in the main text), \emph{(ii)} stimulated de-excitation (diagram (3b)) and \emph{(iii)} stimulated relaxation (diagram (2c)). All the other diagrams are exponentially suppressed because do not benefit from Bose enhancement. In particular:

\vspace{-2mm}
\begin{itemize}[leftmargin=0.2in] \setlength\itemsep{0.15em}
	\item For the excitation and de-excitation processes ${k}_1+100 \leftrightarrow 200+{k}_2$, central panel of Figure~\ref{fig:diagrams2level}, energy conservation dictates $k_1=\sqrt{k_2^2+3m^2\alpha^2/4}\gg m v_{\rm dm}$, so the diagrams (1b,2b,4b), proportional to $f(\mathbf{k}_1)$, are exponentially suppressed.
	\item Similarly, the relaxation and de-relaxation processes $200+200\leftrightarrow 100+{k}$ in diagrams (1c,3c) are proportional to $f(\mathbf{k})$ and by energy conservation $k=m\alpha/\sqrt{2}\gg m v_{\rm dm}$, so the resulting integrals are small.
\end{itemize}
The unsuppressed processes (2a,3b,2c) only induce level %
transitions from 200 to 100. As we now show in detail, if populated, the 200 state will be therefore depleted into the 100 state, and will increase its growth%
.
At $t=0$, $N_{100}=N_{200}=0$ and the only term that contributes to a change in the number of bound particles is the (suppressed) direct capture, as all the other terms are proportional to further powers of $N_{100}$ and $N_{200}$. Ignoring all the terms that are exponentially suppressed except for direct capture, in the $2\pi\alpha/v_{\rm dm}\gg1$ limit Eqs.~(\ref{eq:2leveln1}-\ref{eq:2leveln2}) are well approximated by: 
\begin{align}
\dot{N}_{100}&=C_{100}/m+\Gamma_{100} N_{100}+D N_{100}N_{200}+R N_{100}N_{200}^2 \, ,  \label{eq:Ndot2red1}\\
\dot{N}_{200}&=C_{200}/m+\Gamma_{200} N_{200}-D N_{100}N_{200}-2R N_{100}N_{200}^2  \, ,\label{eq:Ndot2red2}
\end{align}
where the coefficients can be calculated similarly to Section~\ref{ss:light_vs_heavy},\footnote{The integral in the matrix element $\mathcal{M}_{100+k_1\to 200+k_2}$ is a function only of one variable, $\hat{k}_1\cdot \hat{k}_3$, and can be computed numerically; the corresponding integral in $\mathcal{M}_{200+200\to 100+k}$ is computable analytically.} and read
\begin{align} \label{eq:Cn00}
C_{n00}\simeq\frac{16 g^2 \rho_{\rm dm}^3}{m^6 v_{\rm dm}^5} s(\alpha^2/n^2v_{\rm dm}^2)\,, \ \ \  n=1,2\,, \qquad\quad \Gamma_{100}\simeq \Gamma_{200}\simeq\frac{1}{0.34 \tau_{\rm rel}}\simeq\frac{3g^2\rho_{\rm dm}^2}{m^3 v_{\rm dm}^2} \, ,
\end{align}
where to a good approximation $s(x)\simeq e^{-x+\sqrt{4+2x}-2}$, and
\begin{align}
D\simeq\frac{0.035 g^2  \alpha^2 m \rho_{\rm dm}}{v_{\rm dm}}\,,\qquad\quad R\simeq 1.3\cdot 10^{-4} g^2 \alpha^4 m^5 \, .
\end{align}
The solution of Eqs.~(\ref{eq:Ndot2red1}-\ref{eq:Ndot2red2}) with initial conditions $N_{100}=N_{200}=0$ can be understood qualitatively. First, notice that~$\Gamma_{100}$ and~$\Gamma_{200}$ are of the same order (because the matrix elements $M_{k_1+k_2\to k_3+n00}$ with $n=1,2$ are similar), so we expect also $N_{200}$ to increase exponentially, with similar timescale $\tau_{\rm rel}$. The capture coefficients $C_{100}$ and $C_{200}$ are comparable, but since $s(x)$ is exponentially decreasing function of $\alpha/(n v_{\rm dm})$ with $n=1,2$, the latter is less suppressed at large $2\pi\alpha/v_{\rm dm}$. Ultimately, this is because the exponential suppression from $f(\mathbf{k}_3)$, with $k_3=\sqrt{k_1^2+k_2^2+\alpha^2 m^2/n^2}$, is smaller for $n=2$ than $n=1$.  Thus, as mentioned in Section~\ref{ss:light_vs_heavy}, for %
$2\pi\alpha/v_{\rm dm}\gg1$ the mass captured until the time $1/\Gamma$ is larger for the 200 level. 

As a result, initially both levels grow; while %
$N_{100}$ and $N_{200}$ are small, the growth is driven by the capture terms and is linear%
. As soon as $N_{100}$ and $N_{200}$ are large enough, the stimulated capture terms drive the exponential increase of the levels at $t\simeq \tau_{\rm rel}$. Soon after, the stimulated de-excitation and stimulated relaxation terms (last and second-to-last terms in Eq.\,\eqref{eq:Ndot2red2}) become as large as the stimulated capture term, because they are proportional to higher powers of $N_{nlm}$.\footnote{On the other hand, the exponentially-suppressed terms of Figure~\ref{fig:2level}, not reported in Eq.\,\eqref{eq:Ndot2red2}, will be suppressed with respect to these, and therefore not considered in the evolution.} By comparing the capture term with the stimulated de-excitation and the relaxation ones, and using the fact that (before these become relevant) $N_{100}\propto N_{200}\propto e^{t/\tau_{\rm rel}}$, we can estimate that the latter start dominating over the former, inducing the decay of the 200 level, at times $t_{c,D}\simeq \tau_{\rm rel} \log_D$ and $t_{c,R}\simeq 2\tau_{\rm rel} \log_R$, respectively. The factors $\log_D$ and $\log_R$ are of order one and can be estimated from Eq.\,\eqref{eq:Ndot2red2} as $\log_D\simeq\log(15v_{\rm dm}^2/f_2\alpha^2)$ 
and $\log_R\simeq2\log(27v_{\rm dm}^2/f_{2}\alpha^2)$; here $f_2$ is the density of the 200 state in units of the DM density at time $t\simeq\tau_{\rm rel}$. We conclude that, soon after the exponential increase starts, 200 decays into 100. %
Figure~\ref{fig:2level} shows the time evolution of $\tilde{N}_{100}$ and $\tilde{N}_{200}$ for $2\pi \alpha/v_{\rm dm}=10$, from the numerical solution of Eqs.~(\ref{eq:Ndot2red1}-\ref{eq:Ndot2red2}) with initial condition $N_{100}=N_{200}=0$, where the main features described above are reproduced. In particular, the particle number in the $200$ state becomes negligible at $t\gtrsim 2\tau_{\rm rel}$. %
The dynamics of this simplified system should be prototypical of the full case including additional excited states. It could also possibly describe states with nontrivial angular momentum, though a more detailed analysis is required.%

\section{Details on Gross--Pitaevskii simulations}
\label{app:simulations}
In this Appendix we give a brief description of the simulations presented in Section~\ref{sec:simulation} and provide additional results.

\paragraph{Equations of motion and initial conditions.} As mentioned in Section~\ref{sec:simulation}, the combination of the EoM in Eq.\,\eqref{eq:EoM} and initial conditions in Eq.\,\eqref{eq:psi_w} depends only on the three dimensionless parameters $\{\alpha/v_{\rm dm},\tilde{g}\equiv g\rho_{\rm dm}/(m^2v_{\rm dm}^2),\tilde{R}_s\equiv mv_{\rm dm}R_s\}$. This can be easily seen by rewriting $(i\partial_t+\nabla^2/2m-m\Phi)\psi=g|\psi|^2\psi$ in terms of the rescaled variables $\tilde{\mathbf{x}}\equiv m v_{\rm dm} \mathbf{x}$, $\tilde{t}\equiv m v_{\rm dm}^2 t$ and $\tilde{\psi}\equiv\psi/\sqrt{\rho_{\rm dm}/m}$ (note that space and time are normalized to the typical wavelength and period of the DM waves, and $m\psi^2$ to their average density; the relaxation time is simply $mv_{\rm dm}^2\tau_{\rm rel}=1/\tilde{g}^{2}$). In this way, the EoM become
\begin{equation}\label{eq:EoMtilde}
\left(i\partial_{\tilde{t}}+\frac{\tilde{\nabla}^2}{2}- \tilde{\Phi}\right)\tilde{\psi}=\tilde{g}|\tilde{\psi}|^2\tilde{\psi} \, ,
\end{equation}
where $\tilde{\Phi}=-\frac{\alpha/v_{\rm dm}}{\tilde{r}}$ for $\tilde{r}>
\tilde{R}_s$ (i.e. away from the Sun) and $\tilde{\Phi}=-\frac{\alpha/v_{\rm dm}}{2\tilde{R}_s}(3-\tilde{r}^2/\tilde{R}_s^2)$ for $\tilde{r}<\tilde{R}_s$ (inside the Sun). These redefinitions are useful because the initial conditions only depend on $\alpha/v_{\rm dm}$, and read%
\begin{equation}\label{eq:init_con}
\tilde{\psi}_w=\int \frac{d^3\tilde{k}}{(2\pi)^3}\tilde{a}(\tilde{\mathbf{k}})e^{-i\frac{\tilde{k}^2\tilde{t}}{2}}\psi_{\tilde{\mathbf{k}}}(\tilde{\mathbf{x}}) ~,\qquad \langle\tilde{a}^*(\tilde{\mathbf{k}})\tilde{a}(\tilde{\mathbf{k}}')\rangle =\left[8\pi^{3/2}e^{-(\tilde{\mathbf{k}}-\hat{z})^2}\right](2\pi)^3\delta(\tilde{\mathbf{k}}-\tilde{\mathbf{k}}') \, ,
\end{equation}
where we took $\sigma=v_{\rm dm}/\sqrt{2}$ and $\mathbf{k}_{\rm dm}$ parallel to $\hat{z}$, and it is understood that $R_\star\to \frac{v_{\rm dm}}{\alpha}$ in $\psi_{\tilde{\mathbf{k}}}(\tilde{\mathbf{x}})$. In practice, a realization of these initial conditions is (at $t=0$)
\begin{equation}\label{eq:init_con_p}
\tilde{\psi}_{w,0}=A\int \frac{d^3\tilde{k}}{(2\pi)^3}e^{-\frac12{(\tilde{\mathbf{k}}- \hat{z})^2}+i\gamma(\tilde{\mathbf{k}})}\psi_{\tilde{\mathbf{k}}}(
\tilde{\mathbf{x}}) \, ,
\end{equation}
where $\gamma(\mathbf{k})$ is a random function of $\mathbf{k}$ taking values in $[0,2\pi[$. The coefficient $A$ in Eq.\,\eqref{eq:init_con_p} is fixed by the condition $m\langle|\psi_{w,0}|^2\rangle=%
\rho_{\rm dm}$, i.e. $\langle|\tilde{\psi}_{w,0}|^2\rangle=1$.

\vspace{-3mm}
\paragraph{Evolution and systematics.} We solve Eq.\,\eqref{eq:EoMtilde} with initial conditions in Eq.\,\eqref{eq:init_con_p} in a cubic box of length $L$ with $N_x^3=64^3\div 128^3$ points.\footnote{In practice, to compute the initial conditions we used the plane wave form $\psi_{\tilde{\mathbf{k}}}(\tilde{\mathbf{x}})=e^{i\tilde{\mathbf{k}}\cdot\tilde{\mathbf{x}}}$, which is correct except close to $\tilde{r}=0$. Given that plane waves are not hydrogen atom eigenstates, this leads to a small population of the bound states in the initial conditions, which however does not affect the timescale of the exponential growth, given the results of Section~\ref{sec:analytic}.} 
The coefficient $A$ in Eq.\,\eqref{eq:init_con_p} is in this case $A=8(\pi\tilde{L})^{3/2}$, where $\tilde{L}\equiv mv_{\rm dm}L$ is the length in units of the inverse momentum of the waves. The evolution is carried out with periodic boundary conditions and a standard second-order pseudo-spectral method, employed e.g. in~\cite{Levkov:2018kau,Chen:2020cef,Chen:2021oot,Edwards:2018ccc,Kirkpatrick:2021wwz%
}, to which we refer for all the details. (Note that the two additional parameters $\{N_x,\tilde{L}\}$ need to be specified in a simulation.)

The EoM imply the conservation of the total mass, $m\int d^3x |\psi|^2$, (which is automatic with the pseudo-spectral method we use) and of the nonrelativistic energy $\mathcal{E}\equiv\int d^3x\epsilon$, where $\epsilon$ is the energy density in Eq.\,\eqref{eq:energy_density}. The time-independence of $\mathcal{E}$ is therefore a nontrivial check of the numerical evolution. We compute $\mathcal{E}$ throughout all the simulations, check its conservation and rely on simulations where this is conserved better than  $\delta\mathcal{E}/\mathcal{E}<10^{-4}$ (this requires the time-step to be small enough, see footnote~\ref{footnotetime}). 

As anticipated in Section~\ref{sec:simulation}, there are multiple competing requirements for a simulation to capture the formation of the %
gravitational atom 
and be free from systematic uncertainties. 
\vspace{-1mm}
\begin{enumerate}[leftmargin=0.2in,label=(\alph*)] \setlength\itemsep{0.15em}
	\item The lattice spacing $\Delta\equiv L/N_x$ should be small enough to resolve the wavelength of the DM waves and the Sun radius. We checked that this requires $\Delta m v_{\rm dm}=\tilde{L}/N_x\lesssim1$ and $\Delta/R_s=\tilde{L}/(N_x\tilde{R}_s)\lesssim1$ respectively, i.e. there should be at least approximately one grid point per inverse DM momentum and per Sun radius.
    \item The length $L$ should be large enough for the box to be in the infinite volume limit, which requires $L/(2\pi/mv_{\rm dm})=\tilde{L}/2\pi\gg 1$. In other words, the number of wavelengths per box side should be large; $\tilde{L}/2\pi\gtrsim \mathcal{O}(10)$ is enough for this to happen. In particular, this means that the mass of the 
    gravitational atom is always a small fraction of the total mass in the box.
    \item Finally, $R_s$ should be much smaller than $R_\star$ for the point-like approximation of the Sun to hold, i.e. $R_s/R_\star=(\alpha/v_{\rm dm})\tilde{R}_s \ll 1$.

\end{enumerate}

In the simulations of the main text (Figure~\ref{fig:rho_sim} in Section~\ref{sec:simulation}), we used $|\tilde{g}|=0.006$ ($\tilde{g}>0$ for the two lowest values of $2\pi\alpha/v_{\rm dm}$ and $\tilde{g}<0$ otherwise) and $\tilde{R}_s=0.8$ (except for the two highest values of $2\pi\alpha/v_{\rm dm}$, for which $\tilde{R}_s=0.4$). We chose $\tilde{L}=40$ and performed simulations with both $N_x^3=64^3$ and $N_x^3=128^3$ to check the lattice spacing requirement (a). Finally, the function $\gamma(\tilde{\mathbf{k}})$ in the initial conditions in Eq.\,\eqref{eq:init_con_p} has been fixed to be the same (random) function of $\tilde{\mathbf{k}}$ for all simulations in Figure~\ref{fig:rho_sim}.\footnote{\label{footnotetime} We use time-step $\Delta_t m v_{\rm dm}^2\simeq (\Delta m v_{\rm dm})^2/ 4$, which is small enough for the energy $\mathcal{E}$ to be conserved as described above. Note that a much larger time-step is sufficient for $\mathcal{E}$ to be conserved at the $10^{-4}$ level before the 
gravitational atom forms. However, after formation, energy conservation is worsened because the shorter timescales associated to the ground state need to be resolved for large $2\pi\alpha/v_{\rm dm}$.} %

\begin{figure}%
\begin{center}
        \includegraphics[width=0.44\textwidth]{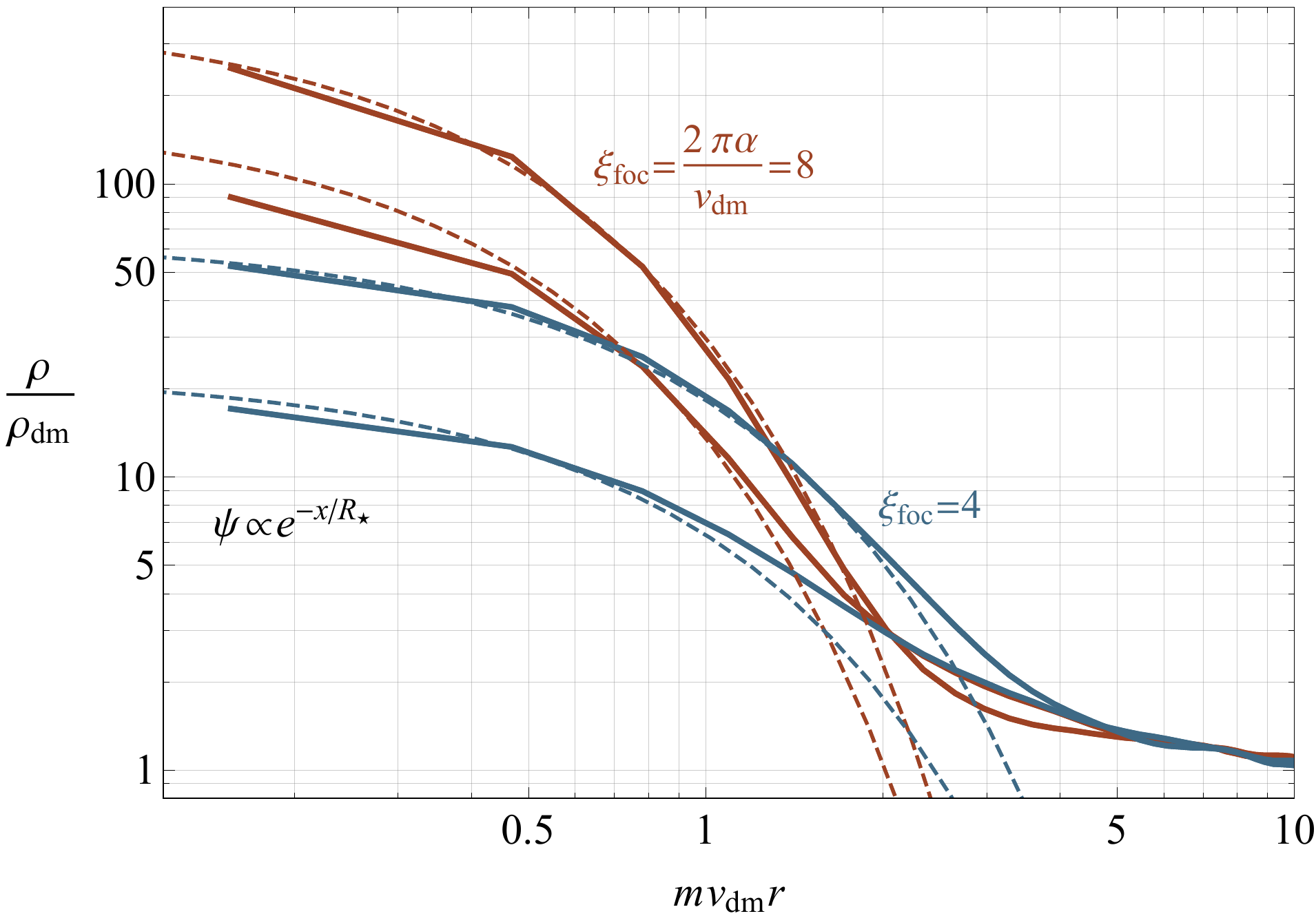}
	\caption{\small{The density profiles of Figure~\ref{fig:rho_sim} for two subsequent times during the exponential increase, corresponding to $t/\tau_{\rm rel}\simeq \{0.83,0.90\}$ and $t/\tau_{\rm rel}\simeq \{1.36,1.50\}$ for $\xi_{\rm foc}=8$ and $\xi_{\rm foc}=4$ respectively. Dashed lines represent the ground-state  profile. 
 }
	}\label{fig:rho_sim_time}
	\end{center}
	\vspace{-3mm}
\end{figure} 

Note that the conditions (a) and (b) are both satisfied with this choice of parameters. As for (c), although $R_s<R_\star$ for all simulations, for the largest values of $2\pi\alpha/v_{\rm dm}$ the ratio $R_s/R_\star$ is not too much smaller than 1 (the worst case is $R_s/R_\star\simeq0.8$, for $2\pi\alpha/v_{\rm dm}=6.5$). However this does not seem to have a significant effect %
on either the exponential-increase time and the gravitational atom profile; see also Figure~\ref{fig:rho_sim}.

\begin{figure}%
\begin{center}
	\includegraphics[width=0.5\textwidth]{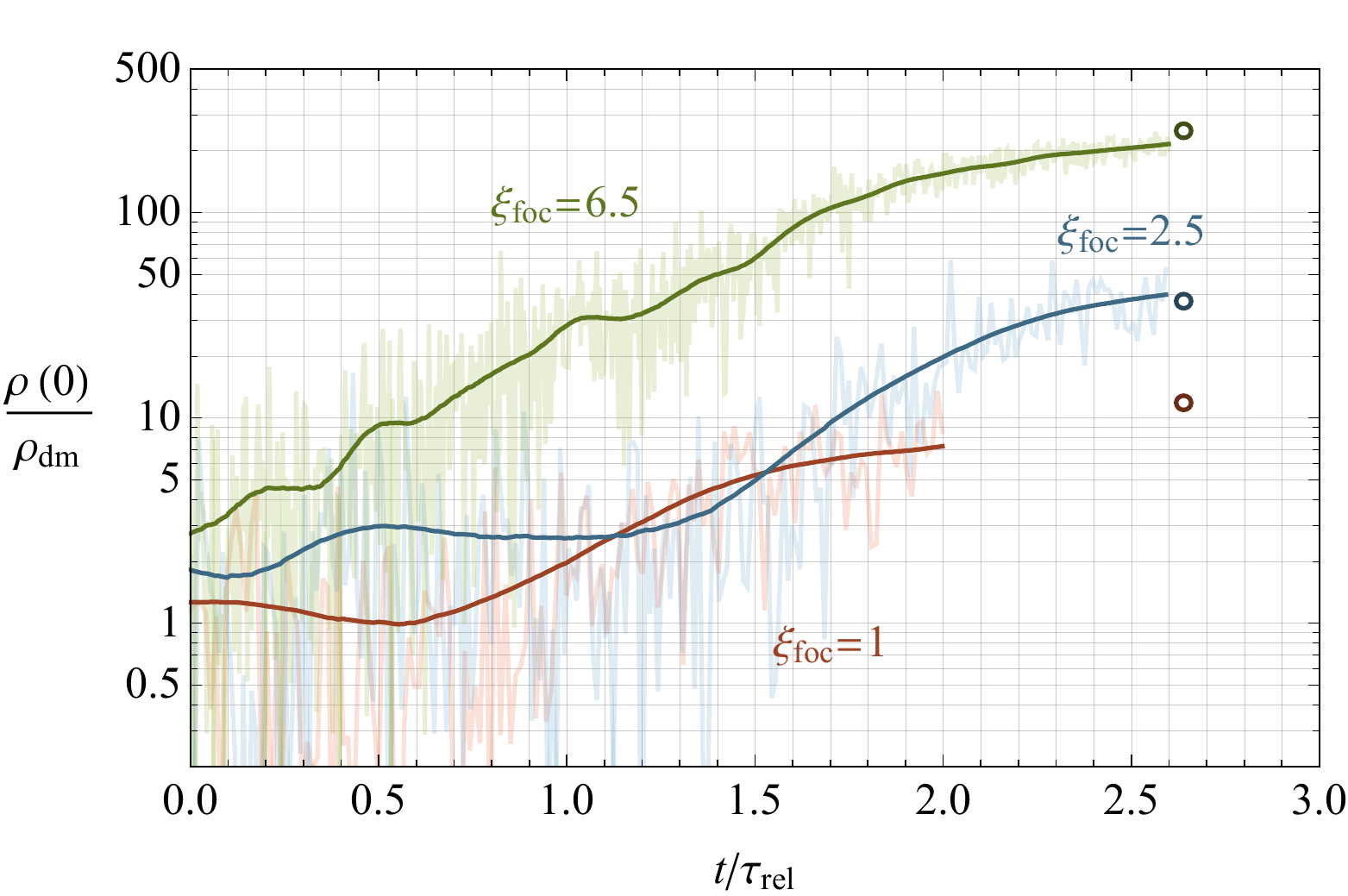}  
	\  \ \  	\includegraphics[width=0.44\textwidth]{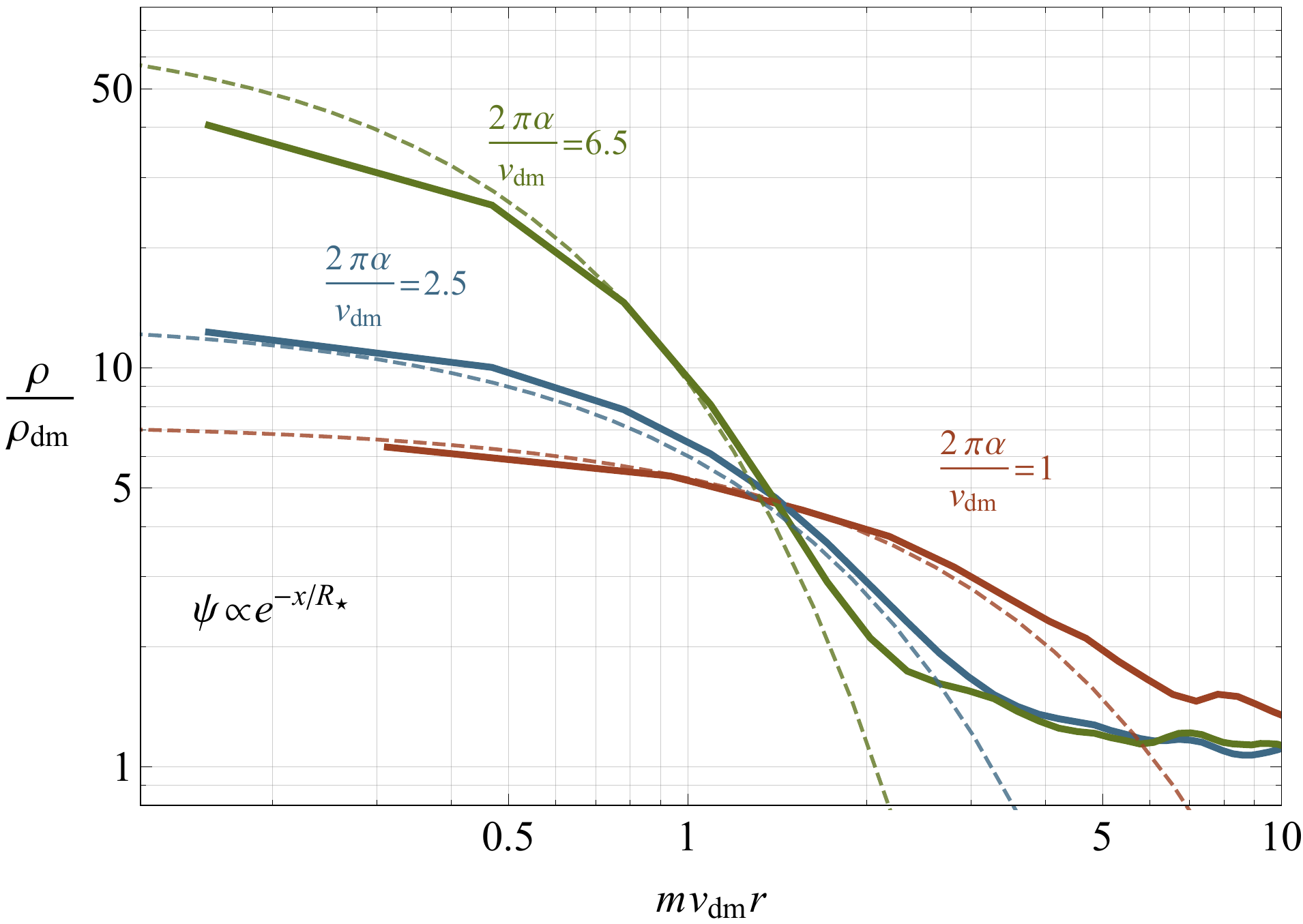}
	\caption{\small{
	\textbf{\emph{Left:}} The time evolution of the dark matter overdensity $\rho(0)/\rho_{\rm dm}$ at the position of the Sun for $\xi_{\rm foc}=2\pi\alpha/v_{\rm dm}=\{6.5,2.5,1\}$ and repulsive self-interactions with $\tilde{g}=0.006\cdot\{1,1,0.5\}$. After an exponential increase, the density tends to saturate at the predicted critical value~$\rho_{\rm crit}$, represented by an empty disk. \textbf{\emph{Right:}} The overdensity profiles during the exponential increase, at times $t/\tau_{\rm rel}\simeq \{1.5,1.7,1\}$; dashed lines represent the ground-state profile of Eq.\,\eqref{eq:psi100}.}
	}\label{fig:rho_sim_small_xi}
	\end{center}
	\vspace{-3mm}
\end{figure}

\vspace{-3mm}
\paragraph{Density profiles and repulsive self-interactions.} 

In Figure~\ref{fig:rho_sim_time} we show the time evolution of the overdensity profile for the simulations of Figure~\ref{fig:rho_sim} in Section~\ref{sec:simulation}, for the two values $\xi_{\rm foc}=2\pi\alpha/v_{\rm dm}=4,8$. This is obtained by taking the spherical average of the field $|\tilde{\psi}|^2$ around the center of the Sun, $r=0$. We show two times during the exponential increase; these times can be extracted from the value of the density in Figure~\ref{fig:rho_sim} (left). Note that %
the profile oscillates in time (especially when the maximum density is not much larger than the average density)%
. Thus, as in Figure~\ref{fig:rho_sim} (right), we plot the profiles averaged over times longer than $1/m v^2_{\rm dm}$, so that the oscillations can average out. From Figure~\ref{fig:rho_sim_time}, it is clear that the profile resembles the ground state during the whole increase, in agreement with our analytic expectations.

Finally, in Figure~\ref{fig:rho_sim_small_xi} we study more closely the system for repulsive self-interactions, $g>0$. We show the overdensity $\rho(0)/\rho_{\rm dm}$ at the position of the Sun as a function of time (left) and the density profiles (right) in three simulations with $2\pi\alpha/v_{\rm dm}=\{6.5,2.5,1\}$ and $\tilde{g}=0.006\cdot\{1,1,0.5\}$.\footnote{As before, $\tilde{L}=40$ and we use $\tilde{R}_s=\{0.3,0.3,0.8\}$ and $\tilde{N}_x=\{128,128,64\}$.} First, the exponential increase of the density is evident, given the logarithmic scale in the $y$-axis. Soon after $t\simeq \tau_{\rm rel}$, the overdensity tends to saturate to the expected critical value $\rho_{\rm crit}/\rho_{\rm dm}=(2\pi\alpha/v_{\rm dm})^2/(2\pi^2\tilde{g})$ in Eq.\,\eqref{eq:rhocrit}, which is shown (multiplied by $0.7$) by an empty disk%
. Additionally, the corresponding density profiles before saturation are well reproduced by the ground state (instead, the profile is deformed after saturation happens, and is flatter). These simulations also show that the exponential increase in the density, i.e. $\Gamma>0$, is likely to occur for values of $2\pi\alpha/v_{\rm dm}$ even as small as 1, which is outside the regime of validity of our analytic calculations.

As mentioned in the main text, we tested numerically values of $|\tilde{g}|$ and $\tilde{R}_s$ that are much larger than those relevant for our Solar System. Thus, these cannot be directly used to justify the formation of the solar halo. However, by performing simulations with various values of $\tilde{g}$ and $\tilde{R}_s$, we checked that the time (relative to $\tau_{\rm rel}$) when the density increases exponentially is independent of %
these parameters, as long as $R_s\lesssim R_\star$%
, which confirms our analytic analysis of Section~\ref{ss:light_vs_heavy}. An example of this is the red curve of Figure~\ref{fig:rho_sim_small_xi},  for which $\tilde{g}$ is half in size than all the other simulations. %
The results of this Appendix and  Section~\ref{sec:analytic} are therefore a convincing evidence of the analytic picture developed in the main text, which can be used in our Solar System for the relevant values of $|{g}|$ and ${R}_s$.

\newpage

\bibliography{references_halo}
\bibliographystyle{JHEP}

\end{document}